Simulation-Based Analytics for Fabrication Quality-Associated Decision Support

by

Wenying Ji

A thesis submitted in partial fulfillment of the requirements for the degree of

Doctor of Philosophy
in
Construction Engineering and Management

Department of Civil and Environmental Engineering
University of Alberta

**ABSTRACT**

Computer-based quality management systems have been widely implemented throughout the construction industry as per the requirement of the International Organization for Standardization (ISO) 9000. Although these systems have facilitated the collection of vast amounts of quality management data, conversion of this data into useable information remains challenging for many practitioners. Automated, data-driven quality management systems, which facilitate the transformation of data into useable information, are often implemented to enhance decision-making processes. However, for a data-driven quality management system to be successful, it must accurately estimate process uncertainty. Integration of accurate, reliable, and straightforward approaches that measure uncertainty of inspection processes are instrumental for the successful implementation of automated, data-driven quality management systems.

This research has addressed these limitations by exploring and adapting Bayesian statistics-based analytical solution and Markov Chain Monte Carlo (MCMC)-based numerical solution for fraction nonconforming posterior distribution derivation purposes. Using these accurate and reliable inputs, this research further develops novel, analytically-based approaches to improve the practical function of traditional pipe welding quality management systems. Multiple descriptive and predictive analytical functionalities are developed to support and augment quality-associated decision-making processes. These include (1) operator quality performance measurement, (2) project quality performance forecast, (3) product complexity measurement, and (4) rework cost estimation and control. Multi-relational databases (e.g., quality management system, engineering design system, and cost management system) from an industrial company in

Edmonton, Canada, are investigated and mapped to implement the proposed novel approaches, and case studies are conducted to demonstrate their feasibility and applicability.

This research has contributed to the academic literature by: (1) providing a novel Bayesian-based approach for fraction nonconforming uncertainty modelling to address hard issues in simulation input model updating; (2) creating an MCMC-based numerical solution for complex probability distribution approximation; (3) developing a dynamic simulation environment that utilizes real-time data to enhance simulation predictability; (4) advancing uncertain data clustering techniques using Hellinger distance-based similarity measurement; (5) providing a systematic approach for analyzing product complexity using the indicator of product quality performance; and (6) creating a novel absorbing Markov chain model for simulating construction product fabrication processes associated with rework.

The industrial contributions of this research are identified as: (1) developing a simulation-based analytics decision-support system to enhance quality-associated decision-support processes; (2) creating reliable and interpretable decision-support metrics for quality performance measurement, complexity analysis, and rework cost management to reduce the data interpretation load of practitioners and to uncover valuable knowledge and information from available data sources; and (3) generating meaningful simulation results to assist practitioners in performing quality and rework cost risk analysis during both the project planning and execution phases of a project.

**PREFACE**

This thesis is an original work by Wenying Ji. This thesis is organized in a paper-based format.

A version of Chapter 2 has been published as Ji, W., and AbouRizk, S.M. (2017). "Credible interval estimation for fraction nonconforming: Analytical and numerical solutions." *Automation in Construction*, 83, 56–67. Dr. AbouRizk was the supervisory authority and was involved with concept formation and manuscript composition.

A version of Chapter 3 has been accepted for publication as Ji, W., and AbouRizk, S.M. "Simulation-based analytics for quality control decision support: A pipe welding case study" by the *Journal of Computing in Civil Engineering* on October 23, 2017. Dr. AbouRizk was the supervisory authority and was involved with concept formation and manuscript composition.

A version of Chapter 4 has been submitted for publication as Ji, W., Li, Y., and AbouRizk, S. M. "Integrated data-driven approach for analyzing pipe welding operator quality performance" to *Advanced Engineering Informatics* on October 27, 2017. Miss Li assisted with data processing and visualization work for the A/B testing analysis. Dr. AbouRizk was the supervisory authority and was involved with concept formation and manuscript composition.

A version of Chapter 5 has been submitted for publication as Ji, W., AbouRizk, S. M., Zaïane, O. R., and Li, Y. "Complexity analysis approach for prefabricated construction products using uncertain data clustering" to the *Journal of Construction Engineering and Management* on September 29, 2017. Dr. AbouRizk was the supervisory authority and was involved with concept formation and manuscript composition. Dr. Zaïane was involved in discussions regarding study methodology development and assisted with manuscript edits. Miss Li assisted with literature review and graph generation.

A version of Chapter 6 has been submitted for publication as Ji, W., and AbouRizk, S. M. "Data-driven simulation model for quality-induced rework cost estimation and control" to the *Journal of Construction Engineering and Management* on November 6, 2017. Dr. AbouRizk was the supervisory authority and was involved with concept formation and manuscript composition.

The research project, of which this thesis is a part, received research ethics approval from the University of Alberta Research Ethics Board, "Implementing data analytics and simulation in understanding, monitoring, and forecasting pipe spool welding quality performance," Pro00069663, approved on November 27, 2016.

## ACKNOWLEDGEMENTS

To my supervisor: Dr. Simaan M. AbouRizk.

To my committee members: Dr. Osmar R. Zaïane, Dr. Aminah Robinson Fayek, Dr. Yasser Mohamed, Dr. Wei Victor Liu, and Dr. Baabak Ashuri.

To my colleagues: Dr. Catherine Pretzlaw, Mr. Stephen Hague, and Ms. Brenda Penner.

To my beloved wife: Ms. Jing Liu.

To my parents: Mr. Zifu Ji and Ms. Jianfeng Fu.

**TABLE OF CONTENT**

# LIST OF TABLES

# LIST OF FIGURES

# LIST OF ABBREVIATIONS

| Abbreviation | Meaning |
| --- | --- |
| ACF | Autocorrelation function |
| BIM | Building information modelling |
| BW | Butt weld |
| CL | Center line of a control chart |
| Cplx | Complexity |
| DDDAS | Dynamic data-driven application system |
| DLL | Dynamic-link library |
| DMAIC | Define, Measure, Analyze, Improve, and Control |
| ERP | Enterprise resource planning |
| GUI | Graphical User Interface |
| ISO | International Organization for Standardization |
| LCL | Lower control limit |
| MAE | Mean Absolute Error |
| MCMC | Markov Chain Monte Carlo |
| NDE | Nondestructive examination |
| NPS | Nominal pipe size |
| ODBC | Open Database Connectivity |
| Q1 | 25% quantile |
| Q3 | 75% quantile |
| RAM | Random-access memory |
| RMSE | Root Mean Square Error |
| RODBC | R package for Open Database Connectivity |
| RT | Radiographic test |
| SQL | Structured Query Language |
| STD | Standard |
| UCL | Upper control limit |

# 1    CHAPTER 1: INTRODUCTION

## 1.1    Problem Statement

The International Organization for Standardization (ISO) has developed a series of quality management standards (ISO 9000), issued in 1987, which have been implemented throughout the world to increase consistency and to ensure minimum quality standards in construction (Chini and Valdez 2003). According to ISO 9000, quality is defined as "the degree to which a set of inherent characteristics fulfills requirements" (Hoyle 2001). In the construction industry, quality performance is listed as a key performance indicator essential for successful project delivery (Bassioni et al. 2004; Chan et al. 2004). From a practical perspective, poor quality performance often leads to penalties, cost and schedule overruns, and productivity loss (Battikha 2002). Consequently, unsatisfactory quality performance can negatively impact companies' reputations and market competitiveness (Jaafari 2000; Yates and Aniftos 1997).

As the concept of lean construction becomes increasingly implemented, a greater number of construction components are being standardized and manufactured in fabrication shops (Salem et al. 2006). Examples include pipe spools for oil refinery plants (Wang et al. 2009) and wall panels for residential buildings (Shewchuk and Guo 2012). Prefabrication of construction products allows for the implementation of more rigorous quality control processes. Specifically, many components are inspected as either conforming or nonconforming to specified quality standards (Montgomery 2007). For example, a failure of nondestructive examination (NDE) for a pipe weld is recorded as nonconforming and requires rework. In practice, construction companies use the indicator of project fraction nonconforming (i.e., the number of nonconforming items over the number of inspected items) to measure their fabrication quality performance. However, uncertainties, such as a vast number of designs and variable operator quality performance, make this measurement more complex. Improper management of the quality-associated processes negatively impacts the overall project performance.

As required by ISO 9000, computer-based quality management systems have been widely implemented for quality management purposes within the construction industry (Battikha 2002; Chin et al. 2004). For this purpose, the construction industry relies primarily on commercial Enterprise Resource Planning (ERP) systems, quality management software, or in-house computer solutions. Although these systems have facilitated the collection of vast amounts of quality management data, conversion of these data into usable information remains challenging for many practitioners (Dean 2014). Other types of information, such as engineering design and cost management information, are typically stored in companies' isolated management systems and are not efficiently utilized for quality-associated decision support.

Therefore, a novel, automated, data-driven decision-support system is needed to improve quality-associated decision-making processes. For the new generation decision support system to be successful, it must:

- Source data from multiple sources;
- Incorporate advanced analytical techniques; and
- Generate reliable and accurate decision-support metrics.

By fusing multiple data sources, the following quality-associated challenges are expected to be addressed through the developed decision-support system in this research:

- The quantitative measurement of nonconforming quality performance uncertainty, which can combine previous knowledge with observed real-time data;

- Data-driven nonconforming quality performance prediction models at product-, project-, and operator-levels;

- Construction product complexity assessment through the indicator of quality performance; and

- Quality-induced fabrication rework cost estimation and control through real-time updated quality and cost performance data.

## 1.2 Research Objectives

The overall goal of this research is to develop a novel, analytics-based decision support system to enhance quality-associated practices of the industrial construction sector. This research intends to achieve the following objectives:

**Objective 1.** Measure fraction nonconforming performance through the integration of historical and real-time information to improve the way fraction nonconforming is determined in current practice. The following activities will be undertaken to achieve this objective:

- Create novel models, capable of both dynamically incorporating real-time data and uncertainty, to measure fraction nonconforming quality performance.

- Create data-driven, fraction nonconforming prediction models for product-, project-, and operator-level forecasting.

**Objective 2.** Develop novel analytical approaches, based on dynamic simulations, for augmenting quality-associated decision support. The following activities will be undertaken to achieve this objective:

- Develop a dynamic simulation environment, which is capable of sourcing multi-relational databases and updating simulation input models with real-time data, to enhance the predictability of simulation models.

- Develop novel analytical models to generate meaningful decision-support metrics for improving quality-associated practices.

## 1.3  Research Methodology

This research develops a simulation-based analytics framework to incorporate analytically-based simulation approaches for enhancing quality-associated performance of construction projects. The proposed framework was derived from state-of-the-art concepts related to data analytics applications (LaValle et al. 2011), dynamic data-driven application systems (Darema 2004), and simulation-based analytics (Dube et al. 2014). The framework makes use of a variety of analytical methods, including data mining, to identify anomalies or missing information, and simulation methods, to populate and fill gaps or to validate data. Algorithms and models that facilitate analytics use the transformed data together with simulation models to generate desired metrics for a given decision-support application within the analysis framework.

At the core of the proposed framework is the dynamic, data-driven application system (DDDAS) concept (Darema 2004). DDDAS refers to a paradigm that strives to seamlessly couple the simulation world with measurements of a real system in real time. This concept facilitates the dynamic addition of new data into simulation models to enhance the accuracy and predictability of the original models at execution. Many of the improvements to the original model are a result of automated internal calibrations; data used for this purpose are typically archival, real-time generated, or from online measurements of actual systems (e.g., sensors, detectors, etc.).

While simulation applications in construction are mature, DDDAS offers potential advancement to the state-of-the-art. The prevalent approach in construction simulation is to use statistical distributions and probabilistic methods (e.g., Markov chain) to model uncertainties. Updating of the input models to realign with new data in a dynamic manner and in real-time represents significant challenges—which this research resolves by building upon the works of DDDAS and Bayesian techniques to update input models (e.g., statistical distributions and probabilities) with new data to provide more accurate predictions.

Here, the concept of simulation-based analytics is developed for achieving the aforementioned objectives using existing quality management, engineering design, and cost management data. The specialized framework, summarized in Figure 1.1, is comprised of five components, namely the data source, data adapter, data analysis module, simulation module, and decision support module. Detailed description and development information for these modules are introduced in Chapter 3, Section 3.3. Notably, this simulation-based analytics system can be generalized and implemented at companies through simple modifications of the data adapter as per the companies' data structure.

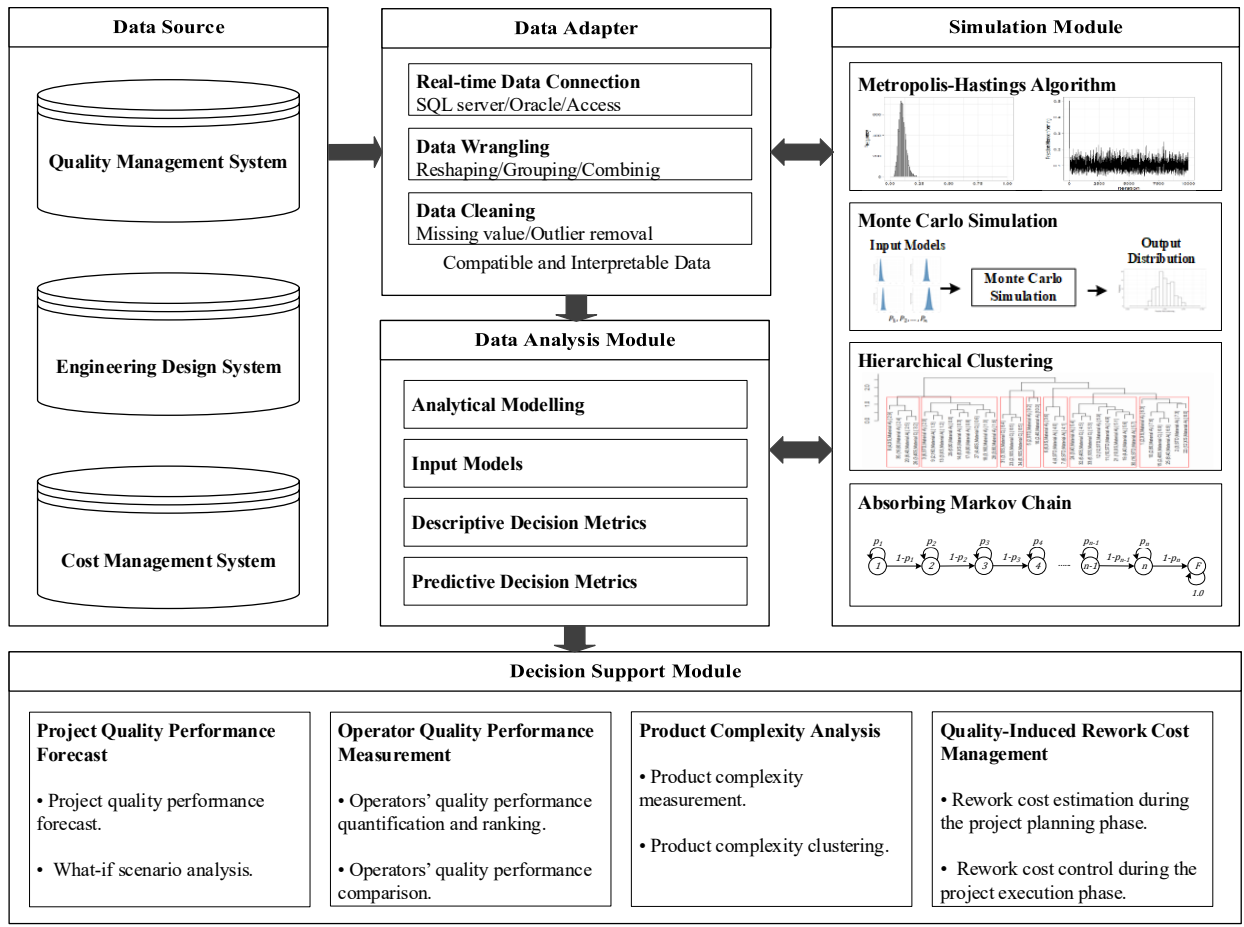

Figure 1.1: Research framework.

R ([https://www.r-project.org](https://www.r-project.org)), a free programming software for statistical computing and graphics, is utilized to drive all components of the entire framework. In addition to basic functionalities, such as data manipulation, statistical calculation, and graphical display, more than 10,000 packages are available on CRAN ([https://cran.r-project.org](https://cran.r-project.org)) for advanced data analytical tasks, such as data mining, machine learning, and simulation. These packages are a collection of R functions that makes it easy to immediately get access to the latest data mining, machine learning, and simulation techniques. R allows users to perform hybrid data analysis processes, such as the transformation of real-time data, the creation of analytical models, and the generation of decision support metrics, under one integrated environment.

## 1.4    Thesis Organization

This thesis is organized following a paper-based format that is consistent with the research framework shown as Figure 1.1. Chapter 2 develops solutions for the dynamically updated simulation inputs, which can be utilized to improve the predictability of the further developed decision-support system. Chapter 3 discusses the development of the specialized simulation-based analytics system for quality-associated decision support. Chapters 4 through 6 develop novel analytical models and meaningful decision-support metrics for achieving the functionalities (i.e., operator quality performance analysis, product complexity analysis, and rework cost management) of the proposed decision-support system. Detailed contents of each chapter are listed as follows.

**Chapter 2:** Credible interval estimation for fraction nonconforming: analytical and numerical solutions.

- Models the nonconforming quality inspection process as a stochastic process.
- Develops Bayesian-based analytical and Markov Chain Monte Carlo (MCMC)-based numerical solutions for deriving posterior distributions and credible intervals of fraction nonconforming for quality performance uncertainty measurement.
- Demonstrates the feasibility and applicability of proposed solutions with a practical case study.

**Chapter 3:** Simulation-based analytics for quality control decision support: a pipe welding case study.

- Creates a dynamic, analytically-based environment that feeds real-time multi-relational data into advanced analytical systems.

- Develops a set of quantitative methods to model, infer, and forecast pipe welding quality control processes.

- Generates accurate and reliable descriptive and predictive decision-support metrics.

**Chapter 4:** Integrated data-driven approach for analyzing pipe welding operator quality performance.

- Fuses and transforms data from separate data sources (i.e., quality management system and engineering design system) into an interpretable dataset.

- Implements an MCMC-based approach to numerically estimate posterior distributions of operators' welding quality performance.

- Utilizes an A/B testing algorithm to compute probabilistic differences between operators' quality performance.

- Proposes potential applications to comprehensively improve pipe welding quality performance for practitioners.

**Chapter 5:** Complexity analysis approach for prefabricated construction products using uncertain data clustering.

- Proposes accurate and reliable measurements of product complexity uncertainty using the product quality performance indicator.

- Derives meaningful assessments of product complexity distribution similarity using the Hellinger distance.

- Generates a reliable and interpretable clustering of products with similar complexity.

**Chapter 6:** Data-driven simulation model for quality-induced fabrication rework cost estimation and control.

- Creates an absorbing Markov chain-based analytical model to perform direct rework cost (e.g., man-hours) estimation and control.
- Generates meaningful and reliable decision support metrics for enhanced decision-making processes.
- Utilizes the previously developed simulation-based analytics framework as the simulation environment to achieve the practical application.

**Chapter 7:** A summary of the research contributions, limitations, and envisioned future work.

# 2 CHAPTER 2: CREDIBLE INTERVAL ESTIMATION FOR FRACTION NONCONFORMING: ANALYTICAL AND NUMERICAL SOLUTIONS[1]

## 2.1 Introduction

Sampling uncertainty must be considered during estimation of a true population variable when data are obtained from a sample rather than an entire population (Weiss 2012). A common tool used to assess uncertainty are interval estimations, which are applied to estimate the margin of sampling error (Casella and Berger 2002). Of the several types of interval estimations, confidence intervals, which are commonly introduced in statistics textbooks, have been widely applied in statistical process control. However, several researchers have outlined the disadvantages of confidence intervals and have contended that confidence intervals are not well-suited to address the needs of scientific research (Morey et al. 2016). Accordingly, due to their straightforwardness (Casella and Berger 2002) and reliability (Gelman et al. 2003), researchers are now advocating for the use of Bayesian credible intervals rather than conventional confidence intervals. In contrast to confidence intervals, an observer can combine previous knowledge with observed data to estimate parameters of interest when using Bayesian statistics (Berger 1985; Gelman et al. 2003). In a Bayesian treatment, prior distributions of the parameters are introduced and posterior distributions are computed, based on Bayes' theorem, from observed data (Bishop 2007). After obtaining posterior distributions, uncertainty can be quantified by providing certain tail quantiles of the posterior distribution (Weaver and Hamada 2016). For example, a 95% credible interval can be specified by the 0.025 and 0.975 quantiles of the posterior distribution.

Calculation of sampling uncertainty in quality management systems is further complicated for quality characteristics that cannot be appropriately represented numerically. Often, quality characteristics are assessed as either conforming or nonconforming to specified quality standards. In contrast to data that is represented numerically, sampling uncertainty must instead be assessed from the fraction nonconforming, defined as the ratio of nonconforming items in a population to the total items in that population (Montgomery 2007). To appropriately incorporate uncertainty, it is necessary to obtain a range of values that cover the true population fraction nonconforming (Nicholson 1985). As is common for statistical processes, this range should be wider for unfamiliar items and narrower for familiar items.

The aim of the present study is to introduce a credible interval estimation approach for fraction nonconforming by providing two alternative types of solutions, namely analytical and numerical, to more effectively incorporate uncertainty in fraction nonconforming inferences. The content of this chapter is organized as follows: An overview of the research workflow is provided in the methodology section, and the research methodology is detailed in the following sections. First, the statistical principles underlying fraction nonconforming for mathematically modelling nonconforming quality control processes are discussed. Then, a detailed introduction to credible interval and Bayesian inference is provided. Afterwards, a Bayesian statistics-based analytical solution and an MCMC method-based numerical solution for determining credible intervals and posterior distributions for fraction nonconforming are introduced. To elaborate on the implementation of the proposed solutions, an illustrative example of each solution is provided. Finally, the feasibility, applicability, and consistency of the two proposed solution types are demonstrated following a practical case study of industrial pipe welding quality management. In addition to providing insights for the improvement of uncertainty estimation in automated data-driven quality management systems, findings of this study will also provide valuable insights on the use of MCMC methods to determine posterior distributions for complex variables.

## 2.2   Research Methodology

The research methodology of this study is illustrated in Figure 2.1. First, the problem was abstracted into a mathematical model using a Bernoulli process—an established model from the area of statistical quality control—to estimate the fraction nonconforming (Montgomery 2007). Second, to demonstrate the advantages of implementing Bayesian statistics for incorporating uncertainty in fraction nonconforming estimation, the theoretical background of credible interval estimation and Bayesian inference were thoroughly investigated. From this, it was determined that a credible interval has a more intuitive interpretation than a classic confidence interval when estimating the unknown fraction nonconforming. The results also demonstrated that Bayesian statistics were capable of recalibrating existing statistical distributions with newly updated data. To determine a non-informative prior distribution for fraction nonconforming estimation, selection of the prior distribution was then investigated. Finally, a Bayesian statistics-based analytical solution and an MCMC-based numerical solution were developed to derive the posterior distribution of fraction nonconforming. To reveal how the inherent mathematical mechanism functions, a step-by-step proof with a calculation example was conducted for the analytical solution; a specialized Metropolis-Hastings algorithm and an illustrative simulation example were provided for the numerical solution. Advantages and disadvantages of each method were discussed. Then, the feasibility and applicability of the proposed solutions were evaluated following their application to an industrial case study. Details of the systematic and theoretical analysis of these research steps are detailed as follows.

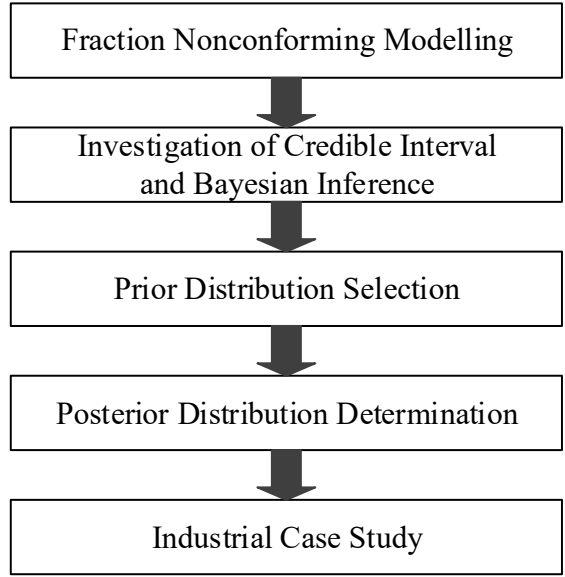

Figure 2.1: Research methodology flow chart.

## 2.3   Fraction Nonconforming Modelling

In the nonconforming quality inspection process, the desired outcome is usually referred to as "success" and the alternative outcome is often referred to as "failure." When an item fails, it must be repaired and inspected until it passes inspection. The inspection outcome $O$ can be treated as a Bernoulli random variable with probability function (Montgomery 2007):

$$P(O) = \begin{cases} p & x = 1 \\ (1-p) = q & x = 0 \end{cases} \tag{2.1}$$

Variable $O$ takes on a value of 1 with probability $p$ and the value 0 with probability $(1-p) = q$. A realization of this random variable is called a Bernoulli trial. The sequence of Bernoulli trials is a Bernoulli process. The number of failed inspections $X$ has a binomial distribution $B(n, \ p)$.

In statistical quality control processes, the fraction nonconforming of the sample is defined as the ratio of the number $X$ of nonconforming items in the sample to the sample size $n$ as Eq. (2.2) (Montgomery 2007).

$$\hat{p} = \frac{X}{n} \qquad (2.2)$$

$\hat{p}$ is a point estimate of the true, unknown value of the binomial variable $p$, which represents the fraction nonconforming of the sampled items. The mean of $\hat{p}$ can be calculated as Eq. (2.3).

$$\mu_{\hat{p}} = p \qquad (2.3)$$

## 2.4    Credible Interval and Bayesian Inference

In statistics, interval estimation is generally defined as the use of sample data to calculate an interval of possible (or probable) values of an unknown population variable (Casella and Berger 2002). Confidence intervals and credible intervals are the most widespread forms of interval estimations. In general, both confidence intervals and credible intervals can be defined for a variable X as $P\{l \leq X \leq u\} = 100(1 - \alpha)\%$. Where l is the lower interval limit, u is the upper interval limit, and $(1 - \alpha)$ is the level of confidence ($\alpha$ is the significance level). However, the interpretation for confidence intervals and credible intervals is conceptually different.

Before introducing the concept of the credible interval, the drawbacks of the confidence interval will be discussed. Generally, a confidence interval is a range of values designed to include the true value of the variable with a tolerance probability of $100(1 - \alpha)\%$ . As the number of failed inspections has a binomial distribution, only confidence intervals for binomial distributions will be discussed here. The Wald's interval, Wilson interval, and Agresti-Coull interval are classical methods for setting confidence intervals for binomial distributions (Brown et al. 2001). Their analytical equations are listed in Table 2.1.

Table 2.1: Classical confidence intervals for binomial distribution.

| Confidence Interval | Formula |
|---|---|
| Wald Interval | $\hat{p} \pm Z_{\alpha/2} \sqrt{\dfrac{\hat{p}(1-\hat{p})}{n}}$ |
| Wilson Interval | $\left( \hat{p} + \dfrac{Z_{\alpha/2}^2}{2n} \pm Z_{\alpha/2} \sqrt{\dfrac{1}{n}\left[ \hat{p}(1-\hat{p}) + \dfrac{Z_{\alpha/2}^2}{4n} \right]} \right) \Big/ \left( 1 + \dfrac{Z_{\alpha/2}^2}{n} \right)$ |
| Agresti-Coull Interval | $\hat{p} \pm Z_{\alpha/2} \sqrt{\dfrac{\hat{p}(1-\hat{p})}{n + Z_{\alpha/2}^2}}$ |

From the confidence interval equations listed in Table 2.1, it is evident that interval endpoints in these intervals depend only on collected data (i.e., the fraction nonconforming $\hat{p}$ and the sample size $n$). However, when the fraction nonconforming $\hat{p}$ is close to zero and sample size $n$ is small, these confidence intervals lose inference accuracy and reliability (Morey et al. 2016). For example, the 95% Wald interval for a new product that has failed inspection one time out of three is (-0.2110, 0.8743), where it is impossible to have a negative lower boundary for the obtained confidence interval. Furthermore, historical data of similar products may indicate that the confidence interval is too large to accurately infer the true fraction nonconforming for the given products. The product real fraction nonconforming should lie within a tighter interval.

Conversely, Bayesian statistics defines the problem philosophically in a different manner. A Bayesian method assumes the variable's value is fixed and has been chosen from an existing probability distribution known as the prior distribution. The Bayesian method incorporates both prior information that is representative of an estimator's belief about the variable before any observation is made, termed the prior distribution, as well as the updated belief about the variable after observation, termed the posterior distribution. Given the estimator's belief, the estimated interval should be considerably tighter than the calculated confidence intervals. In other words, when prior information is taken into consideration, the estimated interval should be more accurate and reliable.

Bayesian statistics is a systematic way of updating information of interest as more observations become available (Gelman et al. 2003). Data are collected and utilized to calculate the probability of different values of the variable based on current data and existing information. This new probability distribution is called the posterior distribution. Subsequently, the uncertainty of the variable can be summarized by providing a range of values based on the posterior distribution that includes $100(1-\alpha)\%$ of the probability. This range is called a $100(1-\alpha)\%$ credible interval. As discussed, the credible interval serves as a summary of the posterior distribution. Its interpretation, therefore, is considerably more meaningful than that of a confidence interval. Also, once the posterior distribution has been generated, it has the advantage of deriving additional statistics such as mean, median, variance, and all quantiles, which can be used during the decision-making process to obtain solutions more directly and intuitively.

## 2.5    The Prior Distribution

In Bayesian statistics, a prior distribution (short for prior) of an uncertain variable is the probability distribution that expresses the estimator's beliefs about this parameter prior to the consideration of any evidence. To determine the prior distribution, it is common to use beta distributions as the standard conjugate priors for inferring variable $p$ in a binomial distribution (Berger 1985). The primary reasons for choosing beta distributions are: (1) the variable fraction nonconforming should be bounded within the range of 0 to 1; (2) beta distributions have the flexibility to provide accurate and representative outputs; and (3) the parameters of beta distributions are intuitively and physically meaningful and easy to estimate from the data.

Suppose the number of failed inspections $X \sim B(n, p)$ and the fraction nonconforming $p$ has a prior distribution $Beta(a, b)$, given by

$$P(p) = Beta(p|a,b) = \frac{\Gamma(a+b)}{\Gamma(a)\Gamma(b)} p^{a-1}(1-p)^{b-1} \qquad (2.4)$$

The parameters $a$ and $b$ are two positive shape parameters that control the distribution shape of the fraction nonconforming $p$. $\Gamma(z)$ is the gamma function. Figure 2.2 depicts six beta distributions with different combinations of shape parameters. These examples are all bounded in the range of 0 to 1, with their diverse shapes catering to different outputs.

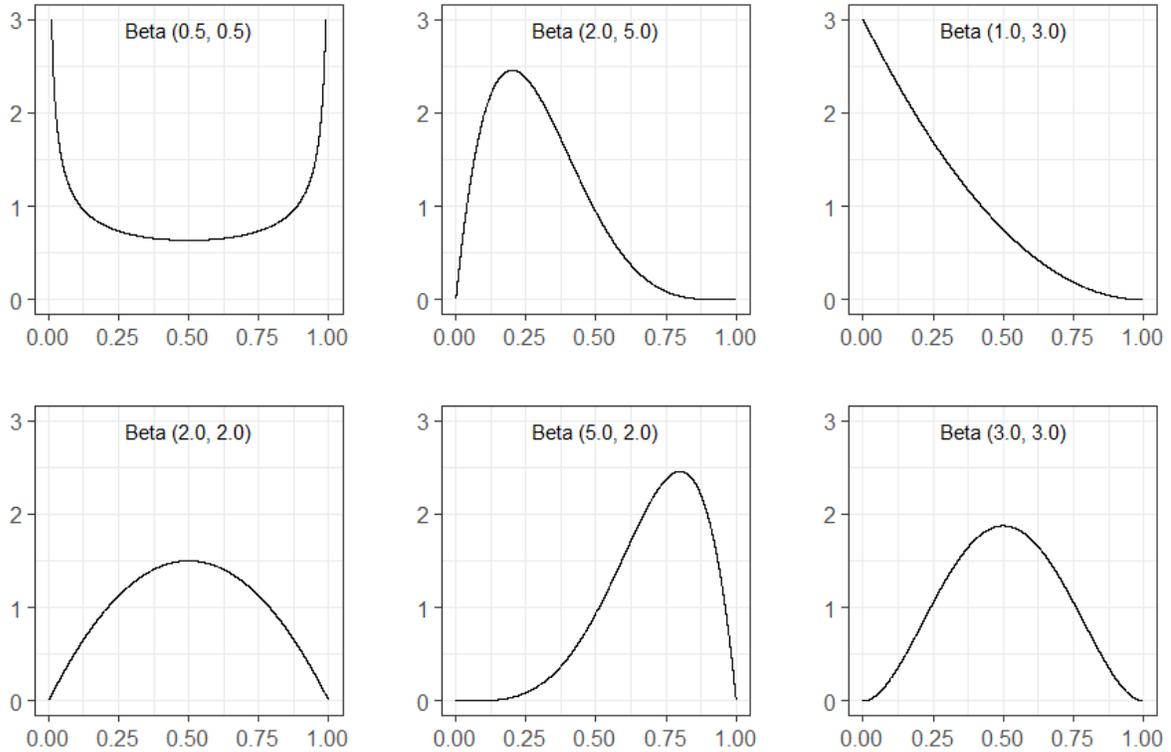

Figure 2.2: Examples for beta distribution.

In general, Bayesian priors can be categorized into informative priors and non-informative priors. An informative prior expresses specific, definite information about a parameter. From the population perspective, the prior distribution can be interpreted as a population of possible variable values. From the subjective knowledge perspective, the guiding principle is that knowledge and uncertainty about the variable must be considered as if its value was randomly chosen from the prior distribution. A non-informative prior expresses vague, flat, and diffuse information. When priors have no population basis, they can be difficult to construct. Identification of priors that are guaranteed to play a minimal role in posterior distributions, therefore, is desired. The reason for using non-informative priors is to remove the effects of external information on current data. For the binomial proportion $p$, the non-informative prior $Beta(1/2,1/2)$ is commonly used in credible interval estimation (Berger 1985). The determination of prior distribution is beyond the scope of this chapter, but it is vital, particularly when the observed data sample is small. In this chapter, the non-informative prior $Beta(1/2,1/2)$ is used to conduct the calculations for demonstration purposes. In practice, the prior distribution can be determined based on historical data, professional experience, and existing knowledge.

## 2.6   The Posterior Distribution

Bayesian inference derives the posterior distribution as a combination of a prior distribution and a likelihood function (Gelman et al. 2003). In this research, the parameter of interest is the fraction nonconforming $p$. The prior distribution of $p$ is $P(p)$, which summarizes what is known about $p$ before the data is observed. The likelihood function $L(X|p)$ provides the distribution of the data, given the fraction nonconforming $p$. $P(X)$ is the marginal distribution of the data $x$. The posterior distribution $P(p|X)$ indicates information in data $x$ together with information expressed in the prior distribution. Based on Bayes' Theorem, the posterior distribution $P(p|X)$ can be obtained as Eq. (2.5).

$$P(p|X) = \frac{L(X|p) \times P(p)}{P(X)} \propto L(X|p)\, P(p) \tag{2.5}$$

In general, analytical solutions for $P(p|X)$ may, or may not, exist. An analytical solution is always preferred, as it has the closed-form equation for posterior distributions. However, when the analytical solution is too complex to derive or does not exist, a numerical solution can instead be used to approximate the true value of interest. Therefore, this chapter will provide both the analytical and numerical solutions for deriving the posterior distributions of the fraction nonconforming $p$. After obtaining posterior distributions, uncertainty can be quantified by certain tail quantiles of the posterior distribution (e.g. credible intervals). For the purpose of estimating credible intervals for fraction nonconforming, the analytical solution is straightforward and accurate. In addition to having no discernable effect on accuracy, the numerical solution offers an alternative approach to solve similar problems and provides insight regarding the implementation of an MCMC method for statistical quality control problems.

### 2.6.1 Analytical Solution

In mathematics, an analytical solution is any formula that can be evaluated in a finite number of standard operations. It is the exact solution derived by a series of logical steps that can be proved correct (Borwein and Crandall 2013). An analytical solution will assist with the understanding of the mechanism behind the modelled problem. In this section, a step-by-step analytical proof of the credible interval estimation is given. The proof steps demonstrate how prior distribution and posterior distribution are mathematically related.

***Step-by-step Proof***

As per the Eq. (2.5), keeping the factors that only depend on $p$, the prior distribution has the form

$$P(p) \propto p^{a-1}(1-p)^{b-1} \tag{2.6}$$

And, similarly, the likelihood function has the form

$$L(X|p) \propto p^{X}(1-p)^{n-X} \tag{2.7}$$

Therefore, the posterior distribution of $p$ can be obtained by multiplying the prior distribution by the likelihood function. Keeping only the factors dependent on $p$, the posterior distribution has the form

$$P(p|X) \propto L(X|p) \, P(p) \propto p^{X+a-1}(1-p)^{n-X+b-1} \tag{2.8}$$

Indeed, the posterior distribution is another beta distribution, and is given by

$$P(p|X) = Beta(X+a, n-X+b)$$
$$= \frac{\Gamma(X+a+n-X+b)}{\Gamma(X+a)\Gamma(n-X+b)} p^{X+a-1}(1-p)^{n-X+b-1} \tag{2.9}$$

Furthermore, the posterior mean is a weighted average of the maximum likelihood estimation and the prior mean and can be calculated as

$$\mu = \frac{X+a}{n+a+b} = \left(\frac{n}{n+a+b}\right)\left(\frac{X}{n}\right) + \left(\frac{a+b}{n+a+b}\right)\left(\frac{a}{a+b}\right) \tag{2.10}$$

Therefore, a $100(1-\alpha)\%$ equal-tailed Bayesian interval is given by Eq. (2.11),

$$[l, u] = [Beta(\alpha/2 \, ; X+a, n-X+b), Beta(1-\alpha/2 \, ; X+a, n-X+b)] \tag{2.11}$$

where $Beta(\alpha; a, b\,)$ denotes the $\alpha$ quantile of a $Beta(a, b)$ distribution. This interval leaves $\alpha/2$ posterior probability in each omitted tail. Indeed, if a beta distribution is used as the prior, then the posterior distribution has a closed-form expression. The posterior distribution depends on both the prior distribution and the data. As the amount of data becomes large, the posterior increasingly approximates the maximum likelihood estimation.

In this chapter, the prior distribution is assumed to be non-informative and is defined as $Beta(1/2, 1/2)$. Therefore, after observing $X$ successes in $n$ trials given the non-informative prior distribution $Beta(1/2, 1/2)$, the posterior distribution for fraction nonconforming $p$ is a beta distribution

$$P(p|X) = Beta(X + 1/2, n - X + 1/2) \tag{2.12}$$

The $100(1 - \alpha)\%$ equal-tailed credible interval is defined as Eq.(2.13).

$$[l, u] = [Beta(\alpha/2; X + 1/2, n - X + 1/2), Beta(1 - \alpha/2; X + 1/2, n - X + 1/2)] \tag{2.13}$$

This credible interval is called the Jefferys Interval (Berger 1985).

***Calculation Example***

For demonstrating the analytical method, a calculation example for $X \sim B(100, 0.1)$ is provided. The lower and upper limits for the credible interval ($\alpha = 5\%$) can be calculated as Eq. (2.14).

$$[l, u] = [Beta(0.05/2; 10 + 0.5, 100 - 10 + 0.5), Beta(1 - 0.05/2; 10 + 0.5, 100$$
$$- 10 + 0.5)] = [Beta(0.025; 10.5, 90.5), Beta(0.975; 10.5, 90.5)] \tag{2.14}$$
$$= [0.0526, 0.1701]$$

This result indicates that the 95% credible interval for 10 non-conformers out of 100 items is $[0.0526, 0.1701]$. The fraction nonconforming is theoretically distributed as $Beta(10.5, 90.5)$. Figure 2.3 shows the posterior distribution and the 95% credible interval for $X \sim B(100, 0.1)$. The posterior distribution is bound between 0 and 1 and is right-skewed. The two tails of each side are not symmetrically distributed.

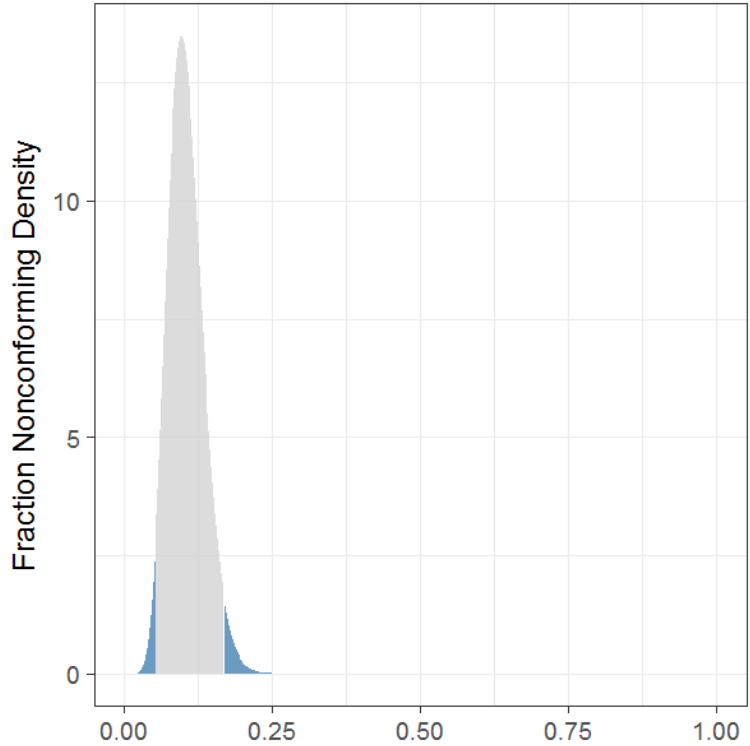

Figure 2.3: Posterior distribution and credible interval ($\alpha = 5\%$) of $X \sim B(100,\ 0.1)$.

### 2.6.2 Numerical Solution

In the previous section, the posterior distribution of the fraction nonconforming has a closed-form solution, which is $Beta(X + a, n - X + b)$. However, in most cases, the closed-form solution is difficult to derive or does not exist. In such cases, the numerical method can instead be used to approximate target distributions. Convergence of the numerical solution lies on the algorithm design and the large number of iterations. In this section, the specialized Metropolis-Hastings algorithm will be designed to approximate the posterior distribution and credible interval for fraction nonconforming.

### *Metropolis-Hastings Algorithm*

The Metropolis–Hastings algorithm is a MCMC method for generating random samples from a probability distribution. It was developed by Metropolis et al. (1953) and subsequently generalized by Hastings (1970). The utilization of this algorithm was primarily limited to the field of physics until the 1990s, when it was exposed to statisticians by Moller and Waagepetersen (2003) and Tierney (1994). In the last decade, the Metropolis-Hastings algorithm has become one of the most popular statistical techniques for distribution approximation (i.e., to generate a histogram) and to compute integrals (Hitchcock 2003; Robert and Casella 2011). Compared to other sampling techniques (e.g. Gibbs sampling), the Metropolis-Hastings algorithm achieves a better numerical approximation in terms of accuracy.

The Metropolis-Hastings algorithm constructs a Markov chain of fraction nonconforming values for $\{p^{(1)}, p^{(2)}, p^{(3)}, ..., p^{(N)}\}$. The next value $p^{(i+1)}$ is chosen by proposing a random move that is conditional on the previous value $p^{(i)}$ and on the ratio of $\frac{P(p^*|X)}{P(p^{(i)}|X)}$, where $p^*$ is a candidate sample from the proposal distribution. This acceptance ratio indicates how probable the new candidate sample is with respect to the current sample. The move is accepted if the new sample is more probable than the existing sample. Otherwise, the move is accepted with the acceptance probability or alternatively, the move is rejected. Given that these conditions are met, the Markov chain of parameter values will remain in the high-density region and will converge to the target distribution $P(p|X)$. As the sampling effort is concentrated in the area with higher posterior density, the time required for obtaining an acceptable convergence is typically reduced compared to other sampling techniques.

The logic of the Metropolis-Hastings algorithm is demonstrated in a step-by-step algorithmic form with the initial value $p^{(0)}$ and repeat for $i = 1, 2, 3, \ldots, N$.

*Step 1.* Choose a new proposed value $p^*$ such that $p^* = p^{(i)} + \Delta p$ , where $\Delta p \sim N(0, \sigma)$.

*Step 2.* Calculate the ratio $\rho = \min\{1, \frac{P(p^*|X)}{P(p^{(i)}|X)}\}$, where $P(p|X)$ is the posterior distribution.

As discussed in the analytical solution, the posterior distribution has the form

$$P(p|X) \propto L(X|p)\, P(p) \propto p^{X+a-1}(1-p)^{n-X+b-1}$$

Therefore, the ratio $\rho$ can be calculated as

$$\rho = \min\{1, \frac{p^{*X+a-1}(1-p^*)^{n-X+b-1}}{p^{(i)X+a-1}(1-p^{(i)})^{n-X+b-1}}\}$$

This equation will be utilized in the code of the proposed Metropolis-Hastings algorithm for deriving the posterior distribution of fraction nonconforming.

*Step 3.* Sample $\mu \sim U_{[0,1]}$, where $\mu$ is a randomly sampled number from the uniform distribution $[0,1]$.

*Step 4.* If $\mu < \rho$

$$p^{(i+1)} = p^*$$

else

$$p^{(i+1)} = p^{(i)}$$

*Step 5.* Return the values $\{p^{(1)}, p^{(2)}, p^{(3)}, ..., p^{(N)}\}$.

The draws $\{p^{(1)}, p^{(2)}, p^{(3)}, ..., p^{(N)}\}$ are regarded as a sample from the targeted distribution $P(p|X)$ only after the chain has passed transient phase and the impact of the initial value can be ignored. After obtaining the fraction nonconforming values $\{p^{(1)}, p^{(2)}, p^{(3)}, ..., p^{(N)}\}$, a frequency histogram plot is generated.

### *Simulation Example*

For consistency with the analytical solution, $X \sim B(100, \ 0.1)$ is also used for illustrating the numerical solution. As the algorithm is a numerical approximation to the true posterior distribution of fraction nonconforming, the outputs have a slight difference for each run. By performing the proposed Metropolis-Hastings algorithm, the empirical posterior distribution based on one run can be obtained in the form of a frequency histogram plot, as illustrated in Figure 2.4. Visually, it has a comparable shape with the derived posterior distribution by the analytical solution shown in Figure 2.3. For estimating the credible interval, the "quantile" function in R Project for Statistical Computing (https://www.r-project.org) is used to find the lower and upper limits for the 95% credible interval using the 0.025 and 0.975 sample quantiles. For the given example $X \sim B(100, \ 0.1)$, the 95% credible interval is $[l, u] = [0.0529, 0.1726]$, which is quantitatively similar to the analytical solution $[0.0526, 0.1701]$.

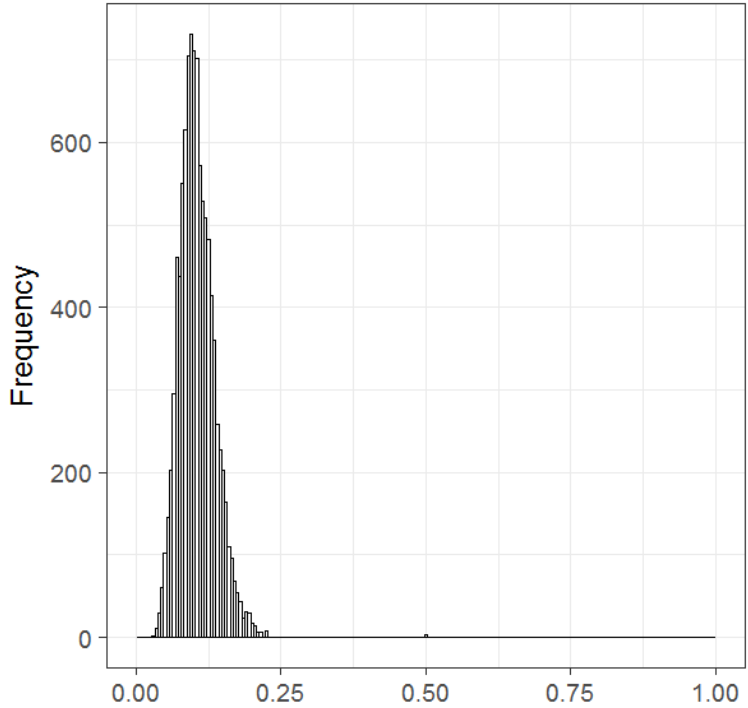

Figure 2.4: Empirical posterior distribution of $X \sim B(100, \ 0.1)$.

For assessing the Metropolis-Hastings algorithm, the trace plot and the autocorrelation function (ACF) plot are commonly adopted. Once the designed algorithm has passed these assessments, it can be used to efficiently and accurately generate a posterior distribution. A trace plot for a parameter is a scatter plot of successive parameter estimates against the number of iterations. The plot provides a straightforward method of examining the convergence behavior of the designed algorithm. An autocorrelation function (ACF) plot can also assist with algorithm diagnosis. If significant values are detected at certain lags, the proposed sampler is not efficiently sampling the posterior distribution.

Figure 2.5 elaborates the 10000 iteration samples in the form of a trace plot for the example of X~B(100, 0.1). The initial value is set at 0.5. The trace plot can be utilized to qualitatively assess the efficiency and accuracy of the proposed algorithm. The current trace plot, observed in Figure 2.5, reveals that the algorithm is efficient, as indicated by the rapid convergence rate, and accurate, as evidenced by the stationary distribution of the values around 0.1. Generally, the fraction nonconforming values generated by the Metropolis-Hastings algorithm construct a Markov chain. This chain has a unique stationary distribution that can always be reached if a large number of iterations is guaranteed. Figure 2.6 includes the trace plots for four scenarios with different initial value $p^{(0)} = \{0.2, 0.4, 0.6, 0.8\}$. The first 200 iterations are drawn for each scenario. Although the four traces begin from different initial values, they reach the same stationary distribution after approximately 50 iterations, indicative of an ideal stationary distribution.

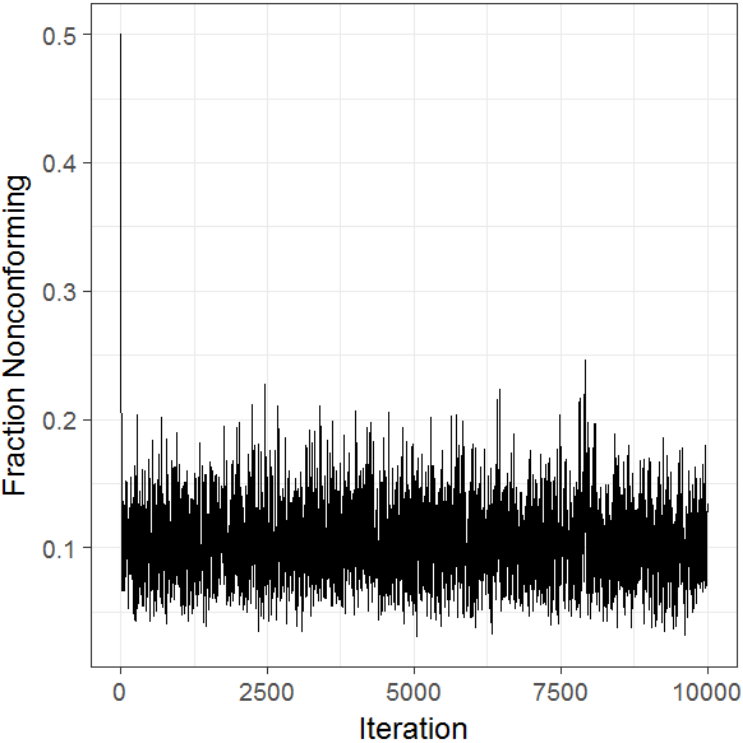

Figure 2.5: Trace plot of $X \sim B(100, \ 0.1)$.

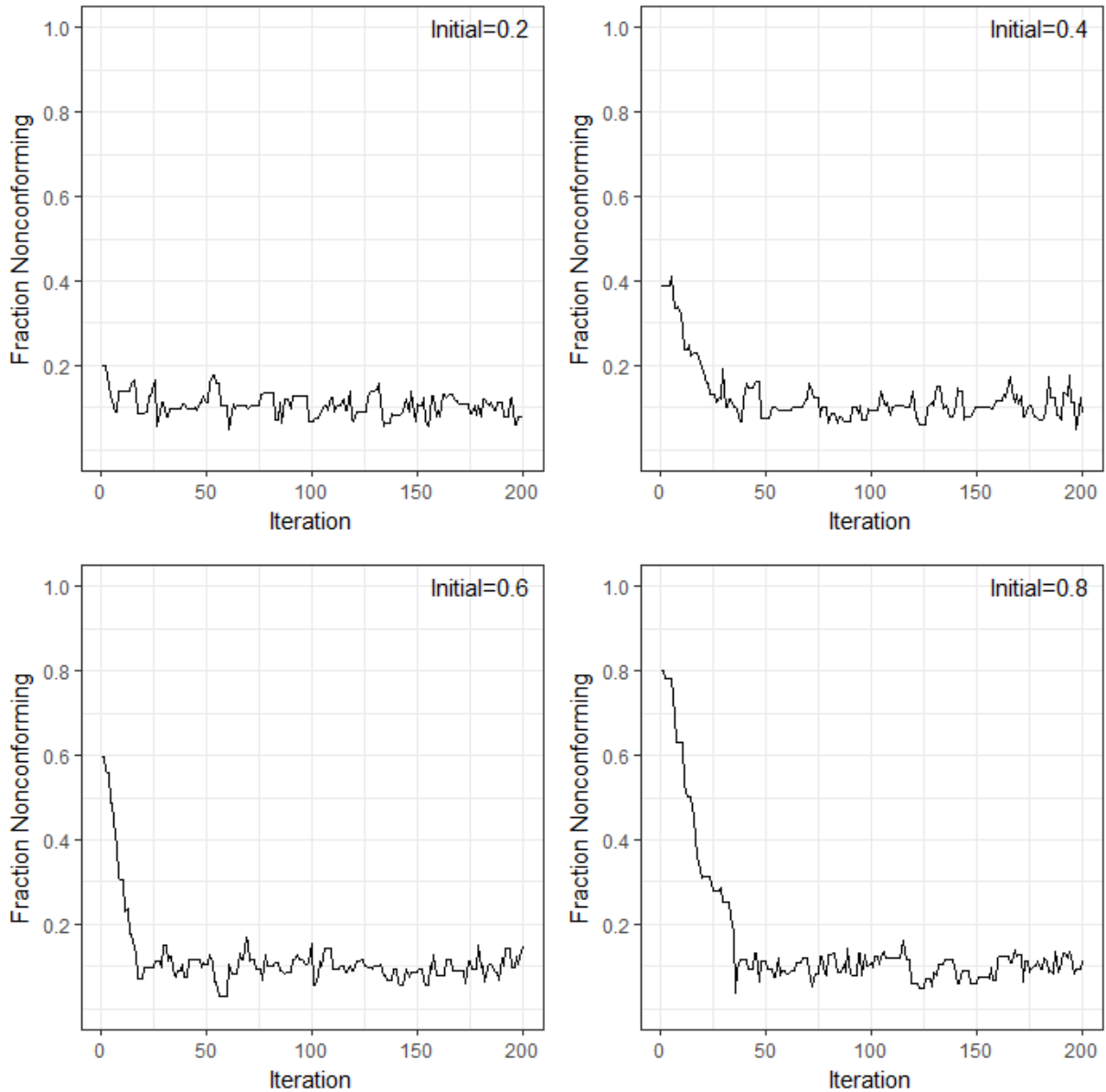

Figure 2.6: Trace plot for different initial value $p^{(0)} = \{0.2, 0.4, 0.6, 0.8\}$.

Figure 2.7 illustrates the ACF plot for the example of X~B(100, 0.1). As shown, the ACF values converge to zero as the number of iterations increases, which indicates that posterior distribution is being efficiently sampled.

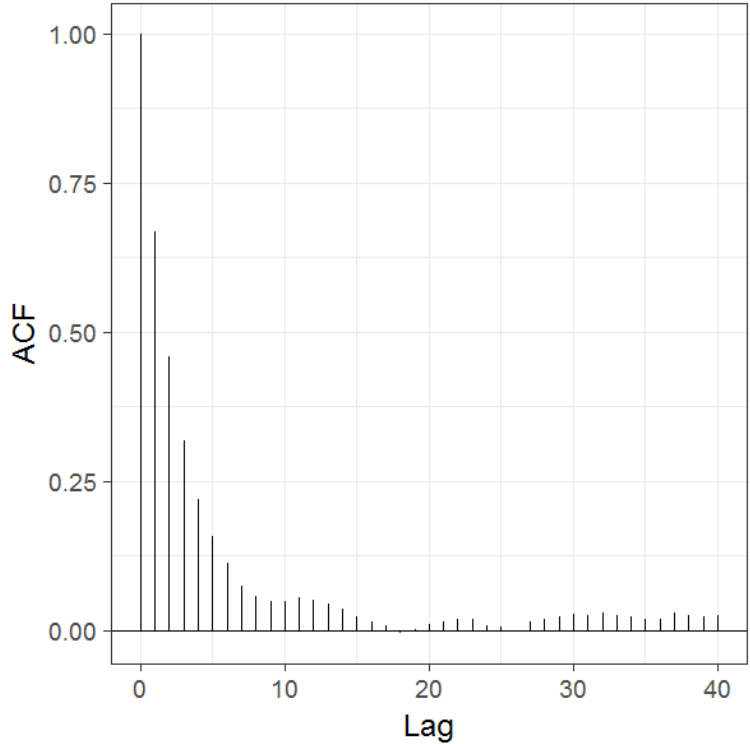

Figure 2.7: ACF plot of $X \sim B(100, \; 0.1)$.

Together, the above assessment results demonstrate that the proposed Metropolis-Hastings algorithm feasibly and reliably performs credible interval analysis for fraction nonconforming. This designed algorithm will be applied in the case study to test the proposed methodology.

### 2.6.3 Discussion

As the concept of credible interval was rarely discussed in previous fraction nonconforming literature, the authors of the present study were interested in investigating the properties of credible intervals for fraction nonconforming. The properties of credible intervals are discussed to demonstrate how changing variables impact credible interval range. The concluded properties can be applied regardless of the type of solution used. According to Eq. (2.11), the credible interval range depends on two variables: the sample size $n$ and the fraction nonconforming $p$. To improve investigation of these relationships, one of the variables is fixed for each experiment.

***Credible Interval vs. Sample Size***

Figure 2.8 depicts the relationship of the credible interval and sample size when fraction nonconforming is fixed at 0.25. As shown, the credible interval is rapidly reduced in the initial phase and converges to 0.25 when $n$ is increased. This indicates that as the amount of data expands, control over uncertainty increases.

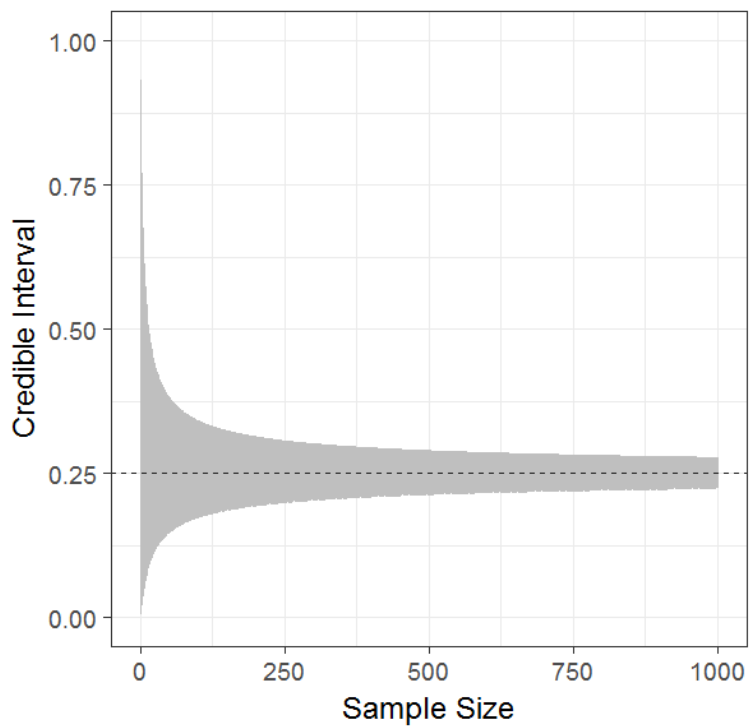

Figure 2.8: The relationship between credible interval and sample size for $X \sim B(n, \ 0.25)$.

*Credible Interval vs. Fraction Nonconforming*

Figure 2.9 depicts the relationship of credible intervals and fraction nonconforming when the sample size is fixed at 100. The credible interval is narrower when fraction nonconforming is approaching 0 or 1. When fraction nonconforming equals 0.5, the credible interval has a maximum range, indicating that actual fraction nonconforming (e.g. quality performance) can affect credible interval range.

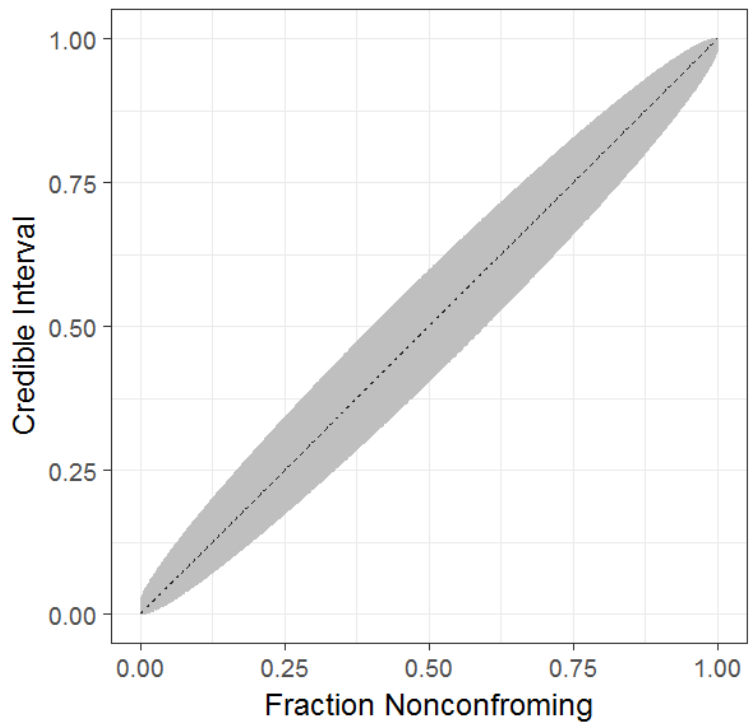

Figure 2.9: The relationship between credible interval and fraction nonconforming for

$$X \sim B(100, \ p).$$

In the following subsection, the analytical solution and the numerical solution are compared from the perspective of solution accuracy, development effort, and computational effort. Conclusions are summarized in Table 2.2.

Table 2.2: Comparison of analytical and numerical solutions.

| Features | Analytical Solution | Numerical Solution |
|---|---|---|
| Accuracy | Exact | Approximate |
| Development Effort | High | Low |
| Computational Effort | Low | High |

*Accuracy.* The analytical solution is an exact solution generated by series of logical steps that can be proven correctly in a "closed form" (Borwein and Crandall 2013). In contrast, the numerical solution approximates the exact solution and requires thousands of iterations to achieve acceptable accuracy.

*Development and Computational Effort.* An analytical solution requires rigorous mathematical deduction to derive and may not even exist. After derivation, an analytical solution may be calculated by hand. In contrast, a numerical solution can be easily achieved, without spending time on complex mathematical deduction, by using advanced computational tools. However, as it requires thousands of iterations to reach a stationary distribution, a considerable amount of computational time is required to achieve the numerical solution.

*Overall Recommendation.* Due to its relative accuracy and low computational effort, implementation of an analytical solution is recommended—particularly if it has already been developed. However, in cases where analytical solution development is not feasible or is unavailable, a numerical solution can be successfully adopted. Practitioners are encouraged to consider the advantages and disadvantages of each solution and to choose the methodology that is best suited to their particular situation.

## 2.7  Case Study

Industrial construction, used to build facilities such as petroleum refineries, petrochemical plants, nuclear power plants, and oil/gas production facilities, is a method of construction involving the large-scale use of offsite prefabrication and preassembly (Barrie and Paulson 1992). Pipe spool fabrication is crucial for the successful delivery of these industrial construction projects. Typically, pipe spools are built in a fabrication shop according to engineering designs and must be cut, fit, welded, and inspected (Song et al. 2006). Pipe spool fabrication is heavily dependent on welding, which must be sampled and inspected to ensure that welding quality requirements are met. Due to various combinations of pipe attributes (e.g. nominal pipe size (NPS), pipe schedule, and material), it is difficult for practitioners to interpret currently-available inspection data to estimate welding quality performance. In this section, the tracked pipe welding inspection data from a pipe fabrication company is used to demonstrate the practical application of the proposed credible interval estimation approach.

### 2.7.1  Data Description

Inspection data were connected and extracted from the company's Structured Query Language (SQL) server of their quality management system and were processed using R Project for Statistical Computing software. For illustration purposes, only records of radiographic tests (RT) of butt welds were used in this chapter. As shown in Table 2.3, inspection records were grouped by combination of pipe attributes, namely NPS, pipe schedule, and material, where pipe schedule defined wall thickness of the pipe, and NPS defined the outside diameter of a pipe. Materials were categorized into material A – plain carbon steel, material B – alloy steel, material C – stainless steel, and material D – others. Numbers of inspected and repaired welds for each type of pipe weld are listed. Notably, the 35 types of pipe welds listed represent the most common welding products in the studied company, encompassing 80% of their business.

Table 2.3: Pipe welds inspection records.

| Pipe Type | Schedule | NPS | Material | Inspected Welds | Repaired Welds |
|-----------|----------|-----|----------|-----------------|----------------|
| 1 | XS | 2 | Material A | 7475 | 249 |
| 2 | STD | 3 | Material A | 4495 | 173 |
| 3 | STD | 6 | Material A | 3518 | 43 |
| 4 | STD | 4 | Material A | 3078 | 66 |
| 5 | STD | 2 | Material A | 4722 | 400 |
| 6 | XS | 6 | Material A | 3705 | 70 |
| 7 | STD | 8 | Material A | 2302 | 51 |
| 8 | XS | 4 | Material A | 1774 | 28 |
| 9 | 160 | 2 | Material A | 2302 | 26 |
| 10 | 80 | 2 | Material A | 1055 | 41 |
| 11 | STD | 10 | Material A | 1131 | 30 |
| 12 | STD | 12 | Material A | 1069 | 34 |
| 13 | XS | 3 | Material A | 1484 | 16 |
| 14 | XS | 8 | Material A | 1318 | 10 |
| 15 | 40S | 2 | Material C | 555 | 21 |
| 16 | 40 | 2 | Material A | 271 | 38 |
| 17 | 80 | 4 | Material A | 638 | 5 |
| 18 | 160 | 3 | Material A | 510 | 5 |
| 19 | 40 | 4 | Material A | 592 | 17 |
| 20 | 40 | 6 | Material A | 333 | 5 |
| 21 | XS | 10 | Material A | 529 | 14 |
| 22 | XS | 12 | Material A | 666 | 31 |
| 23 | 10S | 2 | Material C | 175 | 12 |
| 24 | 40 | 3 | Material A | 217 | 6 |
| 25 | 40 | 8 | Material A | 452 | 17 |
| 26 | 40S | 3 | Material C | 364 | 6 |
| 27 | 40S | 4 | Material C | 271 | 2 |
| 28 | 80 | 3 | Material A | 512 | 6 |
| 29 | 80 | 6 | Material A | 572 | 3 |
| 30 | STD | 16 | Material A | 422 | 13 |
| 31 | 10S | 3 | Material C | 149 | 9 |
| 32 | 40S | 6 | Material C | 171 | 4 |
| 33 | 10S | 6 | Material C | 154 | 4 |
| 34 | 10S | 8 | Material C | 204 | 13 |
| 35 | 80 | 16 | Material A | 634 | 9 |

### 2.7.2 Credible Interval Estimation

As the inspected welds are not fully sampled from the population of each type of weld, estimation of an interval for fraction nonconforming (e.g. percentage repair rate) that considers uncertainties is required. The 95% credible intervals, calculated by implementing the proposed analytical and numerical credible interval estimation approaches, are listed in Table 2.4. The closed-form posterior distributions for the analytical solution are also listed. However, as the posterior distributions for the numerical solution are empirical distributions randomly generated by the Metropolis-Hastings algorithm, only credible intervals are listed for the numerical solution. The numerical solutions are based on the average of 30 runs of the proposed approach. As the analytical solutions are the exact solutions, they can be utilized to validate the accuracy and reliability of the numerical solution. Here, the numerical solutions are approximately equal to the analytical solutions, demonstrating that the MCMC method is valid for this estimation.

Table 2.4: Credible intervals of analytical and numerical solutions.

| Pipe Type | Posterior Distribution $Beta(Shape1, Shape2)$ | | 95% Credible Interval | | 95% Credible Interval | |
|---|---|---|---|---|---|---|
| | Shape1 | Shape2 | Lower Limit | Upper Limit | Lower Limit | Upper Limit |
| 1 | 249.5 | 7226.5 | 0.0294 | 0.0376 | 0.0297 | 0.0376 |
| 2 | 173.5 | 4322.5 | 0.0332 | 0.0444 | 0.0329 | 0.0441 |
| 3 | 43.5 | 3475.5 | 0.0090 | 0.0163 | 0.0088 | 0.0164 |
| 4 | 66.5 | 3012.5 | 0.0168 | 0.0270 | 0.0167 | 0.0278 |
| 5 | 400.5 | 4322.5 | 0.0770 | 0.0929 | 0.0767 | 0.0918 |
| 6 | 70.5 | 3635.5 | 0.0149 | 0.0237 | 0.0153 | 0.0239 |
| 7 | 51.5 | 2251.5 | 0.0167 | 0.0288 | 0.0167 | 0.0285 |
| 8 | 28.5 | 1746.5 | 0.0107 | 0.0224 | 0.0111 | 0.0227 |
| 9 | 26.5 | 2276.5 | 0.0076 | 0.0162 | 0.0073 | 0.0164 |
| 10 | 41.5 | 1014.5 | 0.0284 | 0.0518 | 0.0284 | 0.0517 |
| 11 | 30.5 | 1101.5 | 0.0183 | 0.0371 | 0.0182 | 0.0368 |

The table has an overarching header: "Analytical Solution" spans the Posterior Distribution and first 95% Credible Interval columns; "Numerical Solution" spans the second 95% Credible Interval columns.

| 12 | 34.5 | 1035.5 | 0.0225 | 0.0436 | 0.0222 | 0.0442 |
|----|------|--------|--------|--------|--------|--------|
| 13 | 16.5 | 1468.5 | 0.0064 | 0.0170 | 0.0065 | 0.0177 |
| 14 | 10.5 | 1308.5 | 0.0039 | 0.0134 | 0.0040 | 0.0140 |
| 15 | 21.5 | 534.5 | 0.0243 | 0.0562 | 0.0251 | 0.0558 |
| 16 | 38.5 | 233.5 | 0.1028 | 0.1853 | 0.1041 | 0.1863 |
| 17 | 5.5 | 633.5 | 0.0030 | 0.0171 | 0.0033 | 0.0177 |
| 18 | 5.5 | 505.5 | 0.0038 | 0.0214 | 0.0038 | 0.0214 |
| 19 | 17.5 | 575.5 | 0.0175 | 0.0446 | 0.0181 | 0.0442 |
| 20 | 5.5 | 328.5 | 0.0058 | 0.0326 | 0.0066 | 0.0337 |
| 21 | 14.5 | 515.5 | 0.0152 | 0.0428 | 0.0156 | 0.0431 |
| 22 | 31.5 | 635.5 | 0.0325 | 0.0646 | 0.0331 | 0.0650 |
| 23 | 12.5 | 163.5 | 0.0380 | 0.1132 | 0.0384 | 0.1111 |
| 24 | 6.5 | 211.5 | 0.0116 | 0.0561 | 0.0121 | 0.0582 |
| 25 | 17.5 | 435.5 | 0.0229 | 0.0582 | 0.0228 | 0.0600 |
| 26 | 6.5 | 358.5 | 0.0069 | 0.0337 | 0.0073 | 0.0338 |
| 27 | 2.5 | 269.5 | 0.0015 | 0.0235 | 0.0016 | 0.0251 |
| 28 | 6.5 | 506.5 | 0.0049 | 0.0240 | 0.0049 | 0.0246 |
| 29 | 3.5 | 569.5 | 0.0015 | 0.0139 | 0.0014 | 0.0146 |
| 30 | 13.5 | 409.5 | 0.0174 | 0.0506 | 0.0179 | 0.0528 |
| 31 | 9.5 | 140.5 | 0.0303 | 0.1073 | 0.0320 | 0.1103 |
| 32 | 4.5 | 167.5 | 0.0079 | 0.0547 | 0.0084 | 0.0551 |
| 33 | 4.5 | 150.5 | 0.0088 | 0.0606 | 0.0093 | 0.0606 |
| 34 | 13.5 | 191.5 | 0.0362 | 0.1035 | 0.0374 | 0.1066 |
| 35 | 9.5 | 625.5 | 0.0070 | 0.0257 | 0.0072 | 0.0259 |

### 2.7.3   Error Analysis

In this section, visual representation and quantitative estimation are conducted to test the accuracy of the numerical solution with respect to the analytical solution. To visually compare the credible intervals for both analytical and numerical solutions, results are plotted in Figure 2.10. For each type of pipe weld, two confidence intervals, based on different solutions, are given. The analytical solution is represented by the letter A, while the numerical solution is represented by the letter N. The dashed line (fraction nonconforming = 0.0296) represents the average fraction nonconforming for all 35 types of pipe welds. This value is used as the initial fraction nonconforming value for running the designed Metropolis-Hastings algorithm. As evidenced in Figure 2.10, the two solution types are visually consistent.

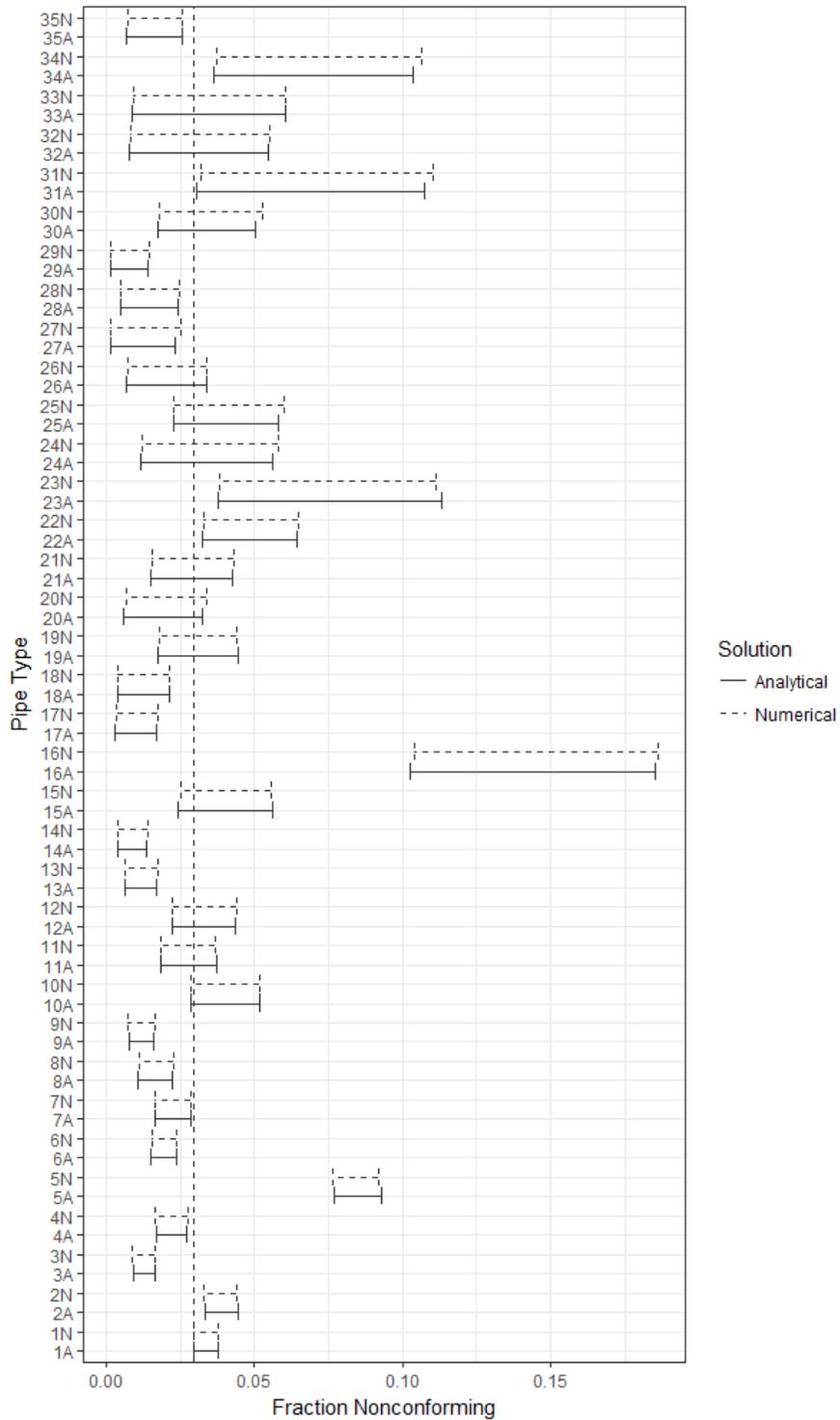

Figure 2.10: Credible intervals for analytical and numerical solutions.

Quantitatively, two indicators, Mean Absolute Error (MAE) and Root Mean Square Error (RMSE) are used to assess the efficacy of the numerical approximation with respect to the analytical solution. MAE describes the average magnitude of differences between numerical and analytical solutions (Willmott and Matsuura 2005), and RMSE represents the standard deviation of differences between numerical and analytical solutions (Willmott and Matsuura 2005). In general, MAE measures the accuracy of the mean approximation, while RMSE reflects the dispersion degree of the approximation performance. For this case study, MAE and RMSE were calculated by Eq. (2.15) and Eq. (2.16),

$$MAE = \frac{1}{n} \sum_{i=1}^{n} |f_N - f_A| \tag{2.15}$$

$$RMSE = \sqrt{\frac{1}{n} \sum_{i=1}^{n} (f_N - f_A)^2} \tag{2.16}$$

where $f_N$ is the numerical approximation, $f_A$ is the analytical solution, n is the number of weld types.

Table 2.5 summarized the residual values for the numerical and analytical solutions. For lower limits, the calculated MAE is 0.0004, and the RMSE is 0.0005. For upper limits, the calculated MAE is 0.0008, and the RMSE is 0.0011. The relatively low values indicate that the numerical approximation is accurate and can serve the credible interval estimation purpose with acceptable errors. Notably, the posterior distributions are right-skewed and more sensitive within the upper tails, resulting in MAE and RMSE values for upper limits that are slightly larger than those for lower limits. When numerical solutions are applied to unfamiliar products, an inappropriate initial value $p^{(0)}$ may result in greater error. This unexpected impact can be eliminated by increasing the number of iterations or removing these values in the initial unstable stage.

Table 2.5: Residual values of numerical solutions with respect to analytical solutions.

| Pipe Type | Analytical Solution 95% Credible Interval | | Numerical Solution 95% Credible Interval | | Residual Value | |
|---|---|---|---|---|---|---|
| | Lower Limit | Upper Limit | Lower Limit | Upper Limit | Lower Limit | Upper Limit |
| 1 | 0.0294 | 0.0376 | 0.0297 | 0.0376 | 0.0003 | 0.0000 |
| 2 | 0.0332 | 0.0444 | 0.0329 | 0.0441 | -0.0003 | -0.0003 |
| 3 | 0.0090 | 0.0163 | 0.0088 | 0.0164 | -0.0002 | 0.0001 |
| 4 | 0.0168 | 0.0270 | 0.0167 | 0.0278 | -0.0001 | 0.0008 |
| 5 | 0.0770 | 0.0929 | 0.0767 | 0.0918 | -0.0003 | -0.0011 |
| 6 | 0.0149 | 0.0237 | 0.0153 | 0.0239 | 0.0004 | 0.0002 |
| 7 | 0.0167 | 0.0288 | 0.0167 | 0.0285 | 0.0000 | -0.0003 |
| 8 | 0.0107 | 0.0224 | 0.0111 | 0.0227 | 0.0004 | 0.0003 |
| 9 | 0.0076 | 0.0162 | 0.0073 | 0.0164 | -0.0003 | 0.0002 |
| 10 | 0.0284 | 0.0518 | 0.0284 | 0.0517 | 0.0000 | -0.0001 |
| 11 | 0.0183 | 0.0371 | 0.0182 | 0.0368 | -0.0001 | -0.0003 |
| 12 | 0.0225 | 0.0436 | 0.0222 | 0.0442 | -0.0003 | 0.0006 |
| 13 | 0.0064 | 0.0170 | 0.0065 | 0.0177 | 0.0001 | 0.0007 |
| 14 | 0.0039 | 0.0134 | 0.0040 | 0.0140 | 0.0001 | 0.0006 |
| 15 | 0.0243 | 0.0562 | 0.0251 | 0.0558 | 0.0008 | -0.0004 |
| 16 | 0.1028 | 0.1853 | 0.1041 | 0.1863 | 0.0013 | 0.0010 |
| 17 | 0.0030 | 0.0171 | 0.0033 | 0.0177 | 0.0003 | 0.0006 |
| 18 | 0.0038 | 0.0214 | 0.0038 | 0.0214 | 0.0000 | 0.0000 |
| 19 | 0.0175 | 0.0446 | 0.0181 | 0.0442 | 0.0006 | -0.0004 |
| 20 | 0.0058 | 0.0326 | 0.0066 | 0.0337 | 0.0008 | 0.0011 |
| 21 | 0.0152 | 0.0428 | 0.0156 | 0.0431 | 0.0004 | 0.0003 |
| 22 | 0.0325 | 0.0646 | 0.0331 | 0.0650 | 0.0006 | 0.0004 |
| 23 | 0.0380 | 0.1132 | 0.0384 | 0.1111 | 0.0004 | -0.0021 |
| 24 | 0.0116 | 0.0561 | 0.0121 | 0.0582 | 0.0005 | 0.0021 |
| 25 | 0.0229 | 0.0582 | 0.0228 | 0.0600 | -0.0001 | 0.0018 |
| 26 | 0.0069 | 0.0337 | 0.0073 | 0.0338 | 0.0004 | 0.0001 |
| 27 | 0.0015 | 0.0235 | 0.0016 | 0.0251 | 0.0001 | 0.0016 |
| 28 | 0.0049 | 0.0240 | 0.0049 | 0.0246 | 0.0000 | 0.0006 |
| 29 | 0.0015 | 0.0139 | 0.0014 | 0.0146 | -0.0001 | 0.0007 |
| 30 | 0.0174 | 0.0506 | 0.0179 | 0.0528 | 0.0005 | 0.0022 |
| 31 | 0.0303 | 0.1073 | 0.0320 | 0.1103 | 0.0017 | 0.0030 |
| 32 | 0.0079 | 0.0547 | 0.0084 | 0.0551 | 0.0005 | 0.0004 |
| 33 | 0.0088 | 0.0606 | 0.0093 | 0.0606 | 0.0005 | 0.0000 |
| 34 | 0.0362 | 0.1035 | 0.0374 | 0.1066 | 0.0012 | 0.0031 |
| 35 | 0.0070 | 0.0257 | 0.0072 | 0.0259 | 0.0002 | 0.0002 |

**2.8    Conclusion**

Interval estimation has been identified as an efficient tool in the DMAIC (Define, Measure, Analyze, Improve, and Control) process, which is a structured problem-solving procedure for quality and process improvement (Montgomery 2007). Here, the authors demonstrate that credible interval estimations provide a more accurate, reliable, and interpretable estimation of the uncertainty of nonconforming quality control processes than conventional confidence interval estimations. This chapter also proposes an analytical and numerical solution for fraction nonconforming inference from the Bayesian statistics perspective.

For the analytical solution, this chapter applies the concept of the credible interval for binomial distribution and implements this concept into a nonconforming quality control process. The mathematical logic underlying the proposed method was revealed by the step-by-step proof. In this solution, a beta distribution was selected as the conjugate prior distribution because of the flexibility of its shape. Consequently, the posterior distribution retained the same beta distribution form in its analytical solution. Although an analytical solution is always preferred, due to its closed-form equation for posterior distributions, many researchers have claimed that when an analytical solution is too complex to derive or does not exist, a numerical solution is suitable for approximation of the true value of interest (Hitchcock 2003). The advances in Bayesian computing have increased the popularity of MCMC-based numerical methods. In this chapter, a Metropolis-Hastings algorithm-based numerical solution to approximate the empirical posterior distribution and credible interval for fraction nonconforming was developed. In addition to its ability to estimate sampling uncertainty, the proposed numerical method also provides valuable insight on the implementation of MCMC in drawing samples from complex and arbitrary probability distributions.

An industrial case study was used to validate the feasibility and applicability of the proposed credible interval estimation solution for pipe welding quality control processes. The achieved error analysis results indicate that, with respect to analytical solutions, the numerical solutions were accurate and reliable.

Integration of accurate, reliable, and straightforward approaches that measure uncertainty of inspection processes, such as those presented here, are instrumental to the successful implementation of automated data-driven quality management systems. Results of the present study indicate that practitioners can improve quality performance, and consequently enhance their market reputation and competitiveness, by implementing the credible interval estimation methodology in their automated quality management and decision-making processes.

Although the present method improves upon current interval estimation techniques, this research is limited by the assumption of a fixed prior distribution. Prior determination is a complex problem that requires incorporation of historical data, professional experience, and existing knowledge. To improve current computing performance, a systematic approach of determining the prior distribution should be further developed. Future work examining this topic include the (1) development of an automated data-driven quality performance forecast system and (2) quantification of quality-induced rework cost.

## 2.9 Acknowledgements

This research was funded by the NSERC Industrial Research Chair in Construction Engineering and Management (IRCPJ 195558-10) and Falcon Fabricators & Modular Builders, Ltd. I would like to acknowledge Rob Reid, Doug McCarthy, Jason Davio, and Christian Jukna for sharing their knowledge and expertise of pipe fabrication quality management.

## 2.10 References

Barrie, D. S., and Paulson, B. C. (1992). *Professional Construction Management: Including CM, Design-Construct, and General Contracting*, McGraw-Hill Science/Engineering/Math.

Battikha, M. (2002). "QUALICON: computer-based system for construction quality management." *Journal of Construction Engineering and Management*, 128(2), 164-173.

Berger, J. O. (1985). *Statistical Decision Theory and Bayesian Analysis*, Springer Science & Business Media.

Bishop, C. (2007). *Pattern Recognition and Machine Learning*, Springer, New York.

Borwein, J. M., and Crandall, R. E. (2013). "Closed forms: what they are and why we care." *Notices of the AMS*, 60(1), 50-65.

Brown, L. D., Cai, T. T., and DasGupta, A. (2001). "Interval estimation for a binomial proportion." *Statistical Science*, 101-117.

Casella, G., and Berger, R. L. (2002). *Statistical inference*, Duxbury Pacific Grove, CA.

Chen, L., and Luo, H. (2014). "A BIM-based construction quality management model and its applications." *Automation in construction*, 46, 64-73.

Dean, J. (2014). *Big Data, Data Mining, and Machine Learning: Value Creation For Business Leaders And Practitioners*, John Wiley & Sons.

Gelman, A., Carlin, J. B., Stern, H. S., and Rubin, D. B. (2003). *Bayesian Data Analysis*, Chapman and Hall/CRC.

Hastings, W. K. (1970). "Monte Carlo sampling methods using Markov chains and their applications." *Biometrika*, 57(1), 97-109.

Hitchcock, D. B. (2003). "A history of the Metropolis-Hastings algorithm." *The American Statistician*, 57(4), 254-257.

Metropolis, N., Rosenbluth, A. W., Rosenbluth, M. N., Teller, A. H., and Teller, E. (1953). "Equation of state calculations by fast computing machines." *The Journal Of Chemical Physics*, 21(6), 1087-1092.

Moller, J., and Waagepetersen, R. P. (2003). *Statistical Inference and Simulation for Spatial Point Processes*, CRC Press.

Montgomery, D. C. (2007). *Introduction to Statistical Quality Control*, John Wiley & Sons.

Morey, R. D., Hoekstra, R., Rouder, J. N., Lee, M. D., and Wagenmakers, E.-J. (2016). "The fallacy of placing confidence in confidence intervals." *Psychonomic Bulletin & Review*, 23(1), 103-123.

Nicholson, B. J. (1985). "On the F-distribution for calculating bayes credible intervals for fraction nonconforming." *IEEE Transactions on Reliability*, R-34(3), 227-228.

Robert, C., and Casella, G. (2011). "A short history of Markov Chain Monte Carlo: subjective recollections from incomplete data." *Statistical Science*, 102-115.

Song, L., Wang, P., and AbouRizk, S. (2006). "A virtual shop modeling system for industrial fabrication shops." *Simulation Modelling Practice and Theory*, 14(5), 649-662.

Tierney, L. (1994). "Markov chains for exploring posterior distributions." *the Annals of Statistics*, 1701-1728.

Weaver, B. P., and Hamada, M. S. (2016). "Quality quandaries: A gentle introduction to Bayesian statistics." *Quality Engineering*, 1-7.

Weiss, N. A. (2012). *Introductory Statistics*, Pearson Education USA.

Willmott, C. J., and Matsuura, K. (2005). "Advantages of the mean absolute error (MAE) over the root mean square error (RMSE) in assessing average model performance." *Climate research*, 30(1), 79-82.

# 3    CHAPTER 3: SIMULATION-BASED ANALYTICS FOR QUALITY CONTROL DECISION SUPPORT: A PIPE WELDING CASE STUDY[2]

## 3.1    Introduction

Pipe fabrication is essential for the construction of industrial projects such as petroleum refineries, petrochemical plants, and oil/gas production facilities. A fundamental process in pipe fabrication is pipe welding. High quality weld performance, or weld integrity, is essential for reducing rework costs, for maintaining project safety, and for ensuring overall product quality. Weld integrity can be controlled through the implementation of quality control systems, which sample and test pipe welds to ensure that quality requirements are satisfied.

Computer-based quality management systems, as required by ISO 9000, have been widely implemented to facilitate the process of quality control within the construction industry (Battikha 2002; Chin et al. 2004). While these systems have facilitated the collection of vast amounts of project records, transformation of these data into valuable information remains challenging for construction companies. In particular, current quality management systems (1) cannot collect data from multiple sources in real-time; (2) place a heavy interpretation load on practitioners; and (3) are unable to generate meaningful decision-support outputs. An approach capable of addressing these shortcomings, such as an analytically-based decision-support system, could positively impact quality management decision-making processes within this industry.

The primary objective of this research is to develop a novel, analytically-based approach to improve the practical functionality of traditional pipe welding quality management systems. This is achieved by (1) creating a dynamic, analytically-based environment that feeds real-time multi-relational data into advanced analytical systems; (2) developing a set of quantitative methods to model, infer, and forecast pipe welding quality control processes; and (3) generating accurate and reliable descriptive and predictive decision-support metrics. In this chapter, the framework of the simulation-based analytics approach is introduced, and the current state-of-the art and its challenges are described. Then, the functional modules and inherent advanced analytics of the specialized simulation-based analytics approach for quality control purposes are outlined, and the deployment of a C#-based prototype system at an industrial company is demonstrated. Finally, research contributions, from both an academic and industrial perspective, are discussed.

---

[2] This chapter is adapted from work that has been accepted for publication as "Simulation-based analytics for quality control decision support: A pipe welding case study" by the *Journal of Computing in Civil Engineering*. doi: 10.1061/(ASCE)CP.1943-5487.0000755 and has been reprinted with permission from the American Society of Civil Engineers.

## 3.2    Simulation-Based Analytics

This research proposes that a simulation-based analytics framework could be used to successfully integrate analytically-based forecasting, planning, and optimization approaches to improve construction project performance. The proposed framework was derived from state-of-the-art concepts in data analytics applications (Barton and Court 2012; LaValle et al. 2011), dynamic data-driven application systems (DDDAS) (Darema 2004), and simulation-based analytics (Dube et al. 2014). The philosophy behind the framework, as shown in Figure 3.1, is the dynamic incorporation of data—as they are generated—into to simulations, thereby enhancing the accuracy and predictability of original models. This framework employs a variety of analytical methods including data mining (to identify anomalies or missing information) and simulation methods (to populate, fill, or validate data). Transformed data is then coupled to simulation systems via designed algorithms and models to generate desired metrics and outcomes.

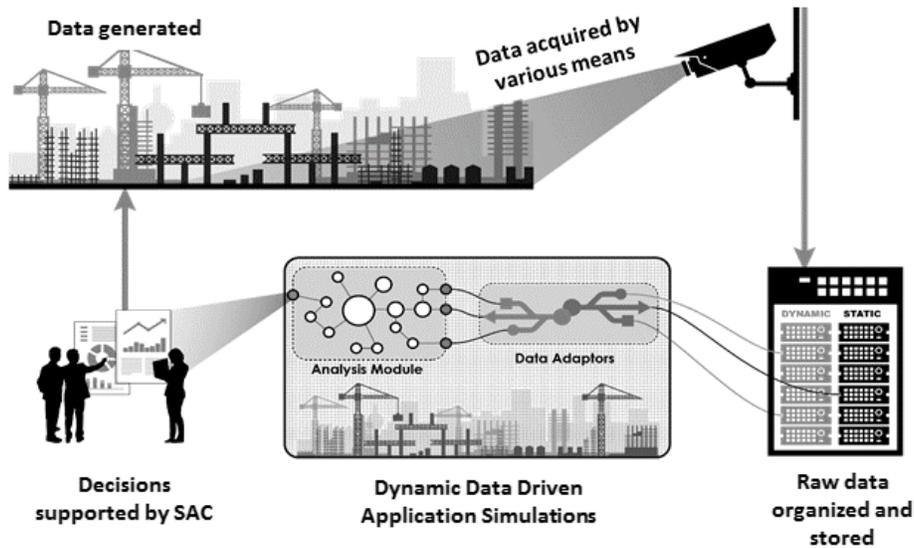

Figure 3.1: Simulation-based analytics for construction.

### 3.2.1 Dynamic Data-Driven Application Systems

At the core of the proposed framework is the concept of a dynamic, data-driven application system (Darema 2004), which strives to seamlessly couple the simulation world with measurements of a real system in real time. A DDDAS (Darema 2004) provides the means for coupling simulation with measurement disciplines (i.e., data collection). Consequently, DDDAS improve upon current models through the incorporation of automated internal calibrations. Data used for this purpose are typically archival or are generated in real-time through collection from sensors or detectors or by input from practitioners.

DDDAS is particularly advanced in the domain of transportation engineering. A research project conducted at the Georgia Institute of Technology attempted to estimate arterial travel time—in real-time—through the integration of simulation with real-time sensor data (Henclewood et al. 2008). Huang and Verbraeck (2009) have also applied DDDAS at a company in the Netherlands to measure tram travel durations using a system called TriTAPT. Data collected were input into simulation models to generate updated schedules that were published for travelers. Similar information was generated to simulate control systems.

Despite the benefits of DDDAS, however, recalibration of input models to realign with dynamic, real-time data is associated with significant challenges. In particular, data collected in real-time tends to be messy, arrives in no particular order, and is often incomplete or erroneous due to transmission problems or sensor malfunctions, which can result in significant interpretation difficulties for construction practitioners.

### 3.2.2 Bayesian-Based Measurements Recalibration

Simulation modellers often use static, statistical distribution-based approaches to represent uncertainties that exist in the systems they are analyzing. In contrast, the proposed framework uses Bayesian techniques to update input models with new data. Bayesian statistics can be used to dynamically update information of interest as additional observations become available (Gelman et al. 2014). Prior distributions of the parameters are introduced and posterior distributions are computed from observed data based on Bayes' theorem (Bishop 2006). After obtaining posterior distributions, uncertainty can be quantified by providing certain tail quantiles of the posterior distribution (Weaver and Hamada 2016). There has been an ample amount of work focused on developing formulations for the application of Bayes' theorem to update normal distributions in literature (Lynch, 2007; Chung et al., 2004). This is partly due to the low dimensionality associated with the parameters of this distribution and the computational challenges associated with evaluating Bayesian formulation for updating probabilistic models.

### 3.2.3    Binary Nature of Weld Quality Management Data

Typically, pipe weld quality is assessed by nondestructive examination (NDE), which detects discontinuities in welds without inducing pipe damage (ASME 2005). In pipe welding quality control processes, the NDE inspection result is recorded as either "conforming" or "nonconforming" to specified quality standards (Montgomery 2007). Given their binary nature, these data cannot be appropriately represented numerically.

In statistical quality control, the terminology "fraction nonconforming" is defined as the ratio of the number of nonconforming items in a population to the total number of items in that population (Montgomery 2007). However, due to the vast possible combinations of pipe attributes (i.e., NPS, pipe schedule, and material) in conjunction with the inappropriate incorporation of inspection sampling uncertainty, understanding and predicting product and project quality performance from fraction nonconforming records remains challenging. Traditional fraction nonconforming estimation methods lack accuracy, interpretability, and capability of incorporating new data. In Chapter 2, both analytical and numerical solutions were developed for the purposes of extending updating capabilities to beta distributions for inferring fraction nonconforming performance. Due to its relative accuracy and low computational effort, the analytical solution is recommended, particularly if it has previously been developed.

### 3.3    Methodology

The proposed simulation-based analytics framework was designed to achieve more efficient quality control decision support. The specialized simulation-based analytics approach consists of five components, namely the data source, data adapter, data analysis module, simulation module, and decision-support module. The data source component extracts quality and design data from quality management and engineering design systems, respectively, which store information required for the analytical process. The data adapter then transforms raw data into a tidy format through data connection, data wrangling, and data cleaning processes. The data analysis module provides a suite of algorithms to analyze the data from the data adapter and/or the simulation module for the establishment of decision-support metrics. The simulation module generates raw data of predictive decision-support metrics by interacting with the data adapter and/or the analysis module. Finally, the decision support module utilizes the outputs from the data analysis module to support decision-making processes. The conceptual framework of the proposed simulation-based analytics approach is illustrated in Figure 3.2.

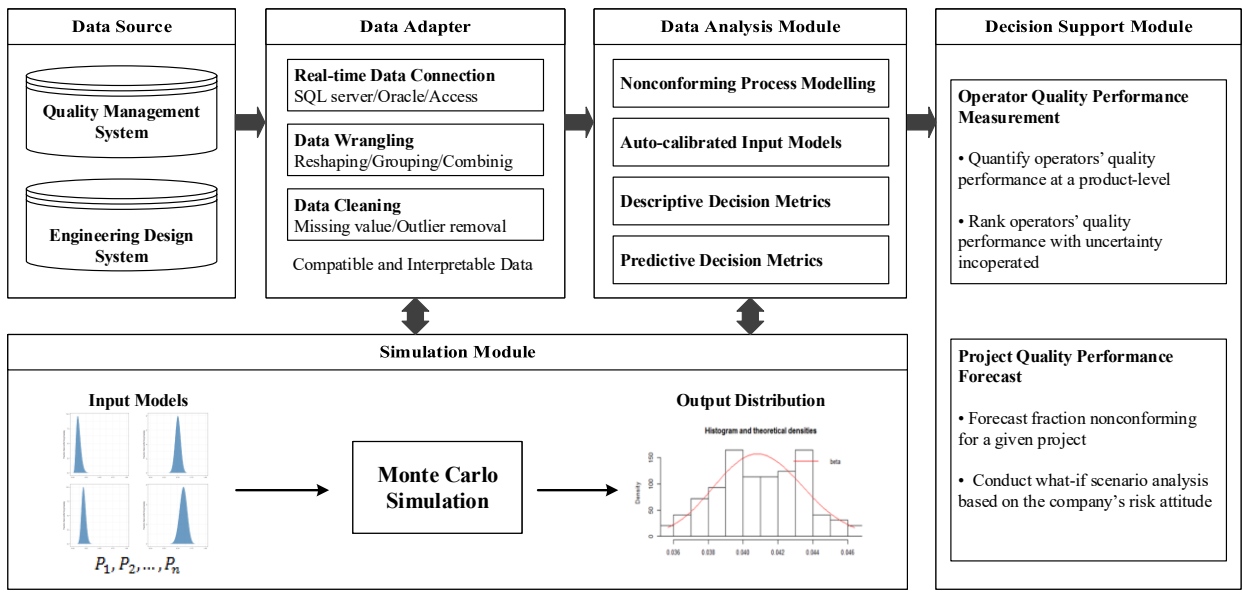

Figure 3.2: The conceptual framework of the proposed simulation-based analytics approach.

R (https://www.r-project.org), a free programming software environment for statistical computing and graphics, was selected to perform all functionalities of the specialized simulation environment. While R was traditionally limited to academia and research, the use of R within private and business analytics sectors is rapidly expanding. R's functionality can be easily expanded with more than 10,000 packages available on CRAN (https://cran.r-project.org). These packages are a collection of R functions, which provide users immediate access to advanced data mining, machine learning, and simulation techniques. Notably, other statistical suites may also be used.

### 3.3.1 Data Sources

Construction companies handle large amounts of raw data during operations, which includes data from two primary sources, namely (1) dynamic data that is obtained from information stored in Enterprise Resource Planning systems and standalone applications (e.g., time-sheets, quantities, and safety incidents) and (2) static data that is passed on from engineering in the form of drawings (e.g., 2D, 3D, and BIM models).

Here, dynamic data were comprised of quality management data and operator information, and static data were comprised of engineering design data. NDE inspection results, weld type, and operator information for each pipe weld were stored in the quality management system, Structured SQL Server (ArcuTrack), of a pipe fabrication company located in Edmonton, Canada. Over the last 10 years, the system has collected vast amounts of data—for butt welds alone, the system has tracked around 250,000 NDE inspection records that include more than 680 combinations of pipe attributes. In this research, the data were split into a modelling set and a testing set in chronological project order. Projects conducted in the first eight years were selected as the modelling set, and subsequent projects were utilized as the testing set.

Pipe weld attributes, such as NPS (i.e., outside diameter), pipe schedule (i.e., wall thickness), and material for all welds were extracted from the engineering design system. Materials were categorized into material A – plain carbon steel, material B – alloy steel, material C – stainless steel, and material D – others. Pipe weld type was defined by the pipe format: NPS, schedule, material, weld type, where pipe (40, 2, A, BW) represents butt welds with NPS of 40, schedule of 2, and material A. Design information was matched to each weld through primary key and foreign key relationships. Notably, data were stored in multi-relational databases, which resulted in the dispersion of useful information across various tables and databases. Detailed information for the two types of data sources are listed in Table 3.1.

Table 3.1: Data sources.

| Data Source | Database | Information |
|---|---|---|
| Quality Management System | ArcuTrack | • NDE inspection result<br>• Weld type (Butt weld, Fillet weld, etc.)<br>• Welder information |
| Engineering Design System | Autodesk Navisworks | • Weld design attributes (NPS and schedule)<br>• Material information (Plain carbon steel, alloy steel, stainless steel, and others) |

### 3.3.2 Data Adaptor

In practice, most information cannot be adequately represented by independent data tables; rather, multiple types of objects are linked together through various types of linkages. Such data are usually stored in relational databases. While multi-relational databases can provide richer information for data mining, most existing data mining algorithms cannot be applied to relational data unless the relational data are first converted into a single table. A data adapter integrates real-time information from various data sources into one, centralized dataset. This is particularly important for data that are collected from a variety of sources or databases. In this case, multi-relational data were dispersed across quality management and engineering design systems. A data adapter was developed to transform raw data, through the processes of data connection, data wrangling, and data cleaning, into compatible and interpretable formats. R was used to perform data connection, data wrangling, and data cleaning tasks.

For real-time data connection, the R package for Open Database Connectivity (RODBC) was used to connect to the SQL Server of the quality management and engineering design systems. The RODBC package provides access to databases (including Microsoft Access, Microsoft SQL Server, or Oracle Database) through an Open Database Connectivity (ODBC) interface. RODBC has two group of functions: the main internal odbc* commands, which implement low-level access to the ODBC functions, and the sql* functions, which operate at a higher level to read, save, copy, and manipulate data between data frames and SQL tables. The data connection through RODBC was periodically redone to ensure that accessed data was up-to-date, thereby achieving real-time computation.

The term "tidy data" has been used to refer to data that are maintained in a table form where each variable is saved in its own column and each observation is saved in its own row (Wickham 2014). Tidy datasets are easy to manipulate, model, and visualize (Wickham 2014). Here, the dplyr and tidyr packages were used to perform data wrangling tasks. By making use of multiple processors, these packages can perform data wrangling tasks in relatively little time, which is critical for processing large-sized datasets, such as the ones used in the present study (Wickham 2017; Wickham and Francois 2017).

Omitted values, noise, and inconsistencies render data inaccurate and unreliable. Data cleaning, which can be used to improve data accuracy and reliability, is an essential step of the proposed approach. Although many mining routines have developed procedures for dealing with these types of data, these procedures are not always robust (Han et al. 2011). Therefore, specific data cleaning rules should be specified for each particular case.

### 3.3.3   Data Analysis Module

A data analysis module is comprised of a suite of algorithms that facilitates the establishment of the metrics required for decision support. A hybrid data analysis module was developed to (1) analytically model nonconforming quality inspection processes; (2) generate auto-calibrated input models (i.e. posterior distributions for product fraction nonconforming); (3) determine the descriptive decision support metric for quantifying operator quality performance at a product-level; and (4) generate the predictive metric for forecasting nonconformance for a given project.

*Nonconforming Process Modelling*

In statistical quality control, fraction nonconforming is defined as the ratio of the number of nonconforming items in a population to the total number of items in that population (Montgomery 2007). For details of nonconforming process modelling, please see Chapter 2, Section 2.3.

*Auto-Calibrated Input Models*

To consider sampling uncertainty, Chapter 2 proposed a Bayesian statistics-based analytical solution to derive the posterior distribution of the fraction nonconforming during estimation of the population fraction nonconforming variable when data are obtained from a sample. This Bayesian statistics-based method is capable of recalibrating the posterior distribution of fraction nonconforming at a product-level by combining both previous knowlede (i.e. prior distribution) and real-time observed data.

Beta distributions are commonly used as the standard conjugate priors for inferring variable $p$ in a binomial distribution (Berger 2013). The primary reasons for choosing beta distributions are that they (1) bind the variable fraction nonconforming between the range of 0 to 1; (2) have the flexibility to provide accurate and representative outputs; and (3) have parameters that are intuitively and physically meaningful and that are easy to estimate from the data (Ji and AbouRizk 2017).

If the fraction nonconforming has a prior distribution $Beta(a,\ b)$, then the posterier distribution would be defined in Eq. (3.4).

$$P(p|X) = Beta(X + a, n - X +\ b) \qquad (3.4)$$

In practice, the prior distribution can be determined based on historical data, professional experience, and existing knowledge. In this research, a non-informative prior distribution $Beta(1/2,\ 1/2)$ was used to remove the effects of external information on current data (Berger 2013). After observing $X$ successes in $n$ trails, and given the non-informative prior $Beta(1/2, 1/2)$, the posterior distribution for fraction nonconforming was defined by the beta distribution:

$$P(p|X) = Beta(X + 1/2, n - X +\ 1/2) \qquad (3.5)$$

### Descriptive Decision Support Metric: Operator Quality Performance

Practitioners are interested in identifying welding operators with exceptional quality performance (i.e. low fraction nonconforming) for particular weld types. Since welder information is tracked by the quality management system, a welder's performance can be determined for each weld type. The derived posterior distribution for fraction nonconforming is implemented to measure welder quality performance at a product-level. For illustration purposes, data for one welding operator with pipe (STD, 2, A, BW) was utilized to demonstrate the metric. After connecting and processing the data, it was determined that 8 out of 51 welds failed inspection for the selected welding operator with pipe (STD, 2, A, BW) at a certain time point. As per Eq. (3.5), the theoretical distribution was determined to be $Beta(8.5, 43.5)$, which is illustrated in Figure 3.3. The obtained posterior distribution represents the fraction nonconforming density of the operator's quality performance (i.e. fraction nonconforming), and the vertical axle depicts the occurrence of nonconforming items for a certain fraction nonconforming value. Based on the derived posterior distribution, inspection sampling uncertainty was incorporated to quantify the welder's quality performances for the particular weld type. For interpretation purposes, the obtained posterior distribution is transformed to a boxplot with a five-point summary. Details of the boxplot are discussed in the Decision Support Module section.

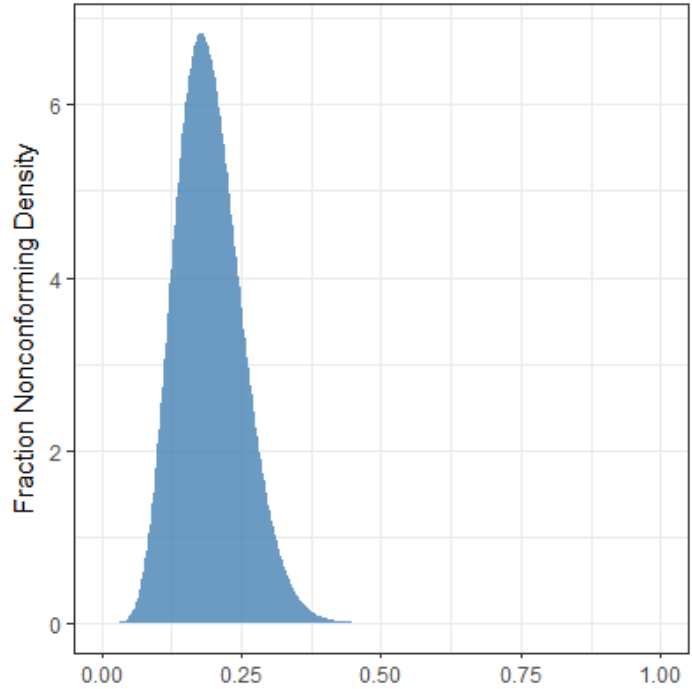

Figure 3.3: Theoretical posterior distribution of fraction nonconforming for an operator with pipe

(STD, 2, A, BW).

***Predictive Decision Support Metric: Project Fraction Nonconforming Estimation***

Raw data for the predictive decision-support metrics are generated by the simulation module. Generated data can be processed through the data adapter and data analysis module for further decision-support purposes.

Pipe welds with various combinations of design attributes each have their own posterior distributions for fraction nonconforming. Therefore, for estimating fraction nonconforming of the overall project, a mixture of distributions must be derived to determine the distribution of the overall fraction nonconforming. The static model can be described as in Eq. (3.6):

$$F(x) = \frac{1}{n} \sum_{i=1}^{k} P_i(x), \quad k > 1 \tag{3.6}$$

where $n$ is the total number of welds for the project and $P_i(x)$ is the cumulative density function of the $i$-th posterior distribution, which represents the fraction nonconforming distribution of this type of pipe weld.

This equation forecasts project fraction nonconforming by randomly sampling a fraction nonconforming rate for each pipe weld from the project. This predictive decision-support metric can be used to predict future project quality performance in a real-time manner from detailed design information. In the next section, the generation of the project fraction nonconforming histogram is described in detail.

### 3.3.4   Simulation Module

A simulation module generates raw data of desired predictive metrics for a given decision-support application using transformed data from the data adapter and/or the analysis module. In construction research, Monte Carlo Simulation has been widely used to conduct uncertainty analysis for achieving improved scheduling (Jun and El-Rayes 2011; Lu and AbouRizk 2000) and cost estimation performance (Firouzi et al. 2016; Touran 1993).

Here, Monte Carlo Simulation was implemented in the simulation module to forecast project fraction nonconforming with incorporated uncertainty. The accuracy and feasibility of this Monte Carlo simulation model have been previously validated through the comparison of simulated project quality performance and actual project quality performance of 35 historical projects by Ji and AbouRizk (2016). Briefly, the steps of the Monte Carlo Simulation module are

(1) identifying input models (i.e. posterior distributions) generated from the data analysis module in a real-time manner; (2) generating the static model that is described by Eq. (3.6); (3) sampling random variables with multiple iterations; and (4) analyzing results for decision-making purposes.

For each type of pipe weld, the posterior distribution for fraction nonconforming can be derived using Eq. (3.5) as an input model for the Monte Carlo Simulation. The project fraction nonconforming is a mixture of distributions of all pipe weld types. After multiple iterations of the Monte Carlo Simulation, a histogram for project fraction nonconforming is generated. Once the project fraction nonconforming histogram has been generated, statistics of interest are obtained using R. A detailed example is provided in the following section. In this research, 100 iterations are performed to achieve the predictive decision-support metric. The duration of each iteration depends on the number of pipe welds a project involves. For simulating all historical projects (250,000 pipe welds), the duration for the entire analytical process would be less than 30 seconds.

This proposed simulation module can serve the fraction nonconforming estimation purpose for both the project planning and project control phases. In the project planning phase, the overall project fraction nonconforming can be simulated from new project weld design information. In the project control phase, the predicted results can be periodically updated by incorporating real-time inspection data.

### 3.3.5   Decision Support Module

The decision support module contains two functionalities to support decision-making processes, namely operator quality performance measurement and project quality performance forecast. These two functionalities can improve day-to-day decision-making at the operational management level and support job planning at the tactical management level. Details of the functionalities are discussed as follows.

***Functionality 1: Operator Quality Performance Measurement***

The main outputs of this functionality include both the quantification and ranking of welding operators' quality performances for each weld type. Based on the proposed approach, useful information can be generated in a real-time manner, which dramatically reduces the data interpretation load for decision makers. Under this functionality, a boxplot is utilized to visualize the welding operator quality performance distribution. In statistics, a boxplot is an efficient graphical representation of a five-number summary (minimum value, 25% quantile (Q1), Median, 75% quantile (Q3), and maximum value) of a distribution (De Veaux et al. 2005). The upper whisker extends from the hinge to the highest value that is within 1.5 * IQR of the hinge, where the IQR equals to the inter-quartile range or distance between the first and third quartiles. The lower whisker extends from the hinge to the lowest value within 1.5 * IQR of the hinge. Data beyond the end of the whiskers are outliers and are plotted as individual points.

In this section, data regarding pipe (STD, 2, A, BW), the most common work type, is used to demonstrate the main outputs of this functionality. 17 operators, who have each had over 100 welds inspected, were selected. Table 3.2 lists the performance records and the five-number summaries of the welding quality performances, with respect to fraction nonconforming, of these operators. Consistent with Figure 3.4, the median value of their performance is sorted in ascending order in Table 3.2. Operator identification numbers were randomly reassigned to maintain employee anonymity.

Table 3.2: Statistical summary of operator welding performance of pipe (STD, 2, A, BW).

| Operator ID | Inspected Welds | Repaired Welds | Min | Q1 | Median | Q3 | Max |
|---|---|---|---|---|---|---|---|
| 1 | 355 | 11 | 0.008 | 0.027 | 0.031 | 0.038 | 0.072 |
| 2 | 147 | 5 | 0.005 | 0.025 | 0.035 | 0.047 | 0.097 |
| 3 | 264 | 9 | 0.013 | 0.029 | 0.035 | 0.043 | 0.080 |
| 4 | 208 | 9 | 0.009 | 0.035 | 0.043 | 0.054 | 0.110 |
| 5 | 123 | 5 | 0.006 | 0.033 | 0.043 | 0.057 | 0.119 |
| 6 | 316 | 16 | 0.020 | 0.043 | 0.051 | 0.059 | 0.097 |
| 7 | 175 | 9 | 0.017 | 0.040 | 0.052 | 0.067 | 0.133 |
| 8 | 120 | 6 | 0.011 | 0.038 | 0.052 | 0.064 | 0.155 |
| 9 | 175 | 11 | 0.023 | 0.053 | 0.064 | 0.076 | 0.138 |
| 10 | 119 | 8 | 0.015 | 0.054 | 0.069 | 0.086 | 0.182 |
| 11 | 104 | 7 | 0.009 | 0.055 | 0.070 | 0.087 | 0.196 |
| 12 | 207 | 16 | 0.028 | 0.068 | 0.080 | 0.093 | 0.159 |
| 13 | 100 | 8 | 0.026 | 0.065 | 0.081 | 0.100 | 0.223 |
| 14 | 307 | 27 | 0.046 | 0.078 | 0.089 | 0.099 | 0.157 |
| 15 | 139 | 13 | 0.036 | 0.080 | 0.096 | 0.113 | 0.193 |
| 16 | 111 | 11 | 0.029 | 0.085 | 0.103 | 0.123 | 0.225 |
| 17 | 175 | 25 | 0.069 | 0.128 | 0.147 | 0.166 | 0.243 |

To visually represent and compare welding operator quality performance, a side-by-side boxplot can be generated, as shown in Figure 3.4. Box width is inversely proportional to the stability of operator quality performance, where a narrower box is indicative of an operator with a stable quality performance for the indicated type of weld. Operators with higher ranks have smaller fraction nonconforming (i.e., better quality performance). The dashed line represents the average welding quality performance of the 17 operators (fraction nonconforming=0.062). The range of the median is relatively wide (0.031 to 0.147), indicating operator performance varies considerably for the same type of welding work.

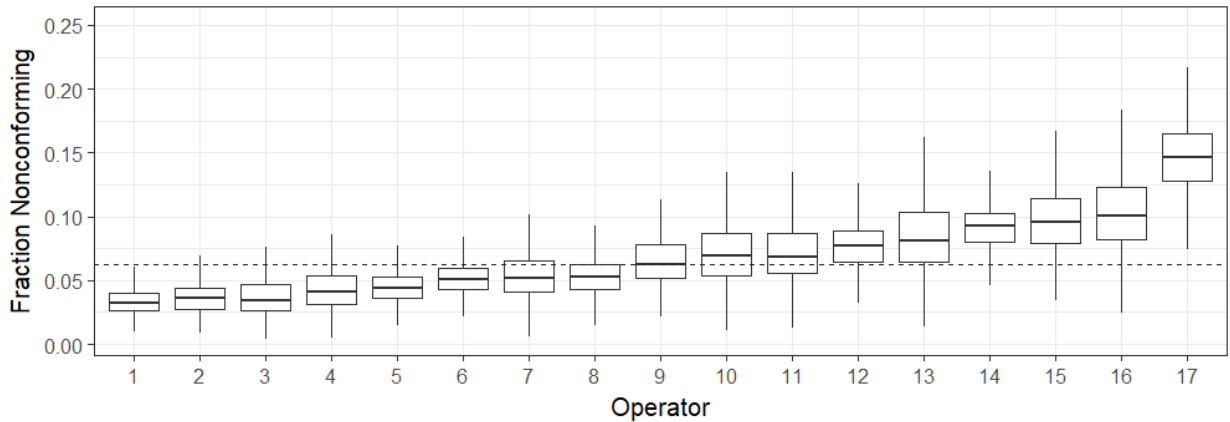

Figure 3.4: Sorted side-by-side boxplot of operators' quality performance with pipe (STD, 2, A, BW).

From this visualized output, practitioners can (1) infer operators' skill level for a specified type of work; (2) identify operators who consistently produce high-quality welds; and (3) support future decision-making processes aimed at improving production planning, employee training, and strategic recruiting. Informing decision-makers with results such as these are expected to positively impact company quality and productivity performance.

***Functionality 2: Project Quality Performance Forecast***

In pipe fabrication projects, the quality performance metric, fraction nonconforming, is measured at the project-level as specified in contracts by clients. Under this functionality, given the project design information, an empirical histogram for project fraction nonconforming can be generated through Monte Carlo Simulation. Practitioners can determine a proper project fraction nonconforming based on their risk attitude (i.e., commitment to achieving quality) by selecting an appropriate quantile. The quantile represents the probability of achieving the corresponding project fraction nonconforming as determined from the simulated project fraction nonconforming histogram. For instance, practitioners may choose 10%, 50%, and 90% quantiles of the histogram to represent an aggressive, neutral, or conservative risk attitude, respectively.

For illustrative purposes, historical project information is extracted to demonstrate the simulation result. Figure 3.5 depicts the empirical histogram (100 iterations) generated by the simulation module for a given project. Table 3.3 summarizes all 10% quantiles for the generated empirical histogram. The median value of this empirical histogram is 0.049, which indicates that the project will achieve a 4.9% fraction nonconforming with a neutral risk attitude. This project quality performance forecast functionality allows practitioners to gain insights into the complexity of the project and to perform risk analysis in terms of quality performance. Furthermore, the simulated result can assist companies to prepare strategic bids by assisting with the estimation of quality-induced rework costs. In practice, companies can specify their own quantiles of interest to support their decision-making processes.

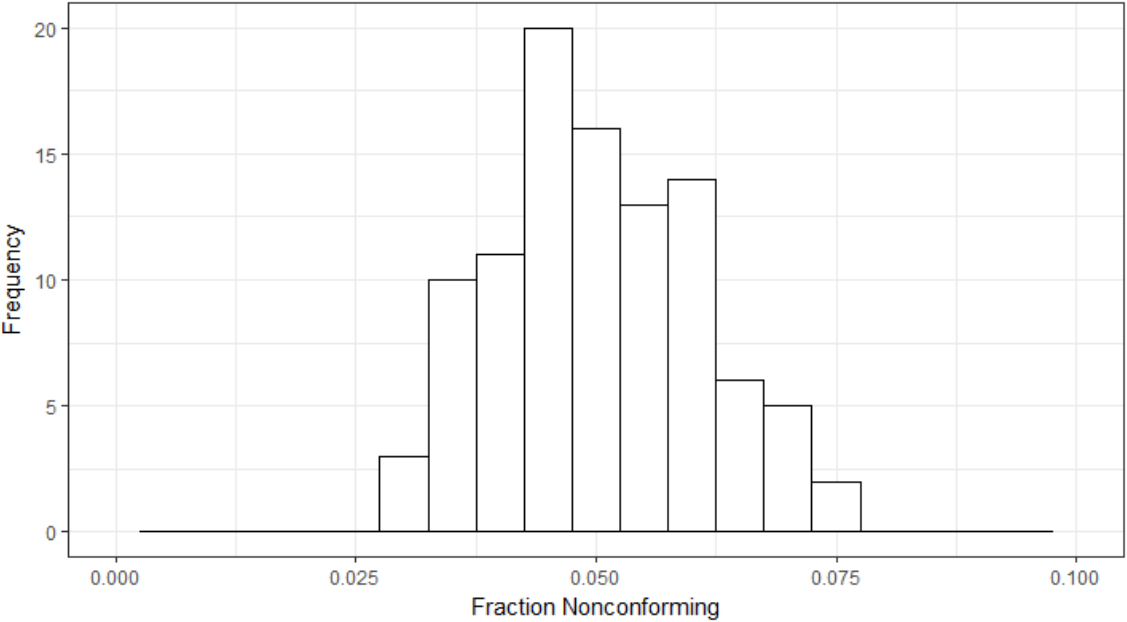

Figure 3.5: The empirical histogram for a historical project.

Table 3.3: Statistical summary of the simulated result of a historical project.

| Quantiles | 0% | 10% | 20% | 30% | 40% | 50% | 60% | 70% | 80% | 90% | 100% |
|---|---|---|---|---|---|---|---|---|---|---|---|
| Fraction Nonconforming | 0.031 | 0.037 | 0.040 | 0.044 | 0.046 | 0.049 | 0.052 | 0.057 | 0.061 | 0.064 | 0.076 |

The accurate and reliable forecasting of project fraction nonconforming provided by these functionalities can assist fabricators to better control quality performance uncertainty and to better estimate quality-induced rework costs. In addition, improved quality performance could benefit clients by enhancing health and safety during the operation phase of the project.

## 3.4    Prototype System Development

Based on the proposed research, a C#-based prototype system was developed to facilitate decision-making processes for a pipe fabrication company in Edmonton, Canada. All of the aforementioned functionalities were incorporated into the company's current quality management system, ArcuTrack, through the developed prototype. During actual system implementation, all historical data and real-time collected data are incorporated using the auto-calibrated input models.

Development of prototype system replicates the same processes of the original work performed using R but is, instead, reprogrammed in C# language. Simphony core service (AbouRizk et al. 2016) is utilized for statistical modelling and simulation purposes during the development process. The workflow of the prototype system is shown in Figure 3.6. The developed C# dynamic-link library (DLL) can connect and process data from the SQL server and can conduct all modules of the original research. A C# Graphical User Interface (GUI) is developed for creating user-friendly interfaces to achieve the previously described functionalities.

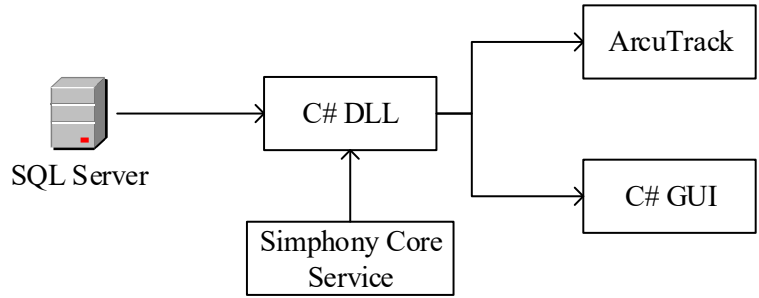

Figure 3.6: Relationship between the developed C# DLL and the original research.

The developed decision-support prototype system for the current quality management system makes use of existing quality management and engineering design data, analytics-based techniques (i.e., Bayesian statistics and Monte Carlo Simulation), and computer programs (i.e., C# and Simphony) to reliably forecast pipe welding fraction nonconforming during both the planning and control phases of a project.

Execution of the welding operator quality performance measurement functionality was selected to demonstrate the performance capabilities of the prototype system. The user interface and the decision-support metric are shown in Figure 3.7. Practitioners can specify the design attributes of the desired weld types. Pipe (STD, 2, A, BW) is selected as an example. The performance plot ranks operators' quality performances in the format of side-by-side boxplots. On the right side of the performance plot, a numerical summary is provided, which details information such as welder ID, median, number of inspected welds, and number of repaired welds. Based on this output, the fabrication shop superintendent can perform day-to-day production planning activities by allocating weld types to the operators most competent at performing said welds. The performance plot and the numerical summary can also be exported for other reporting purposes.

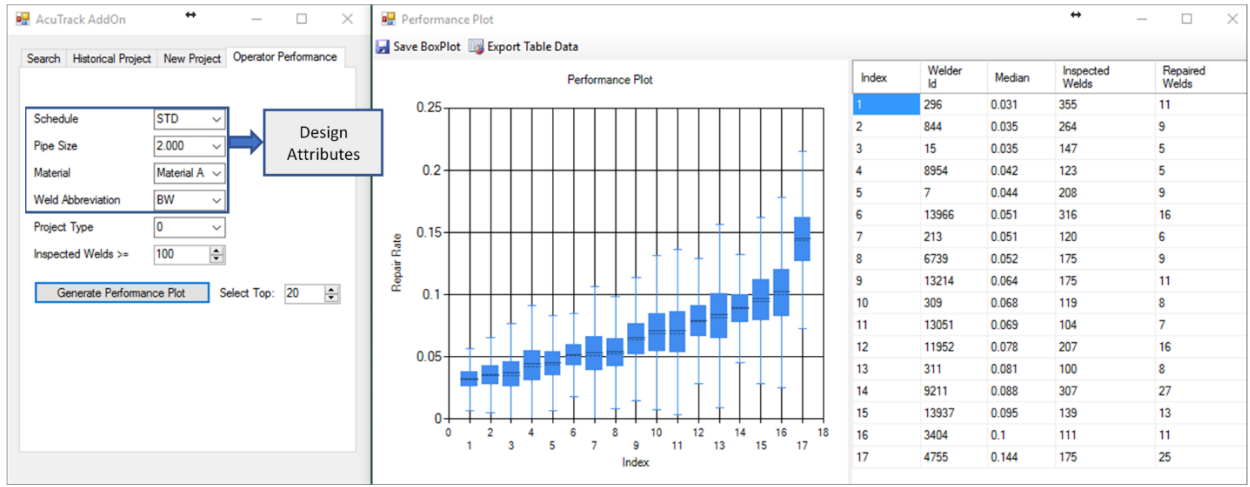

Figure 3.7: Graphical User Interface (GUI) for the functionality of operator quality performance measurement.

## 3.5    Conclusion

Industrial construction companies, much like many manufacturing companies, have difficulty using data collected by their quality management systems to make decisive, purposeful improvements to their production system's performance. Industrial construction companies find it difficult to manage, analyze, and transform data into useful information to improve decision-making and competitiveness. Through the implementation of the proposed simulation-based analytics decision-support framework, this research is expected to assist practitioners to more effectively use quality management and engineering design data to improve pipe welding quality management practices.

The simulation-analytics based system enables real-time data connection, wrangling, and cleaning. In addition, the present research proposes a feasible solution for the issue of input model recalibration to realign with new data in a dynamic manner. In terms of nonconforming quality management, the proposed system provides two functionalities: (1) a visually informative operator quality performance uncertainty measurement (i.e. boxplot) at a product-level in a real-time manner and (2) a project fraction nonconforming performance forecast that can assist practitioners with performing risk analysis during both project planning and control phases. Notably, the proposed system can be generalized for all nonconforming quality-related construction products across a multitude of construction types.

A C#-based prototype system has been developed and implemented by an industrial pipe fabrication company in Edmonton, Canada. This prototype system substantially improves the descriptive and predictive capabilities of the studied company's original quality management system. The Graphical User Interface (GUI) can generate accurate and reliable decision-support metrics in a real-time manner and considerably reduces the data interpretation load of practitioners.

Although the proposed system substantially improves the practice for pipe welding quality control processes, it is limited to quality processes that utilize binary variables. Furthermore, while the proposed system can be used to inform decision-making processes, professional experience and knowledge are still required to ensure effective decisions are made. In the future, product quality performance will be used to quantitatively measure pipe welding product complexity. Additionally, estimation information can be incorporated into the proposed simulation-based analytics system to develop a novel analytical model to estimate quality-induced rework costs for a given project.

## 3.6 Acknowledgements

This research is funded by the NSERC Collaborative Research and Development Grant (CRDPJ 492657). The authors would like to acknowledge Rob Reid, Doug McCarthy, Jason Davio, and Christian Jukna for sharing their knowledge and expertise of industrial pipe welding quality management.

## 3.7 References

AbouRizk, S., Hague, S., Ekyalimpa, R., and Newstead, S. (2016). "Simphony: A next generation simulation modelling environment for the construction domain." *Journal of Simulation*, 10(3), 207-215.

ASME (2005). *Process piping : ASME code for pressure piping, B31*, New York : American Society of Mechanical Engineers.

Barton, D., and Court, D. (2012). "Making advanced analytics work for you." *Harvard business review*, 90(10), 78-83.

Battikha, M. (2002). "QUALICON: Computer-Based System for Construction Quality Management." *J. Constr. Eng. Manage.*, 10.1061/(ASCE)0733-9364(2002)128:2(164), 164-173.

Berger, J. O. (2013). *Statistical decision theory and Bayesian analysis*. Springer Science & Business Media.

Bishop, C. (2006). "Pattern Recognition and Machine Learning" Springer, New York.

Chin, S., Kim, K., and Kim, Y.-S. (2004). "A process-based quality management information system." *Automation in Construction*, 13(2), 241-259.

Darema, F. (2004). "Dynamic data driven applications systems: A new paradigm for application simulations and measurements." *Computational Science-ICCS 2004*, 662-669.

De Veaux, R. D., Velleman, P. F., Bock, D. E., Vukov, A. M., and Wong, A. C. (2005). *Stats: data and models*, Pearson/Addison Wesley Boston.

Dube, P., Gonçalves, J. P., Mahatma, S., Barahona, F., Naphade, M., and Bedeman, M. (2014). "Simulation based analytics for efficient planning and management in multimodal freight transportation industry." *Proc., 2014 Winter Simulation Conf.*, IEEE, Piscataway, NJ,, 1943-1954.

Firouzi, A., Yang, W., and Li, C.-Q. (2016). "Prediction of Total Cost of Construction Project with Dependent Cost Items." *J. Constr. Eng. Manage.*, 10.1061/(ASCE)CO.1943-7862.0001194, 04016072.

Gelman, A., Carlin, J. B., Stern, H. S., and Rubin, D. B. (2014). *Bayesian Data Analysis*. CRC press.

Han, J., Pei, J., and Kamber, M. (2011). *Data mining: concepts and techniques*, Elsevier.

Henclewood, D., Hunter, M., and Fujimoto, R. (2008). "Proposed methodology for a data-driven simulation for estimating performance measures along signalized arterials in real-time." *Proc., 2008 Winter Simulation Conf.*, IEEE, Piscataway, NJ,, 2761-2768.

Huang, Y., and Verbraeck, A. (2009). "A dynamic data-driven approach for rail transport system simulation." *Proc., 2009 Winter Simulation Conf.*, IEEE, Piscataway, NJ, 2553-2562.

Ji, W., and AbouRizk, S. M. (2016). "A Bayesian inference based simulation approach for estimating fraction nonconforming of pipe spool welding processes." *Proc., 2016 Winter Simulation Conf.*, IEEE, Piscataway, NJ, 2935-2946.

Ji, W., and AbouRizk, S. M. (2017). "Credible interval estimation for fraction nonconforming: analytical and numerical solutions." *Automation in Construction*, 83, 56-67.

Ji, W., and AbouRizk, S. M. (2017). "Implementing a data-driven simulation method for quantifying pipe welding operator quality performance." *Proc., 2017 Joint Conference on Computing in Construction (JC3)*, ITCdl, 79-86.

Jun, D. H., and El-Rayes, K. (2011). "Fast and accurate risk evaluation for scheduling large-scale construction projects." *J. Comput. Civil Eng.,* 10.1061/(ASCE)CP.1943-5487.0000106, 407-417.

LaValle, S., Lesser, E., Shockley, R., Hopkins, M. S., and Kruschwitz, N. (2011). "Big data, analytics and the path from insights to value." *MIT Sloan Management Review*, 52(2), 21.

Lu, M., and AbouRizk, S. M. (2000). "Simplified CPM/PERT simulation model." *J. Constr. Eng. Manage.*, 10.1061/(ASCE)0733-9364(2000)126:3(219).

Montgomery, D. C. (2007). *Introduction to statistical quality control*, John Wiley & Sons.

Touran, A. (1993). "Probabilistic cost estimating with subjective correlations." *J. Constr. Eng. Manage.*, 10.1061/(ASCE)0733-9364(1993)119:1(58), 58-71.

Weaver, B. P., and Hamada, M. S. (2016). "Quality quandaries: A gentle introduction to Bayesian statistics." *Quality Engineering*, 1-7.

Wickham, H. (2014). "Tidy data." *Journal of Statistical Software*, 59(10), 1-23.

Wickham, H., Henry, L., and RStudio (2017). "tidyr: Easily Tidy Data with spread () and gather () Functions.", < https://cran.r-project.org/web/packages/tidyr/tidyr.pdf> (Sept. 14, 2017).

Wickham, H., Francois, R., Henry, L., Müller, K., and RStudio (2017). "dplyr: A grammar of data manipulation.", < https://cran.r-project.org/web/packages/dplyr/dplyr.pdf >  (Sept. 14, 2017).

# 4    CHAPTER 4: INTEGRATED DATA-DRIVEN APPROACH FOR ANALYZING PIPE WELDING OPERATOR QUALITY PERFORMANCE

## 4.1    Introduction

Due to its ability to improve overall project performance, many industrial construction projects have implemented a modularized approach to construction (O'Connor et al. 2016). Essential to modularization is pipe spool fabrication, which primarily involves pipe welding. Pipe weld quality can have a considerable impact on the overall performance of modularized construction projects: exceptional welding performance may enhance the quality and productivity of pipe fabrication processes, while poor quality performance may lead to quality-associated cost and schedule overruns. Reliable assessment and forecasting of pipe weld quality performance, particularly from an operator-by-operator perspective, can lead to improved overall project quality and enhanced management practices.

In practice, pipe weld quality is often assessed by nondestructive examination (NDE), which detects discontinuities in welds without inducing pipe damage (ASME 2001). From this, a percentage repair rate, defined as the number of failed NDE over the number of NDE completed, can be used as an index to represent an operator's quality performance. This repair rate is consistent with the concept of fraction nonconforming in statistical quality control theory, which is defined as the ratio of nonconforming items to the total number of items in a population (Montgomery 2007).

The reliable estimation of operator quality performance, however, remains challenging for companies. Quality performance is impacted by the skill level of the operator but also by the complexity of the weld itself (as determined by the pipe attributes), which is further complicated by the vast possible combinations of pipe attributes (e.g., NPS, pipe schedule, and material). Moreover, quality performance estimation may also be confounded by the inappropriate incorporation of sampling uncertainty, which arises when inspection data are collected from a representative sample rather than an entire population (Ji and AbouRizk 2017). However, methods capable of addressing these limitations have yet to be developed.

The overall objective of this research is to devise an analytical approach for quantifying and comparing operator quality performance by incorporating uncertainties that arise as a consequence of pipe design attributes and inspection sampling. To facilitate the quantification of operator welding quality performance, this study proposes an integrated, data-driven approach that efficiently sources and classifies existing quality management and engineering design data, while concurrently incorporating both pipe design attribute and sampling uncertainty. Specifically, the proposed framework (1) fuses and transforms data from separate data sources (i.e., quality management system and engineering design system) into an interpretable dataset; (2) implements an MCMC-based approach to numerically estimate posterior distributions of operator welding quality performance; (3) utilizes an A/B testing algorithm to compute probabilistic differences between operator quality performance; and (4) proposes potential applications to comprehensively improve pipe welding quality performance for practitioners.

Chapter 2 has advocated for the implementation of a MCMC-based numerical method to incorporate sampling uncertainty during nonconforming quality performance analyses. The proposed approach approximates a Bayesian posterior distribution that covers the true population fraction nonconforming of each pipe weld type. This method is capable of incorporating dynamic data and deriving additional statistical values, such as mean, median, and all quantiles (Weaver and Hamada 2016), which can be used to assess historical welding quality performance and improve decision-making processes.

Once the distributions for operator quality performance have been determined using the aforementioned approach, they can be compared using A/B testing. This method, also called controlled experimentation, has been widely implemented in various industries to guide product development and data-driven decisions (Gui et al. 2015). The goal of A/B testing is to determine if two distributions differ from one another. In this research, an A/B testing algorithm is implemented to quantify distribution differences of quality performance (i.e., fraction nonconforming) between operators, allowing users to quantitatively determine if one operator can be expected to outperform another operator for a certain pipe weld type.

The remainder of this chapter is organized as follows: the framework of the proposed approach is introduced to demonstrate the workflow of this research. Then, a detailed explanation of data source identification and data adapter design are presented, and the data analysis module, which includes fraction nonconforming modelling, MCMC-based numerical approximation, and A/B testing algorithm, is explained. An illustrative example is also provided to demonstrate the sequence of the data analysis module, and main outputs from the proposed approach are discussed as decision-support metrics to support decision-making processes. Then, potential applications of the research outcomes from the perspectives of production planning, employee training, and strategic recruiting are discussed. Finally, conclusions regarding principle findings, contributions, and limitations of this research are summarized.

## 4.2    Research Framework

A framework for an integrated data-driven approach for quantifying operator quality performance is proposed. Framework functionalities are demonstrated using quality management and engineering design data from a pipe spool fabrication company in Edmonton, Alberta. The framework consists of four components, namely the data source, data adapter, data analysis module, and decision support module. A workflow of the methodology developed to achieve the research objective is summarized in Figure 4.1. The data source component maps data from quality management and engineering design systems. Then, raw data are transformed into a single table format to allow for further analytical processes. The data analysis module provides a suite of algorithms to analyze the transformed data from the data adapter. Finally, decision-support metrics are generated by the decision support module to facilitate decision-making processes. Each component of the research workflow is discussed in following sections.

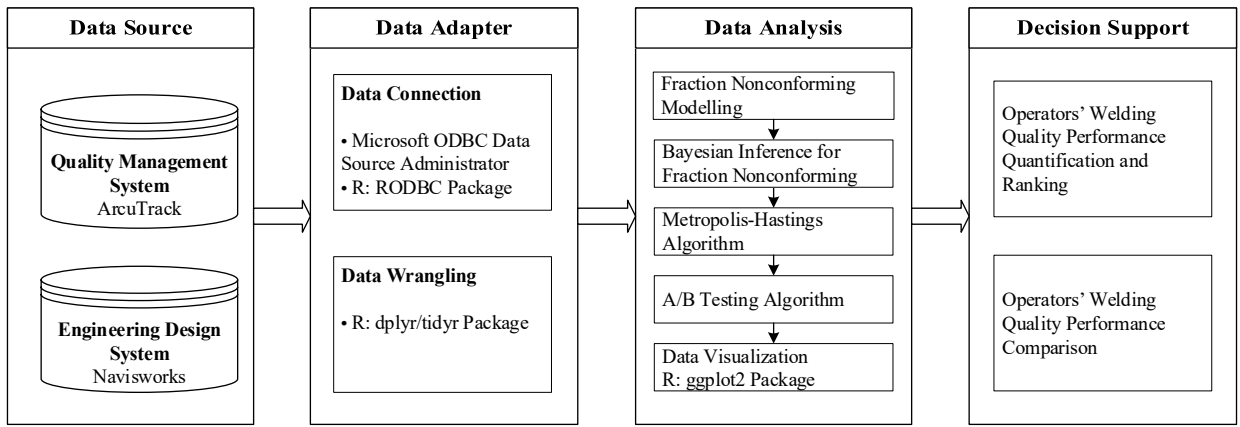

Figure 4.1: A workflow of the proposed methodology.

## 4.3    Data Source

### 4.3.1    Quality Management System

As discussed previously, computer-based quality management systems have been widely implemented for quality management purposes within the construction industry (Chini and Valdez 2003). In practice, these systems are primarily used for quality inspection documentation and quality performance reporting purposes.

Here, data stored in the quality management system SQL server (ArcuTrack) of a pipe fabrication company in Edmonton, Canada were used to demonstrate the application of the proposed methodology. NDE inspection results (e.g., no inspection performed, inspected and passed, or inspected and failed), weld types (e.g., Butt weld, Fillet weld, etc.), and operator IDs for each weld were extracted.

### 4.3.2    Engineering Design System

Building information modelling (BIM) is becoming increasingly implemented for its ability to improve information management of construction projects. Accordingly, detailed engineering models generated to facilitate quantity take-off, material information, and design specifications are accessible for many industrial construction projects. Pipe attributes, such as nominal pipe size (NPS; i.e., outside diameter), pipe schedule (i.e., wall thickness), and material, can be extracted from the engineering design system.

In the present study, pipe attributes were extracted from the engineering design system of the studied company. Here, pipe format (NPS, schedule, material, weld type) is defined to represent one type of pipe weld. For example, pipe (40, 4, B, BW) represents butt welds with an NPS of 40, schedule of 4, and material B. Design information was then mapped to each pipe weld through primary key and foreign key relationships using the designed data adapter.

## 4.4    Data Adapter

The data adapter integrates information from various sources into a single, centralized database. This is particularly important for data that are collected from a variety of sources (e.g., multi-relational databases). A data adapter must be developed to transform raw data, through data connection and data wrangling, into compatible, interpretable formats for further data mining procedures.

In the studied company, required data are dispersed across quality management and engineering design systems. R was used to perform data connection and data wrangling tasks. For data connection, the R package for Open Database Connectivity (RODBC) package was used to connect to the SQL server of both the quality management and engineering design systems. The dplyr/tidyr package was used to perform data wrangling tasks, including data reshaping, grouping, and combining. The completed dataset was transformed into a table format, where each variable was saved in its own column and each observation was saved in its own row.

## 4.5    Data Analysis Module

A data analysis module is comprised of a suite of algorithms that facilitate the establishment of decision-support metrics required by the framework. To achieve functionalities of the decision support module, a series of algorithms have been integrated to model fraction nonconforming, derive posterior distributions of operator quality performance, and compare probabilistic differences of operator quality performance. First, fraction nonconforming is modelled using an established statistical model. Second, the Bayesian inference for fraction nonconforming is introduced, and a Metropolis-Hastings algorithm is implemented to approximate the posterior distributions of operator quality performance. Then, an A/B testing algorithm is developed to compare probabilistic differences between operator quality performance distributions, and operator quality performance is visualized using the ggplot2 package in R. The integrated algorithms and the generated metrics used in the present study are detailed as follows.

### 4.5.1   Fraction Nonconforming Modelling

In statistical quality control, fraction nonconforming is defined as the ratio of the number of nonconforming items in a population to the total number of items in that population (Montgomery 2007). For details of nonconforming process modelling, please see Chapter 2, Section 2.3.

### 4.5.2   Bayesian Inference for Fraction Nonconforming

Inspection is an essential process in pipe spool fabrication to ensure pipe welds meet a specific standard of quality. As mentioned previously, binary inspection results can be modelled and measured by fraction nonconforming, which is a reliable indicator of welding operator quality performance. However, to precisely quantify operator quality performance, all manufactured products would be required for inspection, which would be a time- and cost-intensive process. An alternative solution is to select and inspect a representative sample for quality assurance purposes (Montgomery 2007). Although there are multiple ways to choose representative samples, uncertainties associated with inspection results still exist. To generate a more accurate, reliable, and interpretable result, an MCMC-based numerical approximation of the Bayesian posterior distribution for fraction nonconforming has been developed (Ji and AbouRizk 2017). This approach utilizes Bayes' theorem to update a probability distribution for an interested parameter as more information becomes available.

A Bayesian method involves two important distributions, namely the prior distribution and the posterior distribution. A prior distribution of the interested parameter is the probability distribution without knowing any information, while a posterior distribution is an updated version of the prior distribution that incorporates additional information as it becomes available. Due to its ability to dynamically incorporate new data, a Bayesian Inference-based approach is implemented in this study. Additionally, descriptive statistical information, such as mean, median, and all quantiles, can be derived from the generated posterior distribution, resulting in intuitive, informative outputs that are ready to be used for decision-making purposes.

### *Prior Distribution*

In Bayesian statistics, a prior distribution is a rough estimation of the parameter's probability distribution when no information is available (Berger 2013). Bayesian priors come with informative priors and non-informative priors. Informative priors utilize previous observations, experiences, or knowledge to derive parameters that best approximate the distribution of a studied object. Non-informative priors express vague, flat, and diffuse information, which is commonly used when previous data of the studied object are limited. In this case, the studied object is operator quality performance. The prior distribution of an operator's fraction nonconforming is derivable if there are enough inspections performed in the past. However, due to limited historical performance data, it is difficult to compute a comprehensive prior distribution for each operator. To solve this problem, a non-informative prior is used.

For the prior distribution determination, beta distributions are commonly used as the standard conjugate priors for variable p in a binomial distribution (Gelman et al. 2014). The prior distribution is given by

$$P(p) = Beta(p|a,b) = \frac{\Gamma(a+b)}{\Gamma(a)\Gamma(b)} p^{a-1}(1-p)^{b-1} \qquad (4.4)$$

Here, a and b are two positively shaped parameters that control the shape of the fraction nonconforming distribution. $\Gamma(z)$ is the gamma function. In this research, $Beta\left(\frac{1}{2}, \frac{1}{2}\right)$ is used as the non-informative prior to perform the posterior distribution approximation (Black and Thompson 2001).

***Posterior Distribution***

Bayesian inference calculates a posterior distribution through the integration of a prior distribution and a likelihood function (Black and Thompson 2001). In this research, the fraction nonconforming $p$ is the parameter of interest. The prior distribution of fraction nonconforming $p$ is $P(p)$, which provides an estimation of distribution before any collected observation. The likelihood function $L(X|p)$ describes the distribution of the observation given the fraction nonconforming $p$. $P(X)$ is the marginal distribution of the observation $X$. The posterior distribution $P(p|X)$ represents information in observation x together with information expressed in the prior distribution. In general, a posterior distribution is obtained by updating a prior distribution using observed data. Based on Bayes' Theorem, the posterior distribution $P(p|X)$ can be derived as Eq. (4.5).

$$P(p|X) = \frac{L(X|p) \times P(p)}{P(X)} \propto L(X|p)\,P(p) \qquad (4.5)$$

Keeping the factors that only depend on p, the prior distribution follows

$$P(p) \propto p^{a-1}(1-p)^{b-1} \qquad (4.6)$$

And, the likelihood function follows

$$L(X|p) \propto p^{X}(1-p)^{n-X} \qquad (4.7)$$

Therefore, the posterior distribution of fraction nonconforming $p$ can be obtained by multiplying the prior distribution by the likelihood function. Keeping only the factors dependent on $p$, the posterior distribution has the form

$$P(p|X) \propto L(X|p)\,P(p) \propto p^{X+a-1}(1-p)^{n-X+b-1} \qquad (4.8)$$

In the next section, a Metropolis-Hastings algorithm-based numerical solution for fraction nonconforming estimation will be implemented to approximate the posterior distribution $P(p|X)$.

### 4.5.3 Metropolis-Hastings Algorithm

The Metropolis-Hastings algorithm is an MCMC-based method of generating random samples from a probability distribution. It was developed by Metropolis et al. (Metropolis et al. 1953) and generalized by Hastings (Carlo et al. 1970). In the last decade, the Metropolis-Hastings algorithm has become increasingly popular within the statistical community for approximating distributions (Robert and Casella 2011).

In Chapter 2, a specialised Metropolis-Hastings algorithm has been developed for determining the posterior distribution and credible interval for fraction nonconforming. The accuracy of the MCMC-based algorithm has been previously validated by comparing the derived analytical solution to that obtained using criteria of Mean Absolute Error (MAE) and Root Mean Square Error (RMSE) (Figure 2.10/Table 2.5). Validation details and results have been previously described in Section 2.7.3. In the present study, this specialised algorithm is implemented to quantify operator welding quality performance (i.e., fraction nonconforming).

The Metropolis-Hastings algorithm constructs a Markov chain of fraction nonconforming values for $\{p^{(1)}, p^{(2)}, p^{(3)}, \ldots, p^{(N)}\}$. The value $p^{(i+1)}$ is decided by proposing a random move conditional on the previous value $p^{(i)}$ and on the ratio of $\frac{P(p^*|X)}{P(p^{(i)}|X)}$. This acceptance ratio indicates the probably of the new proposed sample with respect to the current sample. The move is accepted if the new sample is more probable than the existing sample. Otherwise, the move is accepted with the acceptance probability, or the move is rejected. When these conditions are met, the Markov chain of parameter values will remain in the high-density region and will converge to the target distribution $P(p|X)$. As the sampling effort is concentrated in the area with higher posterior density, the time required for obtaining an acceptable convergence is typically shorter than that of other sampling techniques.

Here, the specialised Metropolis-Hastings algorithm is illustrated in a step-by-step algorithmic form with the initial value $p^{(0)}$ and repeated for $i = 1, 2, 3, \ldots, N$.

Step 1. Choose a new proposed value $p^*$, such that $p^* = p^{(i)} + \Delta p$ , where $\Delta p \sim N(0, \sigma)$.

Step 2. Calculate the ratio $\rho = \min\{1, \frac{P(p^*|X)}{P(p^{(i)}|X)}\}$, where $P(p|X)$ is the posterior

distribution. As discussed in the previous section, the posterior distribution has

the form

$$P(p|X) \propto L(X|p) \, P(p) \propto p^{X+a-1}(1-p)^{n-X+b-1}$$

Therefore, the ratio ρ can be calculated as

$$\rho = min\{1, \frac{p^{*X+a-1}(1-p^*)^{n-X+b-1}}{p^{(i)X+a-1}(1-p^{(i)})^{n-X+b-1}}\}$$

Step 3. Sample $\mu \sim U_{[0,1]}$.

Step 4. If $\mu < \rho$

$$p^{(i+1)} = p^*$$

or else

$$p^{(i+1)} = p^{(i)}$$

Step 5. Return the values $\{p^{(1)}, p^{(2)}, p^{(3)}, \dots, p^{(N)}\}$.

---

The draws $\{p^{(1)}, p^{(2)}, p^{(3)}, \dots, p^{(N)}\}$ are regarded as a sample from the targeted distribution $P(p|X)$. Only after the chain has passed the transient phase can the impact of the initial value be ignored. After obtaining the fraction nonconforming values, $\{p^{(1)}, p^{(2)}, p^{(3)}, \dots, p^{(N)}\}$, a frequency histogram plot is generated.

### 4.5.4   A/B Testing Algorithm

A/B testing has been widely used in many consumer-facing web technology companies to guide product development and data-driven decisions (Gui et al. 2015). A/B testing is an important algorithm to compare dissimilarities between two sampled distributions. In this study, A/B testing is used to compare how much greater the fraction nonconforming (FN) of one operator is compared to another. This calculation can be mathematically expressed as a percentage probability $P(FN_A > FN_B)$. The closer a probability is to 100%, the greater the probability that operator A has a greater fraction nonconforming than operator B (i.e., it is more likely that operator B has better quality performance). The closer a probability is to 50%, the smaller the probability that operator A has a greater fraction nonconforming than operator B (i.e., it is less likely that operator B has better quality performance). A detailed explanation is provided in the following paragraph to discuss how to implement an A/B testing algorithm to compare operator quality performance using the obtained empirical distributions.

Typically, there are four ways to compute distributions' dissimilarities using A/B testing: numerical integration, closed-form solution, closed-form approximation, and simulation of posterior draws (Robinson 2017). The first three methods require exact parameters or derivation of parameters to represent distributions. Since the proposed Metropolis-Hastings algorithm is only able to derive a numerical approximation of a fraction nonconforming distribution, an ideal method is to use simulation of posterior draws. Using an A/B testing method to solve $P(FN_A > FN_B)$ essentially asks the question, "if a random draw is picked from operator A's distribution, and a random draw is picked from operator B's distribution, what is the probability operator A's sampled fraction nonconforming is higher than Operator B?" Rather than deriving complicated mathematical equations, a simulation method is used to compare the generated empirical distributions and to calculate a percentage probability. Notably, the simulation method also allows the user to choose the sample size, thereby increasing the accuracy of the probabilistic comparison.

Here, the A/B testing algorithm is illustrated in a step-by-step pseudocode form as follows:

Step 1. Input the run times $n$.

Step 2. Initialize counter $i$ to zero.

Step 3. Initialize comparison result $m$ to zero.

Step 4. While $(i < n)$

       Sample one value $x_i$ from Distribution A

       Sample one value $y_i$ from Distribution B

       If $(x_i \geq y_i)$

           increase $m$ by one

       End If

    End While

Step 5. Calculate $P(FN_A > FN_B) = \frac{m}{n}$.

In summary, sampled values from two distributions, A and B, are compared. $P(FN_A > FN_B)$ is calculated based on the number of instances that a sampled value from distribution A is greater than a sampled value from distribution B divided by total number of comparisons. R is used to implement the A/B testing algorithm to quantify the probabilistic difference between two fraction nonconforming distributions. Practitioners can then use these values to gain a more quantitative and determinate understanding of operator quality performance.

### 4.5.5   Illustrative Example

In this section, data for two welding operators with the same type of pipe weld were used to illustrate the algorithm. After connecting and processing the data, it was determined that, for a certain type of weld, 25 out of 180 welds were inspected as "failure" for operator A, while 10 out of 140 welds were inspected as "failure" for operator B. By performing the specialised Metropolis-Hastings algorithm, the empirical posterior distributions were calculated. The empirical frequency histogram plot and the corresponding boxplot of the results are depicted in Figure 4.2.

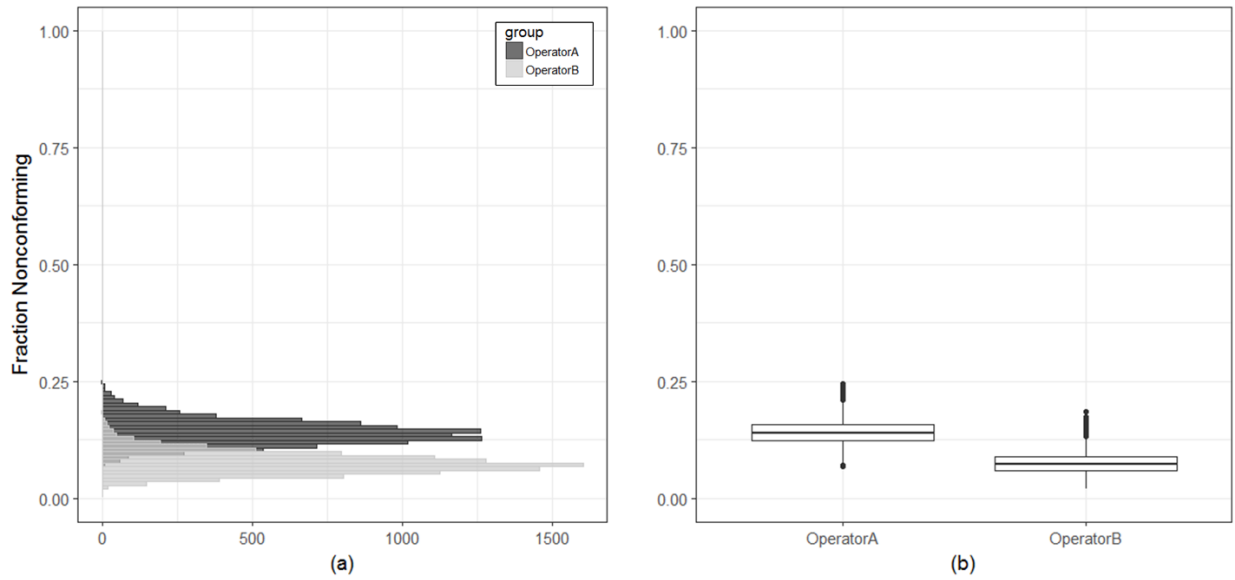

Figure 4.2: Empirical fraction nonconforming distributions for operator A and B: (a) frequency histogram plot; and (b) boxplot.

The frequency histogram plot [Figure 4.2(a)] is plotted based on posterior distributions generated using the Metropolis-Hastings algorithm. Then, the frequency histograms are transformed to boxplots as shown in Figure 4.2(b). A frequency histogram is a graphical representation of a frequency distribution of numerical data, while a boxplot is a graphical representation of a five-number summary [Min, 25% Quantile (Q1), Median, 75% Quantile (Q3), Max]. For a boxplot, the upper whisker extends from the hinge to the highest value that is within 1.5 * IQR of the hinge, where the IQR is the inter-quartile range or distance between the first and third quartiles. The lower whisker extends from the hinge to the lowest value within 1.5 * IQR of the hinge. Data beyond the end of the whiskers are outliers and are plotted as individual points. The minor overlapping region, as shown in Figure 4.2(a), visually indicates that there is a difference between these two distributions. The same pattern can be observed from the boxplot shown in Figure 4.2(b).

As the Metropolis-Hastings algorithm is a numerical approximation to the true posterior distribution for operator welding quality performance, the numerical solution varies slightly during each run of the proposed algorithm. Once the posterior distribution has been generated, statistics such as mean, median, variance, and quantiles can be obtained from the distribution. Moreover, operator quality performance can be compared using the A/B testing algorithm. The statistical software R is utilized to calculate statistical summaries of interest and to perform the A/B testing algorithm. Statistical summaries of the above two operators are listed in Table 4.1.

Table 4.1: Statistical summary of operator A and B quality performance.

| Operator ID | Inspected Welds | Repaired Welds | Min | Q1 | Median | Q3 | Max |
|---|---|---|---|---|---|---|---|
| Operator A | 180 | 25 | 0.060 | 0.123 | 0.139 | 0.158 | 0.247 |
| Operator B | 140 | 10 | 0.015 | 0.060 | 0.075 | 0.090 | 0.181 |

By performing the A/B testing algorithm, $P(FN_A > FN_B)$ is calculated to quantify the probabilistic difference between two operators. It was found that there is 97.5% probability that operator A has higher fraction nonconforming than operator B. Therefore, a conclusion can be drawn that operator A has a greater fraction nonconforming than operator B, indicating that operator B has better performance than operator A for this particular weld type.

## 4.6    Decision Support

In industrial pipe fabrication companies, practitioners are interested in comparing and identifying operators with exceptional quality performance for weld types. By using the proposed approach, analytical outputs, namely quantification, ranking, and comparison of each operator's welding performance for each weld type, can be generated in a near real-time manner, dramatically reducing the data interpretation load of decision makers.

In this section, data regarding pipe (STD, 2, A, BW), the most common work type in the studied company, is used to demonstrate the main outputs of the proposed methodology. 17 operators, who have each had over 100 welds inspected, were selected. Table 4.2 lists the performance records and the five-number summary (Min, Q1, Median, Q3, and Max) of their welding quality performance with respect to fraction nonconforming of these operators. Consistent with Figure 4.2, the median value of their performance is sorted, in descending order, in Table 4.2. Operator IDs were reassigned to maintain employee anonymity.

Table 4.2: Statistical summary of operator welding performance of pipe (STD, 2, A, BW).

| Operator ID | Inspected Welds | Repaired Welds | Min | Q1 | Median | Q3 | Max |
|---|---|---|---|---|---|---|---|
| 1 | 175 | 25 | 0.069 | 0.128 | 0.147 | 0.166 | 0.243 |
| 2 | 111 | 11 | 0.029 | 0.085 | 0.103 | 0.123 | 0.225 |
| 3 | 139 | 13 | 0.036 | 0.080 | 0.096 | 0.113 | 0.193 |
| 4 | 307 | 27 | 0.046 | 0.078 | 0.089 | 0.099 | 0.157 |
| 5 | 100 | 8 | 0.026 | 0.065 | 0.081 | 0.100 | 0.223 |
| 6 | 207 | 16 | 0.028 | 0.068 | 0.080 | 0.093 | 0.159 |
| 7 | 104 | 7 | 0.009 | 0.055 | 0.070 | 0.087 | 0.196 |
| 8 | 119 | 8 | 0.015 | 0.054 | 0.069 | 0.086 | 0.182 |
| 9 | 175 | 11 | 0.023 | 0.053 | 0.064 | 0.076 | 0.138 |
| 10 | 175 | 9 | 0.017 | 0.040 | 0.052 | 0.067 | 0.133 |
| 11 | 120 | 6 | 0.011 | 0.038 | 0.052 | 0.064 | 0.155 |
| 12 | 316 | 16 | 0.020 | 0.043 | 0.051 | 0.059 | 0.097 |
| 13 | 208 | 9 | 0.009 | 0.035 | 0.043 | 0.054 | 0.110 |
| 14 | 123 | 5 | 0.006 | 0.033 | 0.043 | 0.057 | 0.119 |
| 15 | 147 | 5 | 0.005 | 0.025 | 0.035 | 0.047 | 0.097 |
| 16 | 264 | 9 | 0.013 | 0.029 | 0.035 | 0.043 | 0.080 |
| 17 | 355 | 11 | 0.008 | 0.027 | 0.031 | 0.038 | 0.072 |

To visually represent and compare operator welding performance, a side-by-side boxplot was generated, as in Figure 4.2(b), and is shown in Figure 4.3. Box height is inversely proportional to the stability of operator quality performance, where a shorter box is indicative of an operator with stable quality performance for the indicated type of weld. Operators with lower ranks have smaller fraction nonconforming (i.e., better performance). The dashed line represents the overall averaged welding quality performance of the 17 operators (fraction nonconforming = 0.062). The range of the median is relatively wide (0.031 to 0.147), indicating operator performance varies considerably for the same type of welding work. Although root cause analysis may identify factors affecting operator welding performance, it is beyond the scope of this study.

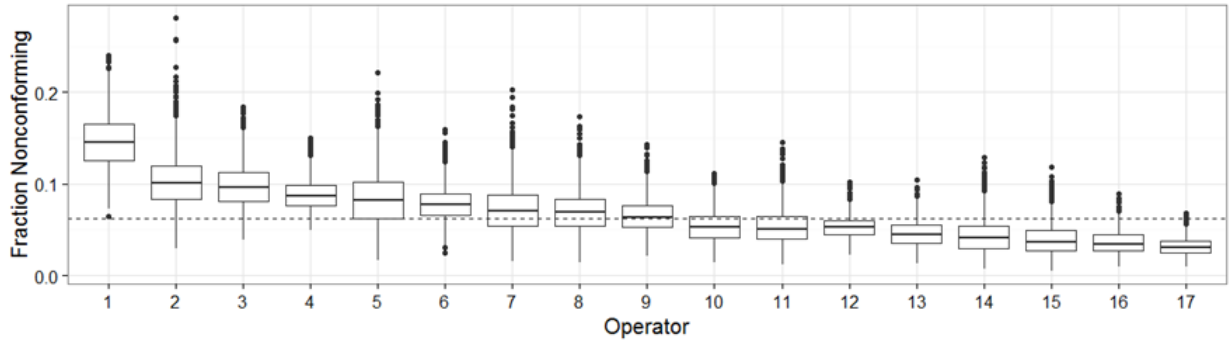

Figure 4.3: Sorted side-by-side boxplot of operators' welding performance with pipe (STD, 2, A, BW).

The side-by-side boxplot only depicts a visual comparison between operators' quality performance and does not serve the purpose of quantitative comparison of performance differences. The A/B testing algorithm is utilized quantify the probabilistic differences between welding operators. Table 4.3 summarizes the probabilistic difference of quality performance between all operators. For example, the first column indicates the probability of operator 1 having better performance than the other operators. Since all operators' fraction nonconforming have been sorted in descending order, all celled numbers are equal to or larger than 50%. For instance, the second row in the first column demonstrates there is a probability of 87% that operator 1 has higher fraction nonconforming (i.e., reduced quality performance) than operator 2. Conversely, it was found that there was a probability of 51% that operator 7 has higher fraction nonconforming than operator 8. These results are consistent with the visualized outcome shown in Figure 4.3 (i.e., operator 7 and operator 8 have near identical boxplots).

Table 4.3: Probabilistic differences of pipe welding quality performance between operators with pipe (STD, 2, A, BW).

| Operator ID (i) | 1 > i | 2 > i | 3 > i | 4 > i | 5 > i | 6 > i | 7 > i | 8 > i | 9 > i | 10 > i | 11 > i | 12 > i | 13 > i | 14 > i | 15 > i | 16 > i | 17 > i |
|---|---|---|---|---|---|---|---|---|---|---|---|---|---|---|---|---|---|
| 1 | 0.50 | | | | | | | | | | | | | | | | |
| 2 | 0.87 | 0.50 | | | | | | | | | | | | | | | |
| 3 | 0.91 | 0.56 | 0.50 | | | | | | | | | | | | | | |
| 4 | 0.97 | 0.64 | 0.57 | 0.50 | | | | | | | | | | | | | |
| 5 | 0.94 | 0.67 | 0.63 | 0.58 | 0.50 | | | | | | | | | | | | |
| 6 | 0.98 | 0.74 | 0.69 | 0.64 | 0.54 | 0.50 | | | | | | | | | | | |
| 7 | 0.98 | 0.80 | 0.77 | 0.74 | 0.65 | 0.63 | 0.50 | | | | | | | | | | |
| 8 | 0.98 | 0.80 | 0.77 | 0.75 | 0.64 | 0.63 | 0.51 | 0.50 | | | | | | | | | |
| 9 | 0.99 | 0.86 | 0.84 | 0.83 | 0.72 | 0.72 | 0.56 | 0.57 | 0.50 | | | | | | | | |
| 10 | 1.00 | 0.94 | 0.92 | 0.93 | 0.84 | 0.85 | 0.72 | 0.73 | 0.69 | 0.50 | | | | | | | |
| 11 | 1.00 | 0.92 | 0.91 | 0.90 | 0.83 | 0.83 | 0.72 | 0.72 | 0.68 | 0.51 | 0.50 | | | | | | |
| 12 | 1.00 | 0.96 | 0.96 | 0.97 | 0.87 | 0.89 | 0.76 | 0.77 | 0.74 | 0.54 | 0.52 | 0.50 | | | | | |
| 13 | 1.00 | 0.97 | 0.97 | 0.98 | 0.91 | 0.93 | 0.81 | 0.83 | 0.80 | 0.63 | 0.61 | 0.62 | 0.50 | | | | |
| 14 | 1.00 | 0.96 | 0.95 | 0.96 | 0.90 | 0.91 | 0.80 | 0.82 | 0.80 | 0.64 | 0.62 | 0.64 | 0.53 | 0.50 | | | |
| 15 | 1.00 | 0.98 | 0.98 | 0.99 | 0.94 | 0.96 | 0.88 | 0.89 | 0.88 | 0.76 | 0.73 | 0.76 | 0.66 | 0.62 | 0.50 | | |
| 16 | 1.00 | 0.99 | 0.99 | 1.00 | 0.96 | 0.98 | 0.92 | 0.93 | 0.92 | 0.81 | 0.79 | 0.82 | 0.70 | 0.66 | 0.54 | 0.50 | |
| 17 | 1.00 | 1.00 | 1.00 | 1.00 | 0.98 | 0.99 | 0.95 | 0.96 | 0.96 | 0.88 | 0.85 | 0.90 | 0.79 | 0.73 | 0.61 | 0.60 | 0.50 |

From these visualized and quantitative outputs, practitioners can (1) infer operators' skill level for a specified type of work; (2) identify operators who consistently produce high-quality welds; (3) understand the probabilistic difference between operators' performances; and (4) support future decision-making processes. These results can have considerable, positive impacts on company quality and productivity performance. Potential applications of the research outcomes are discussed in the following section.

## 4.7    Potential Applications

To strategically improve industrial pipe fabrication companies' competitiveness and reputation within the construction market, operators with exceptional welding quality performance should be effectively and efficiently utilized. Here, three potential applications of the proposed research outcome are identified from the perspectives of (1) production planning; (2) employee training; and (3) strategic recruiting. These applications may lead to considerable improvements in pipe welding quality, while potentially reducing cost and schedule overruns induced by poor quality welds.

### 4.7.1    Production Planning

Operator welding quality performance varies between weld types due to differences in skill level. Development and implementation of an optimal, quality-driven production planning optimization engine, which would allocate welding work to operators with highest quality performance for that particular work/weld type, could directly improve overall welding quality performance.

### 4.7.2    Employee Training

Following the implementation of the proposed methodology, practitioners will be able to determine which operators are most proficient for each weld type. These operators should be invited to demonstrate their production processes and share their professional knowledge for employee training purposes. Companies may also standardize the operation processes as per the high-performance operators' operation processes. Enhanced training programs are crucial for improving the welding quality performance of all operators.

### 4.7.3    Strategic Recruiting

Once high-performance operators are identified, their profiles can be examined to identify critical success factors common to this group. Data mining techniques may be implemented to facilitate complex analytical processes. Human resource teams may strategically conduct data-driven recruiting to hire qualified welding operators who are characterized by these success factors.

### 4.8    Conclusion

Nowadays, huge amounts of project operation data are tracked and stored by construction companies via various enterprise resource planning (ERP) management systems, databases, and software solutions. However, due to the lack of analytically-based decision-support systems, many companies fail to extract valuable information to improve their performance. The construction industry is a labour-intensive industry whose quality performance is heavily dependent on operator performance. A more quantitative understanding of operator quality performance is expected to enhance companies' quality performance and to, in turn, improve market competitiveness.

This chapter proposes an integrated data-driven approach to determine the posterior distribution for quantifying operator welding quality performance for a specific work type and utilizes the A/B testing algorithm to compare the probabilistic differences between operators' quality performances. Real quality management data and engineering design data from a pipe fabrication company, in Edmonton, Canada, were extracted and utilized to demonstrate the applicability and feasibility of the proposed approach. Operator welding quality performances were first represented by posterior distributions to incorporate uncertainty. Then, an A/B testing algorithm was applied to compare operators' performances. Statistics, such as the five-point summary, were easily obtained from the posterior distribution to support decision-making processes.

Practitioners can implement this approach to (1) infer operators' skill level for a specified type of work; (2) calculate probabilistic quality performance differences between operators; (3) identify operators who consistently produce high-quality welds; and (4) support future decision-making processes and improve production planning, employee training, and strategic recruiting.

Although this research provides accurate and interpretable measurements for quantifying welding operator quality performance, it does not consider other types of performance factors, such as productivity performance and safety performance. Investigation of additional types of data to evaluate overall operation performance for welding operators will be required to obtain more comprehensive performance assessments.

## 4.9   Acknowledgements

This research is funded by the NSERC Collaborative Research & Development (CRD) Grant (CRDPJ 492657). The authors would like to acknowledge Rob Reid, Doug McCarthy, Jason Davio, and Christian Jukna for sharing their knowledge and expertise of pipe fabrication quality management.

## 4.10   References

American Society of Mechanical Engineers. (2001). "Power piping: ASME Code for Pressure Piping, B31 (ASME B31.1-2001)." *An American National Standard*.

Berger, J. O. (2013). *Statistical decision theory and Bayesian analysis*. Springer Science & Business Media.

Black, T. C., and Thompson, W. J. (2001). "Bayesian data analysis." *Computing in Science & Engineering*, CRC press Boca Raton, FL, 3(4), 86–91.

Carlo, M., Methods, S., Markov, U., and Applications, T. (1970). "Monte Carlo sampling methods using Markov chains and their applications." *Biometrika*, 57(1), 97–109.

Chini, A. R., and Valdez, H. E. (2003). "ISO 9000 and the U.S. Construction Industry." *Journal of Management in Engineering*, 19(2), 69–77.

Gelman, A., Carlin, J. B., Stern, H. S., Dunson, D. B., Vehtari, A., and Rubin, D. B. (2014). *Bayesian data analysis*. CRC press Boca Raton, FL.

Gui, H., Xu, Y., Bhasin, A., and Han, J. (2015). "Network A/B Testing: : From Sampling to Estimation." *Proceedings of the 24th International Conference on World Wide Web - WWW '15*, ACM Press, 399–409.

Ji, W., and AbouRizk, S. M. (2017). "Credible interval estimation for fraction nonconforming:

Analytical and numerical solutions." *Automation in Construction*, 83, 56–67.

Metropolis, N., Rosenbluth, A. W., Rosenbluth, M. N., Teller, A. H., and Teller, E. (1953).

"Equation of State Calculations by Fast Computing Machines." *The Journal of Chemical Physics*, 21(6), 1087–1092.

Montgomery, D. C. (2007). *Introduction to statistical quality control*. John Wiley & Sons.

O'Connor, J. T., O'Brien, W. J., and Choi, J. O. (2016). "Industrial Project Execution Planning: Modularization versus Stick-Built." *Practice Periodical on Structural Design and Construction*, 21(1), 4015014.

Robert, C., and Casella, G. (2011). "A Short History of Markov Chain Monte Carlo: Subjective Recollections from Incomplete Data." *Statistical Science*, 26(1), 102–115.

Robinson, D. (2017). "Introduction to Empirical Bayes: Examples from Baseball Statistics."

Weaver, B. P., and Hamada, M. S. (2016). "Quality quandaries: A gentle introduction to Bayesian statistics." *Quality Engineering*, 28(4), 508–514.

# 5 CHAPTER 5: COMPLEXITY ANALYSIS APPROACH FOR PREFABRICATED CONSTRUCTION PRODUCTS USING UNCERTAIN DATA CLUSTERING

## 5.1 Introduction

As the implementation of modular construction expands, an increasing number of prefabricated construction products are being engineered and manufactured in fabrication shops. Construction products are heterogeneous in nature and are characterized by various combinations of design attributes, which, in turn, impacts the complexity involved in producing or assembling these products. As product complexity increases, so too do the skills, knowledge, management efforts (e.g. training and quality control), and resource support (e.g. specialized tools and technologies) required for successful performance. Inadequate management of product complexity, therefore, can result in cost and schedule overruns and can hamper overall project delivery.

The heterogeneous nature of construction product design, together with various levels of production knowledge and skill, has made the quantification of product complexity difficult in practice. In recent years, researchers have successfully correlated product complexity with product quality performance in the manufacturing industry, indicating that, for practical purposes, product quality performance can be used as an indicator or surrogate marker of product complexity (Antani 2014; Novak and Eppinger 2001; Williams 1999). Recently, analytically-based quality management systems, which facilitate quantitative quality performance measurements at a product-level with design information associated, have been developed (Ji and AbouRizk 2017a; Ji and AbouRizk 2017b). Integrating these systems to generate a single indicator of product quality performance from which product complexity can be inferred would alleviate the need of practitioners to perform detailed, time-consuming analyses of complex, unreliable, subjective factors (e.g. design information and operator knowledge and skill). A quality performance-based product complexity indicator, however, has yet to be defined or developed within the construction domain.

The aim of the chapter is to develop, validate, and implement an uncertain data clustering approach that is capable of quantifying and clustering quality performance-based product complexity indicators (hereafter referred to as product complexity indicators) from quality management and engineering design information. Specifically, the proposed approach has developed a framework that can provide (1) accurate and reliable measurements of product complexity indicator uncertainty; (2) meaningful assessments of product complexity indicator distribution similarity; and (3) an interpretable clustering of products with similar complexity indicators. The content of this paper is organized as follows: First, a comprehensive literature review is provided to demonstrate the rationale of the proposed research. Then, details of the methodology are introduced. To elaborate on the implementation of the proposed approach, an illustrative example is provided. Finally, the feasibility and applicability of the proposed approach are validated following a practical case study of industrial pipe weld complexity analysis. In addition to providing simplified, interpretable, and informative insights for understanding construction product complexity using quality management and engineering design information, this research also develops a novel Hellinger distance-based hierarchical clustering technique for grouping uncertain data (i.e., probability distributions).

## 5.2   Rationale

### 5.2.1   Product Complexity

In the construction research domain, construction project complexity has been primarily investigated from four perspectives: (1) influencing factors contributing to project complexity;

(2) the impact of project complexity; (3) project complexity measurement methods; and (4) management of project complexity (Luo et al. 2017). Throughout these studies, product complexity has been found to influence overall project complexity (Baccarini 1996; Senescu et al. 2012; Williams 1999). In spite of these findings, product complexity has not been conceptually defined and thoroughly analyzed within construction management literature. In this study, the authors define product complexity as:

*"The level of constructing difficulty based on the product's design and on the knowledge and ability of an operator to construct a product given its specific design information."*

This definition is consistent with informal statements in product design and development literature (Baldwin and Clark 2000; Galvin and Morkel 2001; Novak and Eppinger 2001). For instance, Novak and Eppinger (2001) stated that "[t]he effect of this product design choice on the outsourcing decision can be profound, as greater product complexity gives rise to coordination challenges during product development."

Interviews conducted with five industrial construction companies in Edmonton, Canada, highlighted the difficulty that these organization have with determining product complexity from design information. Currently, complexity is assessed by examining the detailed design information of each product type. Given that there may be hundreds of product types in a single project, establishing product complexity is often a costly, time-intensive endeavor. Practitioners would benefit from the development of a framework that could rapidly generate a simple, reliable indicator of product complexity for estimating purposes. Several researchers in manufacturing literature have indicated that complexity can be reliably estimated from quality performance data (Antani 2014; Novak and Eppinger 2001).

While quality performance data are captured in practice, prefabricated products are often inspected as either conforming or nonconforming to specified quality standards and quality performance data cannot, therefore, be represented numerically (Ji and AbouRizk 2017a). Research conducted by Ji and AbouRizk (2017a) has quantitatively solved this issue by providing a Bayesian statistics-based analytical solution (i.e., a beta distribution) to estimate fraction nonconforming performance uncertainty at a product-level through the investigation of both quality management and engineering design information. The issue of assessing product complexity indicator in construction is further complicated by the myriad of prefabricated products that may be involved during project delivery. Although complexity of each product may be quantified, these data must be reduced into a format that is simple, interpretable, and informative for industry professionals (e.g. design and operations personnel). Notably, however, the uncertain nature of product complexity renders traditional clustering methods inappropriate for solving uncertain data clustering problems. A method capable of rapidly and reliably estimating product complexity and clustering hundreds of products of similar complexity into a manageable number of classification groups would improve the practice of product complexity analysis and management.

### 5.2.2  Uncertain Data Clustering

In data mining, cluster analysis or clustering is the process of partitioning a set of objects in such a way that objects in a cluster are more similar to one another than to the objects in other clusters. An advantage of data clustering is that clustering can, automatically, lead to the discovery of previously unknown groups within data. Clustering as a standalone tool can be implemented to gain insights into the distribution of data and to observe the characteristics of each cluster (Han et al. 2011).

For many application domains, the ability to unearth valuable knowledge from a dataset is impaired by unreliable, erroneous, obsolete, imprecise, and noisy data (Schubert et al. 2015; Züfle et al. 2014)—or, in other words, uncertain data that is commonly described by a probability distribution (Jiang et al. 2013; Pei et al. 2007). Uncertain data are found in modeling situations where a mathematical model only approximates the actual nonconforming quality control process. Clustering uncertain data (i.e., probability distributions) is associated with substantial challenges concerning modeling similarity between uncertain objects and regarding the development of efficient computational methods (Jiang et al. 2013). Traditional clustering methods, such as partitioning-based clustering methods (e.g. k-means) and density-based clustering methods (e.g. DBSCAN), are dependent on geometric distances (e.g. Euclidean distance and Manhattan distance) between observations (Han et al. 2011). Such distances are not capable of grouping uncertain objects that are geometrically indistinguishable, such as products with similar repair rates that vary in terms of quality performance.

Jiang et al. (2013) were the first to use Kullback-Leibler (KL) divergence, which is a special type of $f$-divergence to measure distribution similarity, for uncertain data clustering problems. However, computing KL divergence to measure the similarity between complex distributions is very time-consuming and may even be infeasible (Jiang et al. 2013). The derivation of KL divergence between two beta distributions involves calculations of complicated digamma functions, thereby requiring additional computational efforts.

The Hellinger distance is another type of $f$-divergence that is widely used to quantify the similarity between two probability distributions in the field of statistics. The Hellinger distance, however, has yet to be used for solving uncertain data clustering problems in the data mining domain. In contrast to KL divergence, an analytical solution for measuring the similarity of beta distributions, which largely reduces the computational complexity of uncertain data clustering problems, exists. Therefore, the Hellinger distance is used in this research to model the similarity between distributions for product complexity indicator clustering purposes. The mathematical proof is provided in Appendix 1. Notably, the Hellinger distance-based uncertain data clustering method proposed here can be further generalized to other types of uncertain data (i.e., probability distributions).

## 5.3   Methodology

The proposed methodology is conducted following three steps. First, to measure the prefabricated construction product complexity indicator, a Bayesian statistics-based quality performance measurement (i.e., a posterior distribution of fraction nonconforming), which incorporates uncertainty, is introduced. Second, to develop a systematic product complexity indicator scoring approach, the Hellinger distance is used to measure complexity indicator similarity between various types of products. Finally, to cluster product complexity indicators into homogeneous groups, the agglomerative hierarchical clustering technique is adopted using the obtained Hellinger distance-based similarity measure. Details of the systematic and theoretical analysis of these steps are discussed as follows.

### 5.3.1 Step 1. Quality Performance-based Product Complexity Indicator

To quantitatively measure product complexity, a quality performance indicator termed fraction nonconforming, which represents the ratio of the number of nonconforming items $X$ in the sample to the sample size $n$, is utilized. Fraction nonconforming can be mathematically expressed as Eq. (5.1) (Montgomery 2007).

$$\hat{p} = \frac{X}{n} \qquad (5.1)$$

To appropriately incorporate the sampling uncertainty of the population fraction nonconforming variable $p$ when data are obtained from a sample, a Bayesian statistics-based analytical solution has been developed to determine the posterior distribution of the fraction nonconforming $p$ (Ji and AbouRizk 2017a). The posterior distibution uses a non-informative prior distribution $Beta(1/2,\ 1/2)$. It is given as Eq. (5.2).

$$P(p|X) = Beta(X + 1/2, n - X + 1/2) \qquad (5.2)$$

This Bayesian statistics-based solution, which is capable of updating the posterior distribution by combining previous knowledge and real-time data, has been demonstrated to be more accurate, reliable, and interpretable than the traditional statistical methods (Ji and AbouRizk 2017a).

As discussed previously, product complexity has been found to be positively correlated with product quality performance. In this research, the fraction nonconforming $p$ is used to assess the product complexity indicator, termed $Cplx$. Therefore, the posterior distribution of $Cplx$ is identical to the posterior distribution of the fraction nonconforming, as shown in Eq. (5.3).

$$P(Cplx|X) = P(p|X) = Beta(X + 1/2, n - X + 1/2) \tag{5.3}$$

This posterior distribution measures the product complexity indicator for a certain type of construction product. In the following step, the complexity indicator distribution similarity measurement and the complexity indicator scoring approach are introduced to evaluate product complexity in a systematic and interpretable way.

### 5.3.2  Step 2. Product Complexity Indicator Scoring

In this step, the product complexity indicator is scored by accounting for uncertainty. The Hellinger distance is introduced to measure the distribution similarity of product complexity indicators. The distances obtained for paired products are used to construct a Hellinger distance matrix, which is required for product complexity indicator clustering as follows.

To score the product complexity indicator, the distribution similarity of product complexity indicators between all types of products should be measured. A significant challenge in modeling distribution similarity is that the distribution similarity cannot be captured by geometric distances, such as the Euclidean distance or the Manhattan distance. In statistics, $f$-divergence is a function, $D_f(P||Q)$, that measures the similarity between two probability distributions (Liese and Vajda 2006). The Hellinger distance is a special case of $f$-divergence, which was defined in terms of the Hellinger integral by Ernst Hellinger in 1909 (Hellinger 1909). The reason for choosing Hellinger distance is that, for measuring the similarity between two beta distributions, the Hellinger distance has a closed-form solution, which largely reduces the computational efforts compared to other cases of $f$-divergences.

Conceptually, the Hellinger distance between two distributions, $P = \{p_i\}$ i $\in$ [n] and $Q = \{q_i\}$ i $\in$ [n], is defined as Eq. (5.4) (Hellinger 1909).

$$H(P, Q) = \frac{1}{\sqrt{2}} \left\| \sqrt{P} - \sqrt{Q} \right\|_2 = \frac{1}{\sqrt{2}} \sqrt{\int \left( \sqrt{p_i} - \sqrt{q_i} \right)^2} \qquad (5.4)$$

To measure the product complexity indicator similarity, the specialized Hellinger distance between two Beta distributions, $X_t \sim Beta(a_1, b_1)$ and $Y_t \sim Beta(a_2, b_2)$, is derived as Eq. (5.5).

$$H(X_t, Y_t) = \sqrt{1 - \frac{Beta(\frac{a_1 + a_2}{2}, \frac{b_1 + b_2}{2})}{\sqrt{Beta(a_1, b_1) \times Beta(a_2, b_2)}}} \qquad (5.5)$$

Where $0 < H(X_t, Y_t) < 1$. The Hellinger distance represents the similarity measurement between two product complexity indicator distributions: the larger the distance, the smaller the similarity between the distributions. A detailed mathematical proof for the closed-form solution is provided in Section 5.8.

Using the calculated Hellinger distances for all pairs of products, a two-dimensional distance matrix is constructed. The obtained Hellinger distance matrix ($M = (x_{ij})$ with $1 \leq i, j \leq N$) is a distance matrix containing all complexity indicator similarity measurements. The entry $x_{ij}$ represents the similarity measurement between product types $i$ and $j$. The obtained matrix always adheres to the following properties: (1) the entries on the main diagonal are all zero (i.e. $x_{ij} = 0$ for all $1 \leq i \leq N$); (2) all the off-diagonal entries are in the range of 0 to 1 (i.e. $0 \leq x_{ij} \leq 1$ if $i \neq j$); and (3) the matrix is symmetric ($x_{ij} = x_{ji}$).

This Hellinger distance matrix, however, can only demonstrate the quantitative distribution similarity measurements for each pair of product types. To determine the sequence of the complexity indicator scores of all product types, medians of the posterior distributions of $Cplx_i$ are compared. $P(0.5, Cplx_i|X_i)$ represents the 50% quantile (i.e., median) of the $Cplx$ distribtution. Therefore, the most non-complex product can be searched by indexing $Min(P(0.5, Cplx_i|X_i))$. If multiple distributions possess the same median, the distribution with the smaller variation is considered the less complex product.

Here, it is assumed that the ascendingly sorted product complexity indicator scores follow the sequence $(Cplx\ Score_n)_{n \in N}$, which denotes a sequence whose $nth$ element is given by the variable $Cplx\ Score_n$. $P_n$ is the probability distribution of the $nth$ scored product complexity indicator in the sequence $(Cplx\ Score_n)_{n \in N}$, and the sequence of $Cplx\ Score_n$ is defined by the recurrence relation expressed as Eq. (5.6):

$$Cplx\ Score_n = Cplx\ Score_{n-1} + H(P_n, P_{n-1})$$

$$\text{With seed value } Cplx\ Score_1 = 0$$

(5.6)

Where, $H(P_n, P_{n-1})$ represents the Hellinger distance between the $nth$ and $(n-1)th$ distributions of the sequenced product complexity indicators. Explicitly, the recurrence yields the following equations:

$$Cplx\ Score_2 = Cplx\ Score_1 + H(P_2, P_1)$$

$$Cplx\ Score_3 = Cplx\ Score_2 + H(P_3, P_2)$$

(5.7)

$$Cplx\ Score_4 = Cplx\ Score_3 + H(P_4, P_3)$$

$$\dots$$

All the involved Hellinger distances are available and can be indexed from the obtained Hellinger distance matrix. By using the recurrence relation defined as Eq. (5.6), complexity indicator scores for all types of products can be calculated. After all complex scores are derived, they are scaled to a range from 0 to 10, where a score of 10 represents the most complex product.

While the complexity scoring is used for clustering purposes, it also has a practical benefit. Transformation of uncertain quality performance distributions (i.e., beta distributions) into deterministic numbers, ranging from 0 to 10, can also reduce the interpretation load of practitioners, particularly for non-quality associated industrial personnel.

### 5.3.3 Step 3. Product Complexity Indicator Clustering

A method capable of clustering products of similar complexity indicators would improve product analysis and management, especially when a vast number of product types are involved. A useful summarization tool, which provides an interpretable visualization of the data, is, therefore, needed. Among multiple clustering techniques, hierarchical clustering is selected due to its ease of use and to the interpretability of the results. In addition, compared with partitioning-based clustering methods (e.g., k-means) and density-based clustering methods (e.g., DBSCAN), hierarchical clustering avoids treating data as an outlier. This characteristic is desired in this product complexity clustering problem because each type of product should be clustered into a complexity group rather than be excluded as an outlier. In data mining, hierarchical clustering is a method of cluster analysis that works by grouping similar data objects into a hierarchy or "tree" of clusters (Han et al. 2011). Visualizing this hierarchy provides a useful visual summary of the data. The agglomerative hierarchical clustering method begins by treating each object as an individual cluster and then iteratively merging clusters into larger and larger clusters until all objects are merged into a single cluster. To determine which clusters should be combined, a measure of similarity between sets of clusters is required. This is achieved by selecting an appropriate metric—in this case, the Hellinger distance—and a complete-linkage criterion that specifies the similarity of clusters as a function of the pairwise distances of the observations within the clusters. The complete-linkage criterion considers the distance between two clusters to be equal to the largest distance from any member of one cluster to any member of the other cluster. Complete-linkage tends to find compact clusters of approximately equal diameters and achieves more accurate clustering results.

The hierarchy of clusters can be represented as a tree structure called a dendrogram. Leaves of the dendrogram consist of one item as an individual cluster, while the root of the dendrogram contains all items belonging to one cluster. Internal nodes represent clusters formed by merging clusters of children, and the algorithm results in a sequence of groupings. The user then selects a "natural" clustering from this sequence.

## 5.4   Illustrative Example

To demonstrate the proposed methodology, an illustrative example has been developed. Quality inspection results (i.e., the number of inspected items and the number of repaired items) of eight types of products are detailed in Table 5.1. Each product type represents products with the same combination of design attributes. These data and the proposed approach are used to assess product complexity indicator, score the product complexity indicator level, and cluster product complexity indicators.

Table 5.1: Quality inspection results of eight types of products.

| Product Type | Number of Inspected Items | Number of Repaired Items |
|:---:|:---:|:---:|
| 1 | 200 | 5 |
| 2 | 170 | 4 |
| 3 | 50 | 2 |
| 4 | 48 | 2 |
| 5 | 100 | 2 |
| 6 | 99 | 2 |
| 7 | 98 | 4 |
| 8 | 101 | 4 |

### 5.4.1   Step 1. Product Complexity Indicator

Following Eq. (5.3), theoretical distributions of product complexity indicators and median values of these distributions are derived as indicated in Table 5.2. When comparing the center of non-symmetric distributions, the median is the most appropriate statistical estimation. Accordingly, types 5 and 4 are the least and most complex products, respectively.

Table 5.2: Product complexity distributions and median values.

| Product Type | $P(Cplx\|X)$ | $P(0.5, Cplx_i\|X_i)$ |
|:---:|:---:|:---:|
| 1 | Beta (5.5, 195.5) | 0.0258 |
| 2 | Beta (4.5, 166.5) | 0.0245 |
| 3 | Beta (2.5, 48.5) | 0.0432 |
| 4 | Beta (2.5, 46.5) | 0.0450 |
| 5 | Beta (2.5, 98.5) | 0.0217 |
| 6 | Beta (2.5, 97.5) | 0.0219 |
| 7 | Beta (4.5, 94.5) | 0.0424 |
| 8 | Beta (4.5, 97.5) | 0.0412 |

To visualize the theoretical beta distributions, a side-by-side box plot is developed by calculating the five-number summary (Min, Q1, Median, Q3, Max) and is illustrated in Figure 5.1. Medians of types 1, 2, 5, and 6 are approximately 2%, while product types 3, 4, 7, and 8 are approximately 4%. For paired types 1+2, 3+4, 5+6, and 7+8, each group has similar distribution spreads. Therefore, to account for the uncertainty of these distributions, the expected clusters should be product types 1+2, 3+4, 5+6, and 7+8.

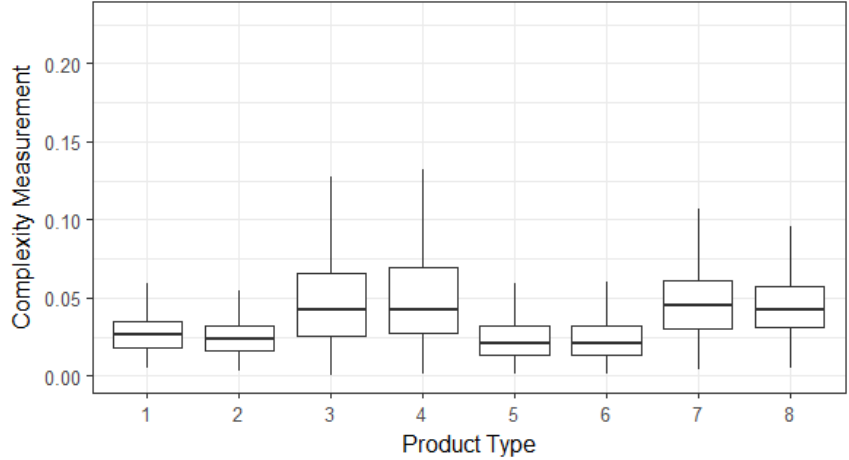

Figure 5.1: Side-by-side boxplot for eight types of product complexity measurements.

### 5.4.2  Step 2. Product Complexity Indicator Scoring

By implementing the derived Hellinger distance equation for the two beta distributions, an $8 \times 8$ symmetric Hellinger distance matrix is constructed, as shown in Eq. (5.8). This matrix corresponds to all the previously discussed properties of the Hellinger distance matrix and is used to perform the product clustering analysis.

$$M = \begin{bmatrix} 0.0000 & 0.0602 & 0.4100 & 0.4290 & 0.2109 & 0.2090 & 0.3900 & 0.3694 \\ 0.0602 & 0.0000 & 0.4023 & 0.4219 & 0.1566 & 0.1552 & 0.3989 & 0.3789 \\ 0.4100 & 0.4023 & 0.0000 & 0.0232 & 0.3737 & 0.3688 & 0.1604 & 0.1674 \\ 0.4290 & 0.4219 & 0.0232 & 0.0000 & 0.3937 & 0.3888 & 0.1703 & 0.1796 \\ 0.2109 & 0.1566 & 0.3737 & 0.3937 & 0.0000 & 0.0057 & 0.4100 & 0.3936 \\ 0.2090 & 0.1552 & 0.3688 & 0.3888 & 0.0057 & 0.0000 & 0.4046 & 0.3881 \\ 0.3900 & 0.3989 & 0.1604 & 0.1703 & 0.4100 & 0.4046 & 0.0000 & 0.0230 \\ 0.3694 & 0.3789 & 0.1674 & 0.1796 & 0.3936 & 0.3881 & 0.0230 & 0.0000 \end{bmatrix} \quad (5.8)$$

As per the medians from $P(0.5, Cplx_i | X_i)$ shown in Table 5.2, type 5 is characterized as the least complex. The product complexity indicator score can then be calculated through the recurrence relation as Eq. (5.6) by indexing the corresponding Hellinger distance matrix. The calculated complexity indicator scores, listed in Table 5.3, are within the range of 0 to 10. Several insights can be extracted from this result. For example, although the complexity indicator distributions of products 2 and 5 have similar medians, their $Cplx$ scores are quite different. This is primarily due to the variability in the spread of their complexity indicator distributions, which indicates that, even though products may possess similar median values, the product complexity may differ. This is also the reason for implementing a Hellinger distance to measure similarities among distributions.

Table 5.3: Complexity scores for eight types of products.

| Product Type | $Cplx\ Score$ | $P(0.5, Cplx_i\|X_i)$ |
|:---:|:---:|:---:|
| 1 | 2.8 | 0.0258 |
| 2 | 2.0 | 0.0245 |
| 3 | 9.7 | 0.0432 |
| 4 | 10.0 | 0.0450 |
| 5 | 0.0 | 0.0217 |
| 6 | 0.1 | 0.0219 |
| 7 | 7.7 | 0.0424 |
| 8 | 7.4 | 0.0412 |

### 5.4.3  Step 3. Product Complexity Indicator Clustering

To generate the dendrogram plot of the hierarchical clustering outcome, the statistical computing and graphics software, R (https://www.r-project.org), is used. Using the obtained Hellinger distance matrix, and following the introduced agglomerative hierarchical clustering algorithm, 8 types of products are partitioned into clusters as shown in the dendrogram. To merge clusters of products, as opposed to individual products, the complete-linkage criterion is used to measure the distance between clusters. Products with small distance differences are grouped together. The heights (horizontal line) at which two clusters are merged represent the dissimilarity between two clusters in the data space. By specifying the expected number of clusters as four, types 1+2, 3+4, 5+6, 7+8 are grouped together (Figure 5.2) in a manner that is consistent with the visually-based prediction.

Given this illustrative example, the inherent mechanism of the proposed hybrid data mining technique is comprehensively illustrated. The outcomes of all steps adequately verify the functionalities of this hybrid data mining approach. In the next section, a practical case study will be conducted to validate the feasibility and applicability of the proposed methodology.

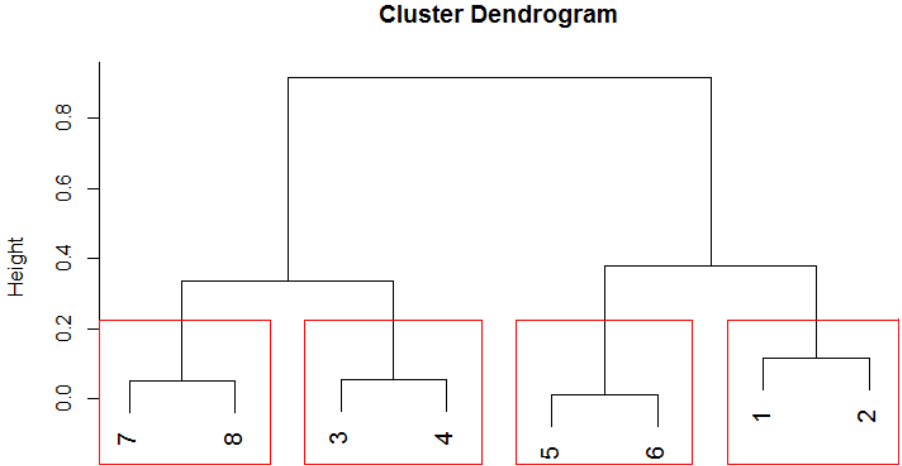

Figure 5.2: Cluster dendrogram of the illustrative example.

## 5.5    Case Study

Industrial construction is a construction method that involves large-scale use of offsite prefabrication and preassembly for building facilities, such as oil/gas production facilities and petroleum refineries. Pipe spool fabrication is crucial for the successful delivery of industrial projects. Pipe spool fabrication is heavily dependent on welding, which must be sampled and inspected to ensure that welding quality requirements are met. Typically, the difficulty (i.e., complexity) of pipe welds depends on various pipe attributes, such as nominal pipe size (NPS;

the outside diameter of a pipe), schedule (wall thickness of a pipe), and material. In this section, an industrial pipe spool fabrication company in Edmonton, Canada, is studied to analyze pipe weld complexity using the proposed uncertain data clustering approach.

The case study is conducted following the data source identification, data adapter design, data analysis, and decision support procedures that are summarized in Figure 5.3. First, multiple data sources are investigated to extract useful information related to pipe weld quality performance and design attributes. Second, a data adapter is designed to efficiently connect data and map data into a single, tidy dataset. Then, the proposed hybrid data mining approach is implemented to perform the product complexity analysis. Finally, main outputs are generated to produce new information and to support decision-making processes. All four procedures are performed using the statistical computing and graphics software, R ([https://www.r-project.org](https://www.r-project.org)).

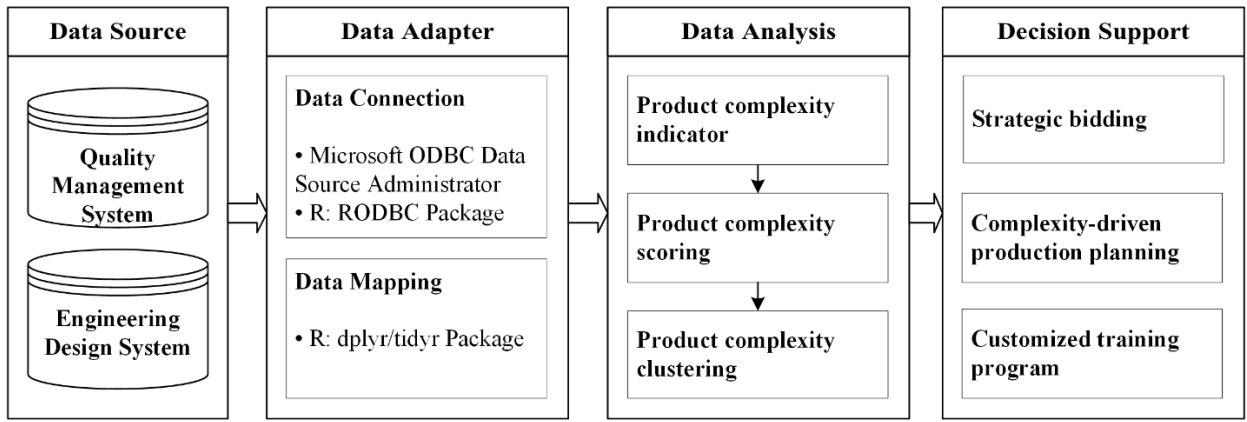

Figure 5.3: Workflow of the case study.

### 5.5.1 Data Source

The quality management and engineering design systems of the studied company were used to extract the non-destructive examination (NDE) inspection and pipe weld design attributes information, respectively. In this chapter, only the inspection records of radiographic tests (RT) of butt welds were extracted from the quality management system. RT inspection results are tracked in three statuses for each pipe weld, namely: 0 – no inspection performed; 1 – inspected and passed; and 2 – inspected and failed. The engineering design system of the studied company stored pipe weld design attributes by the pipe format (NPS, schedule, material). For example, pipe (6, STD, A) represents butt welds with NPS of 6, schedule of STD, and material A.

### 5.5.2   Data Adapter

Since the multi-relational data required were dispersed across quality management and engineering design systems, a data adapter was required to collect useful information from various data sources into a single, centralized, tidy dataset. In this case, a data adapter was also necessary to transform raw data, through data connection and data wrangling, into compatible, interpretable data formats. This is particularly important for data that are collected from a variety of sources or databases.

For data connection, the R package for Open Database Connectivity (RODBC) was used to connect to SQL server of both the quality management and engineering design systems (Ripley et al. 2016). The dplyr/tidyr package was used to perform data wrangling tasks, including data reshaping, grouping, and combining (Wickham et al. 2017; Wickham et al. 2017). The completed dataset was transformed into a table format, where each variable was saved in its own column and each observation was saved in its own row. A sample for the centralized dataset is listed in Table 5.4. This dataset combines the pipe design attributes and quality inspection results.

Table 5.4: A data sample for the centralized dataset.

| Weld Type | NPS | Schedule | Material | Inspection Result |
|-----------|-----|----------|----------|-------------------|
| 1 | 10 | 40S | B | 1 |
| 2 | 2 | 40 | C | 0 |
| 3 | 6 | XS | D | 2 |
| … | … | … | … | … |

For inferring the fraction nonconforming quality performance of each type of weld, all pipe welds were required to be grouped by pipe attribute (i.e., NPS, schedule, and material). Then, the data was summarized to count the total number of welds, inspected welds, and repaired welds for each type of pipe weld. A sample of the wrangled dataset is provided as Table 5.5. Each row represents the historical quality inspection information for a certain type of pipe weld. This table is then used to perform the complexity analysis as follows.

Table 5.5: The wrangled dataset of the top 35 weld types.

| Weld Type | NPS | Schedule | Material | Total Welds | Inspected Welds | Repaired Welds |
|-----------|-----|----------|----------|-------------|-----------------|----------------|
| 1 | 2 | XS | Material A | 37059 | 7475 | 249 |
| 2 | 3 | STD | Material A | 19464 | 4495 | 173 |
| 3 | 6 | STD | Material A | 14866 | 3518 | 43 |
| 4 | 4 | STD | Material A | 13020 | 3078 | 66 |
| 5 | 2 | STD | Material A | 10304 | 4722 | 400 |
| 6 | 6 | XS | Material A | 9916 | 3705 | 70 |
| 7 | 8 | STD | Material A | 8722 | 2302 | 51 |
| 8 | 4 | XS | Material A | 8601 | 1774 | 28 |
| 9 | 2 | 160 | Material A | 6044 | 2302 | 26 |
| 10 | 2 | 80 | Material A | 5854 | 1055 | 41 |
| 11 | 10 | STD | Material A | 4822 | 1131 | 30 |
| 12 | 12 | STD | Material A | 4728 | 1069 | 34 |
| 13 | 3 | XS | Material A | 3733 | 1484 | 16 |
| 14 | 8 | XS | Material A | 3193 | 1318 | 10 |
| 15 | 2 | 40S | Material C | 2431 | 555 | 21 |
| 16 | 2 | 40 | Material A | 2088 | 271 | 38 |
| 17 | 4 | 80 | Material A | 2056 | 638 | 5 |
| 18 | 3 | 160 | Material A | 1676 | 510 | 5 |
| 19 | 4 | 40 | Material A | 1550 | 592 | 17 |
| 20 | 6 | 40 | Material A | 1673 | 333 | 5 |
| 21 | 10 | XS | Material A | 1676 | 529 | 14 |
| 22 | 12 | XS | Material A | 1652 | 666 | 31 |
| 23 | 2 | 10S | Material C | 1261 | 175 | 12 |
| 24 | 3 | 40 | Material A | 1358 | 217 | 6 |
| 25 | 8 | 40 | Material A | 1413 | 452 | 17 |
| 26 | 3 | 40S | Material C | 1441 | 364 | 6 |
| 27 | 4 | 40S | Material C | 1253 | 271 | 2 |
| 28 | 3 | 80 | Material A | 1436 | 512 | 6 |
| 29 | 6 | 80 | Material A | 1407 | 572 | 3 |
| 30 | 16 | STD | Material A | 1406 | 422 | 13 |
| 31 | 3 | 10S | Material C | 1117 | 149 | 9 |
| 32 | 6 | 40S | Material C | 1128 | 171 | 4 |
| 33 | 6 | 10S | Material C | 912 | 154 | 4 |
| 34 | 8 | 10S | Material C | 912 | 204 | 13 |
| 35 | 16 | 80 | Material A | 961 | 634 | 9 |

### 5.5.3   Data Analysis

Prior to performing the comprehensive product complexity indicator clustering analysis, the wrangled dataset was examined and relevant information was extracted and analyzed. A total of 224,298 welds comprised of 631 weld types were conducted over that last ten years. As per the cumulative frequency graph shown in Figure 5.4, the top 35 types of pipe welds represented the most common pipe welding products and accounted for 80% of the company's business. Due to the frame limitation of graphing, only the top 35 types of pipe welds were selected to perform the product complexity analysis.

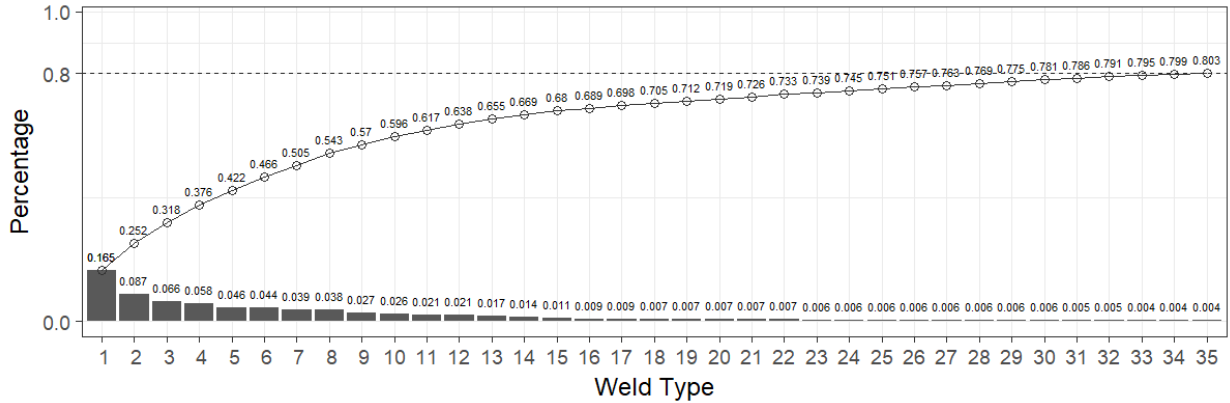

Figure 5.4: Cumulative percentage of the top 35 weld types.

The first step of the proposed methodology was applied to determine the product complexity indicator, with incorporated uncertainty, for each pipe weld type. Here, "Inspected Welds" and "Repaired Welds" from Table 5.5 were used to construct beta distributions as per Eq. (5.3). To be consistent with Figure 5.4, a side-by-side box plot, shown in Figure 5.5, was generated with the same sequence to visualize the distributions of the product complexity indicators. Notably, although some types followed similar distribution patterns, the product complexity indicators of these products varied considerably.

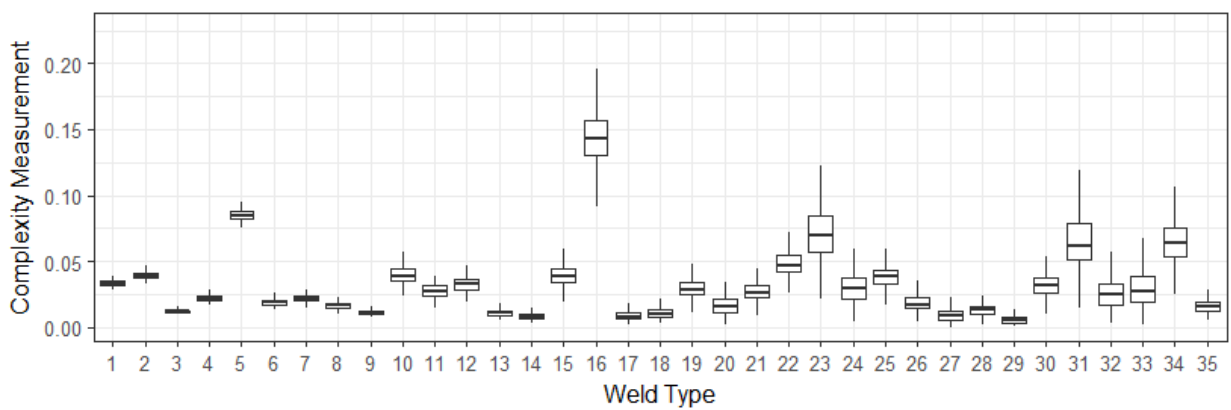

Figure 5.5: Complexity measurements of the top 35 weld types.

The complexity indicator distribution similarity between various groups of welding products was measured using the Hellinger distance metric. After obtaining the Hellinger distance matrix, welding products were scored based on the proposed scoring method. Also, the top 35 products were clustered into seven complexity groups, based on distribution similarity measurements, by using the agglomerative hierarchical clustering method (Figure 5.6). The name of each weld type is formatted as "Type ID.(NPS, schedule, material).[$Cplx\ Score$]" to include all design attributes and complexity indicator score information. Clusters are labelled from A to G based on the corresponding complexity level of that cluster, where Cluster A is the most complex group. The total business percentage of that cluster is also summarized and shown in Figure 5.6. The height at which two clusters are merged represents the dissimilarity between the two clusters in the data space. This type of information is expected to enable practitioners to better understand product complexity in a more informative and thorough manner.

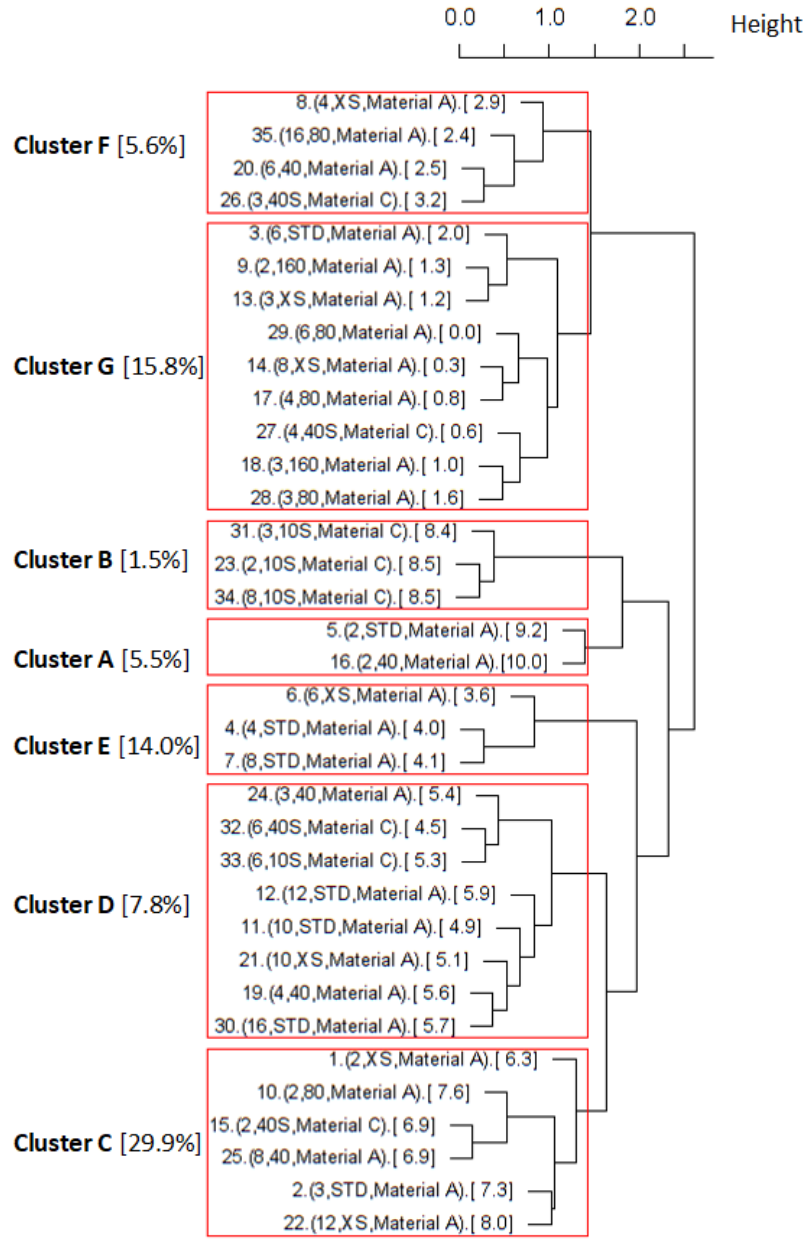

Figure 5.6: Complexity clustering dendrogram of the top 35 weld types.

Once high-complex products are clustered, products design attributes can be examined to identify factors common to this group. For example, a thin NPS is characteristic of both pipe weld types belonging to Cluster A, which is consistent with expert professional experience and knowledge.

5.5.4  **Validation**

To ensure that the proposed method is capable of reliably estimating product complexity, a systematic, expert evaluation-based validation was conducted. Eight welding operators, each with more than five-years working experience, were invited to evaluate the welding difficulty of selected weld types based on their own professional experience and knowledge. The authors excluded any "quality" related words from the evaluation description to eliminate any potential biases. The validation was conducted using the following protocol:

Step 1. Pick one weld type that accounts for the largest business percentage of each cluster (i.e., A to G). This will ensure that welding operators have sufficient experience with each of the chosen welds.

Step 2. Reshuffle the order of the selected weld types by changing the sequence in which they are presented to the welders.

Step 3. Invite welding operators to rank the welding difficulty based on their professional experience and knowledge using integers 1 to 7, where 7 represents the most complex weld type and 1 represents the least complex weld type.

Step 4. Once evaluations are collected, an average difficulty value is calculated for each type of pipe weld.

Step 5. Assign letter levels (i.e., A to G) to the sorted types of pipe welds based on the results.

Step 6. Compare the expert evaluation results with those obtained using the developed approach.

Table 5.6: Detailed validation results.

| Clustered Complexity | Weld Type | Design Attributes | Welding Difficulty Evaluation | | | | | | | | | |
|---|---|---|---|---|---|---|---|---|---|---|---|---|
| | | | Welder 1 | Welder 2 | Welder 3 | Welder 4 | Welder 5 | Welder 6 | Welder 7 | Welder 8 | Average | Letter Level |
| A | 5 | (2, STD, Material A) | 7 | 7 | 7 | 7 | 7 | 7 | 7 | 7 | 7.0 | A |
| B | 23 | (2, 10S, Material C) | 6 | 6 | 6 | 6 | 6 | 6 | 6 | 6 | 6.0 | B |
| C | 1 | (2, XS, Material A) | 5 | 4 | 5 | 5 | 5 | 5 | 3 | 5 | 4.6 | C |
| D | 11 | (10, STD, Material A) | 3 | 5 | 4 | 3 | 4 | 3 | 5 | 3 | 3.8 | D |
| E | 4 | (4, STD, Material A) | 4 | 3 | 3 | 4 | 3 | 4 | 4 | 4 | 3.6 | E |
| F | 8 | (4, XS, Material A) | 1 | 2 | 2 | 2 | 1 | 2 | 2 | 1 | 1.6 | F |
| G | 3 | (6, STD, Material A) | 2 | 1 | 1 | 1 | 2 | 1 | 1 | 2 | 1.4 | G |

Table 5.6 demonstrates the detailed validation results. Although welding operator rankings were variable, the overall evaluated welding difficulty levels followed the same sequence that was obtained using the proposed framework. This validation outcome demonstrates the feasibility and applicability of the proposed product complexity approach and supports the hypothesis that product quality performance is associated with product complexity.

### 5.5.5 Decision Support

The management team from the studied company has confirmed that a data-driven decision support approach that enables the timely transformation of large datasets into useable knowledge is highly desirable in practice. To better support these practical needs, the product complexity analysis functionality has been incorporated into the previously developed simulation-based analytics framework (Ji and AbouRizk, 2017b). Once incorporated into the simulation-based analytics framework, product complexity levels, together with their detailed design information, can be targeted for management, design, and operation professionals who require this type of information in a timely manner.

Results of the proposed research, such as those obtained in the case study, are expected to enhance decision-making with the overall aim of improving the competitiveness and reputation of organizations within the industry. Three detailed perspectives identified through interviews with industrial professionals, by which the proposed method is expected to enhance decision-support processes, are described.

***Strategic Bidding***

A better understanding of product complexity would assist practitioners in reducing uncertainty, which, in turn, could lead to improved cost performance. When bidding for new projects, practitioners may use the proposed framework to derive product complexity measurements in a relatively rapid manner that is conducive to the strict timelines associated with bid preparation. These measurements can then be used to allocate a contingency percentage that is more reflective of product complexity. For example, if a complex product is encompassing a majority of a bid, an organization should increase the contingency in their bid estimate. This would mitigate product complexity uncertainty and enhance the accuracy of bidding performance, thereby enhancing the company's profitability.

### Complexity-driven Production Planning

Previous research has developed a data-driven method to quantitatively identify exceptional operators for specific weld types (Ji and AbouRizk 2017b). Development and implementation of a complexity-driven production planning approach, which would allocate welding tasks to operators with high performance for each particular weld type, could directly improve overall welding quality performance and, consequently, reduce quality-induced rework cost and improve productivity. Such an automated production planning system, which incorporates both product complexity and detailed operator information, is expected to considerably increase the efficiency of industrial construction product prefabrication.

### Customized Training

Enhanced training programs are crucial for improving product complexity management processes and overall project performance. The proposed research transforms vast amounts of data into valuable knowledge in a simplified, interpretable, and informative format that efficiently improves practitioners' understanding of product complexity and reduces the time required to familiarize practitioners with this process. Once high-complexity products are identified, practitioners would be able to determine which operators are most proficient for each high-complex product. These operators should be invited to demonstrate their welding technique and share their professional knowledge for customized training purposes.

## 5.6 Conclusion

Product complexity is a predominant, yet often uncertain, factor that affects the success of construction project delivery. In this research, a novel uncertain data clustering approach was proposed to improve product complexity analysis by extracting hidden, intricate product complexity patterns from product quality performance measures. This approach contributes to the improved understanding of product complexity and, consequently, reduces the interpretation load for practitioners. Systematic procedures were developed for product complexity indicator determination, scoring, and clustering purposes. A pre-established product quality performance measurement, which incorporates uncertainty, is introduced as an indicator of product complexity. To the best of our knowledge, this is the first time that prefabricated construction product complexity is conceptually defined and quantitatively interpreted from the aspect of product quality performance.

The Hellinger distance is implemented to quantify the similarity of product complexity indicator distributions while considering uncertainty. In addition to providing a product complexity indicator score, the obtained Hellinger distance matrix is further utilized to perform the agglomerative hierarchical clustering method for intrinsically grouping products to achieve a better interpretation of product complexity. This novel Hellinger distance-based clustering approach is capable of clustering beta distributions and can be generalized and implemented for other types of uncertain data (i.e., probability distributions) clustering problems.

An industrial case study in Edmonton, Canada, was conducted to demonstrate the feasibility and applicability of the proposed uncertain data clustering approach. The achieved results indicate that the proposed method can appropriately cluster pipe weld types into homogeneous product complexity levels. Practitioners can implement this approach to enhance their product complexity management practices from the perspectives of (1) strategic bidding, (2) complexity-driven production planning, and (3) customized training.

Although this research proposes a novel approach to analyze construction product complexity, product complexity scores are, in fact, quality performance-based indicators of product complexity rather than a direct measure of complexity itself. In the future, additional indicators, such as productivity performance and safety performance, could be incorporated to measure product complexity in a more comprehensive and scientific way. Also, the authors would like to quantitatively correlate product complexity to design attributes and to forecast product complexity from product design information.

## 5.7    Acknowledgements

This research is funded by the NSERC Collaborative Research & Development (CRD) Grant (CRDPJ 492657). The authors would like to acknowledge Rob Reid, Doug McCarthy, Jason Davio, and Christian Jukna at Falcon Fabricators and Modular Builders Ltd. for sharing their knowledge and expertise of industrial pipe welding complexity and quality management.

## 5.8 Analytical Proof - Hellinger distance for two beta distributions

Let $P = \{p_i\}$ $i \in [n]$ and $Q = \{q_i\}$ $i \in [n]$ be two probability distributions supported on $[n]$.

The Hellinger distance between two probability distributions is defined by:

$$H(P,Q) = \frac{1}{\sqrt{2}} \left\| \sqrt{P} - \sqrt{Q} \right\|_2$$

$$= \frac{1}{\sqrt{2}} \sqrt{\sum_i^n (\sqrt{p_i} - \sqrt{q_i})^2}$$

$$= \frac{1}{\sqrt{2}} \sqrt{\int (\sqrt{p_i} - \sqrt{q_i})^2}$$

Let $X_t$ and $Y_t$ be two independent Beta probability distributions where:

$$X_t \sim Beta(a_1, b_1) \quad Y_t \sim Beta(a_2, b_2)$$

$$H(X_t, Y_t) = \frac{1}{\sqrt{2}} \sqrt{\int (\sqrt{X_t} - \sqrt{Y_t})^2 dt}$$

$$H^2(X_t, Y_t) = \frac{1}{2} \int (\sqrt{X_t} - \sqrt{Y_t})^2 dt$$

$$H^2(X_t, Y_t) = \frac{1}{2} \left[ \int \sqrt{X_t}^2 dt - 2 \int \sqrt{X_t}\sqrt{Y_t} \, dt + \int \sqrt{Y_t}^2 dt \right]$$

$$H^2(X_t, Y_t) = \frac{1}{2} \left[ \int X_t \, dt - 2 \int \sqrt{X_t Y_t} \, dt + \int Y_t \, dt \right]$$

The integral of a probability density over its domain equals to 1.

$$H^2(X_t, Y_t) = \frac{1}{2}\left[1 - 2\int \sqrt{X_t Y_t}dt + 1\right]$$

$$H^2(X_t, Y_t) = 1 - \int \sqrt{X_t Y_t}\,dt$$

Probability density function of beta distribution is defined as:

$$f_t = \frac{(t-u)^{a-1}(l-t)^{b-1}}{Beta(a,b)(u-l)^{a+b-1}}$$

Where $a, b$ are shape parameters and $u, l$ are upper and lower boundaries.

Specify lower and upper boundaries, respectively, as 0 and 1, probability density function of Beta distribution becomes:

$$f_t = \frac{t^{a-1}(1-t)^{b-1}}{Beta(a,b)}$$

Beta function is defined as:

$$Beta(a,b) = \int_0^1 t^{a-1}(1-t)^{b-1}\,dt$$

Based on the function of beta distribution, the squared Hellinger distance can be written as:

$$H^2(X_t, Y_t) = 1 - \int_0^1 \sqrt{\frac{t^{a_1-1}(1-t)^{b_1-1}}{Beta(a_1,b_1)} \times \frac{t^{a_2-1}(1-t)^{b_2-1}}{Beta(a_2,b_2)}}\,dt$$

$$H^2(X_t, Y_t) = 1 - \frac{1}{\sqrt{Beta(a_1, b_1) \times Beta(a_2, b_2)}} \int_0^1 \sqrt{t^{a_1-1}(1-t)^{b_1-1} \times t^{a_2-1}(1-t)^{b_2-1}} \, dt$$

$$H^2(X_t, Y_t) = 1 - \frac{1}{\sqrt{Beta(a_1, b_1) \times Beta(a_2, b_2)}} \int_0^1 \sqrt{t^{a_1+a_2-2} \times (1-t)^{b_1+b_2-2}} \, dt$$

$$H^2(X_t, Y_t) = 1 - \frac{1}{\sqrt{Beta(a_1, b_1) \times Beta(a_2, b_2)}} \int_0^1 \sqrt{t^{2(\frac{a_1+a_2}{2}-1)} \times (1-t)^{2(\frac{b_1+b_2}{2}-1)}} \, dt$$

$$H^2(X_t, Y_t) = 1 - \frac{1}{\sqrt{Beta(a_1, b_1) \times Beta(a_2, b_2)}} \int_0^1 t^{(\frac{a_1+a_2}{2}-1)} \times (1-t)^{(\frac{b_1+b_2}{2}-1)} \, dt$$

$$H^2(X_t, Y_t) = 1 - \frac{Beta(\frac{a_1+a_2}{2}, \frac{b_1+b_2}{2})}{\sqrt{Beta(a_1, b_1) \times Beta(a_1, b_2)}}$$

$$H(X_t, Y_t) = \sqrt{1 - \frac{Beta(\frac{a_1+a_2}{2}, \frac{b_1+b_2}{2})}{\sqrt{Beta(a_1, b_1) \times Beta(a_2, b_2)}}}$$

## 5.9    References

Antani, K. R. (2014). "A study of the effects of manufacturing complexity on product quality in mixed-model automotive assembly." Doctor of Philosophy (PhD), CLEMSON UNIVERSITY.

Baccarini, D. (1996). "The concept of project complexity—a review." *International Journal of Project Management*, 14(4), 201-204.

Baldwin, C. Y., and Clark, K. B. (2000). *Design rules: The power of modularity*, MIT press.

Galvin, P., and Morkel, A. (2001). "The effect of product modularity on industry structure: the case of the world bicycle industry." *Industry and Innovation*, 8(1), 31.

Han, J., Pei, J., and Kamber, M. (2011). *Data mining: concepts and techniques*, Elsevier.

Hellinger, E. (1909). "Neue Begründung der Theorie quadratischer Formen von unendlichvielen Veränderlichen." *Journal für die reine und angewandte Mathematik*, 136, 210-271.

Ji, W., and AbouRizk, S. M. (2017a). "Credible interval estimation for fraction nonconforming: analytical and numerical solutions." *Automation in Construction*, 83, 56-67.

Ji, W., and AbouRizk, S. M. (2017b). "Simulation-based analytics for quality control decision support: a pipe welding case study." *Journal of Computing in Civil Engineering*, https://doi.org/10.1061/(ASCE)CP.1943-5487.0000755. [Accepted: Oct. 23, 2017]Jiang, B., Pei, J., Tao, Y., and Lin, X. (2013). "Clustering uncertain data based on probability distribution similarity." *IEEE Transactions on Knowledge and Data Engineering*, 25(4), 751-763.

Liese, F., and Vajda, I. (2006). "On divergences and informations in statistics and information theory." *IEEE Transactions on Information Theory*, 52(10), 4394-4412.

Luo, L., He, Q., Jaselskis, E. J., and Xie, J. (2017). "Construction Project Complexity: Research Trends and Implications." *Journal of Construction Engineering and Management*, https://doi.org/10.1061/(ASCE)CO.1943-7862.0001306, 04017019.

Montgomery, D. C. (2007). *Introduction to statistical quality control*, John Wiley & Sons.

Novak, S., and Eppinger, S. D. (2001). "Sourcing by design: Product complexity and the supply chain." *Management science*, 47(1), 189-204.

Pei, J., Jiang, B., Lin, X., and Yuan, Y. (2007). "Probabilistic skylines on uncertain data." *Proc., Proceedings of the 33rd international conference on Very large data bases*, VLDB Endowment, 15-26.

Ripley, B., Lapsley, M., and Ripley, M. B. (2016). "Package 'RODBC'", <https://cran.r-project.org/web/packages/RODBC/index.html > (Sept. 19, 2017).

Schubert, E., Koos, A., Emrich, T., Züfle, A., Schmid, K. A., and Zimek, A. (2015). "A framework for clustering uncertain data." *Proceedings of the VLDB Endowment*, 8(12), 1976-1979.

Senescu, R. R., Aranda-Mena, G., and Haymaker, J. R. (2012). "Relationships between project complexity and communication." *Journal of Management in Engineering*, https://doi.org/10.1061/(ASCE)ME.1943-5479.0000121, 183-197.

Wickham, H., Henry, L., and RStudio (2017). "tidyr: Easily Tidy Data with spread () and gather () Functions.", < https://cran.r-project.org/web/packages/tidyr/tidyr.pdf> (Sept. 14, 2017).

Wickham, H., Francois, R., Henry, L., Müller, K., and RStudio (2017). "dplyr: A grammar of data manipulation.", < https://cran.r-project.org/web/packages/dplyr/dplyr.pdf > (Sept. 14, 2017).

Williams, T. M. (1999). "The need for new paradigms for complex projects." *International journal of project management*, 17(5), 269-273.

Züfle, A., Emrich, T., Schmid, K. A., Mamoulis, N., Zimek, A., and Renz, M. (2014). "Representative clustering of uncertain data." *Proc., Proceedings of the 20th ACM SIGKDD international conference on Knowledge discovery and data mining*, ACM, 243-252.

# 6 CHAPTER 6: DATA-DRIVEN SIMULATION MODEL FOR QUALITY-INDUCED FABRICATION REWORK COST ESTIMATION AND CONTROL USING ABSORBING MARKOV CHAINS

## 6.1 Introduction

Rework is a predominant, uncertain factor that, when improperly managed, contributes to construction cost and schedule overruns (Love et al. 2010). In construction literature, rework is often represented by various terms such as "nonconformance" (Abdul-Rahman 1995), "quality deviation" (Burati Jr et al. 1992), "defect" (Josephson and Hammarlund 1999), and "quality failures" (Barber et al. 2000) and has been defined as "the unnecessary effort of redoing an activity or process that was incorrectly implemented the first time (Love 2002)." Although definitions of rework vary, it is commonly agreed that rework involves redoing work as a consequence of nonconformance of the original work to predefined requirements (Hwang et al. 2009). In essence, rework occurs when construction products are inspected and recorded as "nonconforming" to specified quality standards.

In practice, quality-induced rework costs are typically estimated by assuming a fixed percentage of the direct cost that is based on estimators' expectations rather than on data-driven facts. However, due to variability in both product complexity and operator quality performance, it is difficult to appropriately and accurately estimate rework costs for construction prefabrication in the planning phase. Similarly, during project execution processes, it is difficult to update the original rework cost estimation periodically to better reflect actual rework cost performance. Practitioners, therefore, require more capable decision-support systems to facilitate rework cost estimation and control purposes for construction prefabrication processes.

With the implementation of the mature data-driven quality control system demonstrated in Chapter 3, it is becoming increasingly feasible to forecast quality performance for a given project in a more reliable way. The accurate product quality performance measurements generated by such systems can provide more reliable inputs for rework cost estimation and control; at the same time, cost management systems (Potts and Ankrah 2014) can track both estimated and actual costs of certain construction activities. By fusing quality- and cost-related information, this chapter aims to propose a novel, data-driven simulation model to estimate and control direct rework costs in a more systematic, objective manner for construction prefabrication processes. The proposed model can be integrated into previously developed simulation-based analytics systems, enhancing the ability of these systems to generate accurate and reliable decision-support metrics in real-time. Specifically, the proposed research has been able to accomplish this by (1) creating an absorbing Markov chain-based analytical model for performing direct rework cost (e.g., man-hours) estimation and control for prefabrication processes; (2) generating meaningful and reliable decision-support metrics (i.e., simulated rework man-hour estimation and dynamically updated rework man-hour control) for enhanced decision-making processes; and (3) integrating previously developed simulation-based analytics framework as the simulation environment to achieve a functional and feasible practical application.

The content of this chapter is organized as follows: First, previous research on simulation-based analytics and Bayesian-based fraction nonconforming modelling are introduced, and the fundamental concept of the absorbing Markov chain is discussed. Then, a specialized absorbing Markov chain model is created to analytically simulate the construction product prefabrication process. Two types of decision-support metrics are developed to assist with rework cost estimation and control during project planning and execution. To elaborate on the functionalities of these decision-support metrics, an illustrative example is provided. Then, the applicability and feasibility of the novel approach are validated by examining a practical pipe welding project. Finally, research contributions, limitations, and future work are disscussed.

## 6.2    Previous Research

In this section, previous research work is discussed to summarize the foundation on which the research described here is built. First, the simulation-based analytics framework, which has advanced real-time, data-driven construction simulation, is introduced. Afterwards, a more capable fraction nonconforming modelling technique for providing real-time auto-calibrated input models to data-driven, quality management systems is demonstrated.

### 6.2.1    Simulation-based Analytics

Static, statistical, distribution-based approaches are widely used to represent uncertainties in construction simulation. To improve reliability and accuracy of simulation outputs, simulation input models should be updated in a real-time manner using dynamically generated data from the actual system. Real-time input model recalibration, however, remains a challenge in construction decision-support system development.

To overcome this limitation, a simulation-based analytics framework has been specialized and developed for pipe welding quality control decision supports (Chapter 3). Specifically, this framework can be used to generate operator quality performance measurements and project quality performance forecasts. This framework consists of five modules, namely the data source, data adapter, data analysis, simulation, and decision-support modules. Feasibility and applicability of the simulation-based analytics approach have been proven through the implementation of the prototyped decision support system (Section 3.5 and 4.6). This approach allows simulation models to be updated by incorporating real-time data, thereby enhancing the predictability of original models. An integrated simulation environment has been created using R (https://www.r-project.org/).

In present research, the specialized, simulation-based analytics decision support environment is enhanced to incorporate cost information (i.e., man-hour) for rework cost estimation and control purposes. Therefore, the scope of this research will focus on introducing the newly developed simulation module and demonstrating its ability to provide powerful decision-support metrics.

### 6.2.2 Bayesian-based Fraction Nonconforming Modelling

As discussed above, achieving real-time parameter updating is the most challenging aspect of the simulation-based analytics framework. Bayesian theorem provides a method for computing the posterior probability distribution, allowing parameters to be updated when new data is acquired (Bishop 2006). Once the posterior distribution is derived, uncertainty can be quantified by measuring certain quantiles of the posterior distribution. The obtained posterior distributions provide more reliable and accurate inputs for enhanced predictability of simulation models.

Construction products are inspected as either conforming or nonconforming to specified quality standards and, therefore, cannot be assessed numerically. In statistical quality control, fraction nonconforming is defined as the ratio of the number of nonconforming items in a population to the total number of items in that population (Montgomery 2007). Conventional fraction nonconforming modelling techniques (i.e., classic statistical models) are incapable of incorporating new data to derive more accurate and reliable distributions. To provide a more accurate, reliable, and interpretable estimation of fraction nonconforming that considers sampling uncertainty, both a Bayesian statistics-based analytical solution and a Metropolis-Hastings algorithm-based numerical solution have been developed to derive posterior distributions and credible intervals of fraction nonconforming (Chapter 2). By fusing engineering design data and quality management data from a pipe fabrication company, the authors have implemented the proposed solutions to measure the quality performance (i.e., fraction nonconforming distributions) of various types of pipe welding products.

In this research, these dynamically updated fraction nonconforming distributions are utilized to provide more accurate and reliable input models for rework cost simulation analyses.

## 6.3    Absorbing Markov Chains

A Markov chain can be described as a set of states, $S = \{s_1, s_2, s_3, \ldots, s_r\}$. A process begins in one of these states and moves, successively, from one state to another. Each move is called a step. If the chain is currently in state $s_i$, it moves to state $s_j$ in the next step with a probability denoted by $p_{ij}$. Notably, the probability does not depend upon which state(s) the chain was in prior to the current state.

If it is impossible to leave the current state, a state $s_i$ of a Markov chain is called an absorbing state (i.e. $p_{ij} = 1$). A Markov chain is absorbing if it has at least one absorbing state and if, from every state, it is possible to proceed to an absorbing state (albeit, not necessarily in one step).

If there are $r$ absorbing states and $t$ transient states in one Markov chain, the transition matrix will have the following canonical form, shown as Eq. (6.1).

$$P = \begin{bmatrix} \mathbf{Q} & \mathbf{R} \\ \mathbf{O} & \mathbf{I} \end{bmatrix} \quad\quad\quad (6.1)$$

$\mathbf{Q} \in [0,1]^{t \times t}$ contains the transition probabilities between any pair of transient states, while $\mathbf{R} \in [0,1]^{t \times r}$ contains the probabilities of moving from any transient state to any absorbing state. $\mathbf{O}$ is the $r \times t$ zero matrix and $\mathbf{I}$ is the $r \times r$ identity matrix.

As per the theorem, for an absorbing Markov chain, the matrix $\mathbf{I} - \mathbf{Q}$ has an inverse $\mathbf{N}$ called fundamental matrix, where $\mathbf{N} = \mathbf{I} + \mathbf{Q} + \mathbf{Q}^2 + \cdots$ (Resnick 2013). Then, the fundamental matrix for the absorbing Markov chain is

$$\mathbf{N} = [\mathbf{I} - \mathbf{Q}]^{-1} \quad\quad\quad (6.2)$$

The fundamental matrix contains $N_{ij}$, where $N_{ij}$ is the expected number of times that the process is in the transient state $j$ if it begins in transient state $i$.

## 6.4    Model Development

In this section, a specialized absorbing Markov chain model, which is capable of considering rework uncertainty, is developed to model the construction product fabrication process. The probabilistic graphical model (Bishop 2006), which is comprised of nodes (also known as vertices) and is connected by links (also called edges or arcs), is depicted in Figure 6.1. For one project, it is assumed that a number of products $n$ need to be manufactured, and these $n$ products construct the $n$ transient states of the absorbing Markov chain. The only absorbing state $F$ represents the completion of the project. The production process begins at transient state 1 and completes at the absorbing state $F$. A transient state $i$ indicates that the $(i-1)^{th}$ product has been completed and that the $i^{th}$ product is being manufactured. A product in transient state $i$ has the probability $p_i$ to remain in this state, indicating that the product has failed the quality inspection, requires rework, and has the probability of $(1 - p_i)$ to move forward to the $(i + 1)^{th}$ transient state.

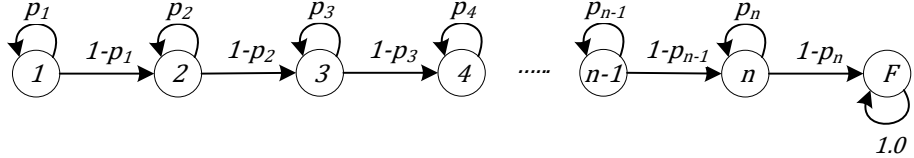

Figure 6.1: The specialized absorbing Markov chain model for construction product prefabrication considering rework.

Here, $p_i$ is the estimation of fraction nonconforming (i.e., percent repair rate) for product $i$, which represents the ratio of the number of nonconforming items $X_i$ in a sample of sample size $n_i$. $\hat{p}_i$ can be mathematically expressed as Eq. (6.3) (Montgomery 2007).

$$\hat{p}_i = \frac{X_i}{n_i} \tag{6.3}$$

The number of nonconforming items $X_i$ and the sample size $n_i$ can be obtained from historical quality inspection records. When new products are manufactured, these parameters are dynamically updated in the quality management system.

To appropriately incorporate the sampling uncertainty of the population fraction nonconforming variable $p_i$ when data are obtained from a sample, a Bayesian statistics-based analytical solution has been developed to determine the posterior distribution of the fraction nonconforming $p_i$ (Ji and AbouRizk 2017c). The posterior distribution uses a non-informative prior distribution $Beta(1/2,\ 1/2)$, which is given in Eq. (6.4).

$$P(p_i|X_i) = Beta(X_i + 1/2, n_i - X_i + 1/2) \qquad (6.4)$$

This Bayesian statistics-based solution, which is capable of updating the posterior distribution by combining previous knowledge and real-time data, is more accurate, reliable, and interpretable than traditional statistical methods that have previously been established (Chapter 2). Once the posterior distribution is derived, a credible interval can be determined.

As shown in Figure 6.1, there are $n$ transient states and one absorbing state in the specialized absorbing Markov chain model. Following Eq. (6.1), the transition matrix $\mathbf{P}$ for the specialized model is derived as Eq. (6.5).

$$\mathbf{P} = \begin{bmatrix} p_1 & 1-p_1 & 0 & 0 & 0 & \dots & 0 & 0 \\ 0 & p_2 & 1-p_2 & 0 & 0 & \dots & 0 & 0 \\ 0 & 0 & p_3 & 1-p_3 & 0 & \dots & 0 & 0 \\ 0 & 0 & 0 & p_4 & 1-p_4 & \dots & 0 & 0 \\ \vdots & \vdots & \vdots & \vdots & \ddots & \ddots & \vdots & \vdots \\ 0 & 0 & 0 & 0 & 0 & p_{n-1} & 1-p_{n-1} & 0 \\ 0 & 0 & 0 & 0 & 0 & 0 & p_n & 1-p_n \\ 0 & 0 & 0 & 0 & 0 & 0 & 0 & 1 \end{bmatrix} \qquad (6.5)$$

The $n \times n$ matrix $\mathbf{Q}$, which contains transition probabilities between transient states, is shown as Eq. (6.6).

$$\mathbf{Q} = \begin{bmatrix} p_1 & 1-p_1 & 0 & 0 & 0 & \dots & 0 \\ 0 & p_2 & 1-p_2 & 0 & 0 & \dots & 0 \\ 0 & 0 & 0 & p_4 & 1-p_4 & \dots & 0 \\ \vdots & \vdots & \vdots & \vdots & \ddots & \ddots & \vdots \\ 0 & 0 & 0 & 0 & 0 & p_{n-1} & 1-p_{n-1} \\ 0 & 0 & 0 & 0 & 0 & 0 & p_n \end{bmatrix} \qquad (6.6)$$

And the $n \times 1$ matrix $\mathbf{R}$, containing transition probabilities from transient states to the absorbing state $F$, is shown as Eq. (6.7).

$$\mathbf{R} = \begin{bmatrix} 0 \\ 0 \\ 0 \\ 0 \\ \vdots \\ 0 \\ 1 - p_n \end{bmatrix} \tag{6.7}$$

$\mathbf{I}$ is the $1 \times 1$ identity matrix that is described as Eq. (6.8).

$$\mathbf{I} = [1] \tag{6.8}$$

Then, the fundamental matrix $\mathbf{N}$ can be derived following Eq. (6.2), shown as Eq. (6.9).

$$\mathbf{N} = [\mathbf{I} - \mathbf{Q}]^{-1} = \begin{bmatrix} \dfrac{1}{1-p_1} & \dfrac{1}{1-p_2} & \dfrac{1}{1-p_3} & \cdots & \dfrac{1}{1-p_n} \\ 0 & \dfrac{1}{1-p_2} & \dfrac{1}{1-p_3} & \cdots & \dfrac{1}{1-p_n} \\ 0 & 0 & \dfrac{1}{1-p_3} & \cdots & \dfrac{1}{1-p_n} \\ \vdots & \vdots & \vdots & \ddots & \vdots \\ 0 & 0 & 0 & 0 & \dfrac{1}{1-p_n} \end{bmatrix} \tag{6.9}$$

As discussed, $N_{ij}$ is the expected number of times that transient state $j$ is occupied when the initial state is the transient state $i$. From the derived matrix $\mathbf{N}$ for the specialized model, the expected number of times for remaining in state $i$ can be expressed as $\frac{1}{1-p_i}$, since $p_i \in [0, 1]$, $\frac{1}{1-p_i} \in [1, +\infty]$. Notably, when fraction nonconforming $p_i$ is 0, no rework occurs for product $i$. Conversely, when fraction nonconforming $p_i$ is 1, rework always occurs and product $i$ will never be completed.

## 6.5    Decision Support Metrics

In this section, decision-support metrics derived from the proposed absorbing Markov chain model are introduced to achieve two objectives: (1) rework cost estimation during the planning phase and (2) rework cost control during the execution phase. Details of these two types of metrics are provided as follows.

### 6.5.1    Rework Cost Estimation

Practitioners are interested in estimating rework cost more reliably during the project planning phase for bidding preparation. A metric, capable of reliably forecasting the impact of rework on cost estimation, could considerably improve contingency estimates for rework. Here, the rework cost is defined as the rework man-hours required to complete a piece of product due to quality-induced issues.

From Eq. (6.9), $N_{1i}$ (i.e., $\frac{1}{1-p_i}$) is the expected number of times product $i$ is manufactured as a consequence of rework. Therefore, the estimated rework man-hours considering rework for each product can be calculated as Eq. (6.10).

$$\eta_i \times t_i \times (N_{1i} - 1) \qquad\qquad (6.10)$$

Where $\eta_i$ is an efficiency factor that equals the required rework man-hours over the original estimated man-hours, and $t_i$ is the estimated man-hours for the $i^{th}$ product. Then, $\eta_i \times t_i$ is the estimated rework man-hours. $N_{1i}$ is the expected number of times that the model remains in transient state $i$. Then, $(N_{1i} - 1)$ is the expected times a piece is reworked for the $i^{th}$ product.

Accordingly, the total estimated rework man-hours can be described as Eq. (6.11).

$$T_{RMH} = \sum_{i=1}^{n} \eta_i \times t_i \times (N_{1i} - 1) \qquad (6.11)$$

Then, as per Eq. (6.9), Eq. (6.11) is derived as Eq. (6.12).

$$T_{RMH} = \sum_{i=1}^{n} \eta_i \times t_i \times (\frac{1}{1 - p_i} - 1) \qquad (6.12)$$

Where $p_i$ is a randomly sampled number from the posterior distribution $P(p_i|X_i)$ of fraction nonconforming (i.e., quality performance). This equation forecasts the total estimated rework man-hours and can be used to predict future project rework man-hours from the historical quality performance and estimated man-hours associated from related product information.

By running the proposed absorbing Markov chain for multiple iterations (e.g. 1000 iterations), a frequency histogram is generated to represent the empirical distribution of the estimated rework man-hours considering uncertainty. From this, companies can specify their interested quantiles to support their decision-making processes. For example, the 50% quantile (i.e., median value) represents a neutral risk attitude of the company. A detailed illustrative example is provided in Section 6.5.2.

### 6.5.2 Rework Cost Control

During the project execution phase, the rework cost should be periodically updated and monitored to accurately reflect the actual rework status of all products. The updated rework cost depends on the actual quality inspection results and the actual man-hours spent on reworked items.

As per International Organization for Standardization (ISO) 9000 requirements, actual quality inspection results can be obtained from the quality management information system, while actual rework man-hours are typically available in the timesheet tracking system.

The updated rework man-hours prediction at state $k$ is comprised of the actual rework man-hours for the first $(k-1)$ states and the estimated man-hours for subsequent states. It can be described by Eq. (6.13).

$$T_{RMH\ at\ State\ k} = T_{RMH\ Actual\ Happened} + \sum_{i=k}^{n} \eta_i \times t_i \times (\frac{1}{1-p_i} - 1) \qquad (6.13)$$

For rework control purposes, the original rework man-hour estimation is utilized to construct a control chart. An example control chart is presented in Figure 6.2. The control chart contains a center line (CL) that represents the median value of the estimated rework man-hours. Two other horizontal lines, named the upper control limit (UCL) and the lower control limit (LCL), construct a control area to monitor the updated rework man-hours. Provided that the updated rework man-hours are within the control area, the production process is assumed to be in control, and no action is required. If the value of the updated rework man-hours is higher than the UCL, the production process is out of control and requires investigations and corrective actions to identify and mitigate the causes of the anomaly. If the value of the updated rework man-hours is lower than the LCL, the actual production performance in terms rework is better than previous historical performance.

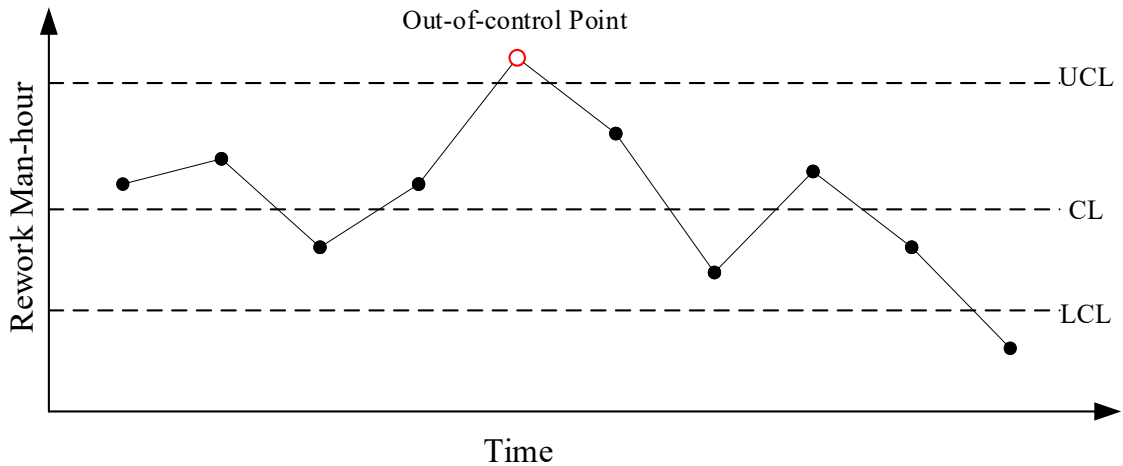

Figure 6.2: Example control chart.

Here, the value of the center line equals the median value of the rework man-hours estimated in the planning phase. As per the classic statistical control theory (Montgomery 2007), the values of LCL and UCL are established as the 95% credible interval (i.e., 2.5% and 97.5% quantiles) of the simulated histogram of the estimated rework man-hours. An illustrative example is provided in the following section to demonstrate the applicability of this rework cost control metric.

## 6.6    Illustrative Example

This example project involves ten construction welds. Here, three types of data information are provided to perform rework cost analyses; they are (1) historical quality performance of the corresponding ten types of welds, (2) estimated man-hours and actual rework man-hours (with three scenarios) of ten welds, and (3) actual quality inspection results of these ten welds (with three scenarios). Detailed information of these types of data are discussed as follows.

Historical quality performance (i.e., number of inspected items, number of repaired items, and fraction nonconforming distributions) of the corresponding ten types welds are summarized in

Table 6.1. The column of fraction nonconforming $P(p_i|X_i)$ is derived using Eq. (6.4) based on the summarized historical quality inspection records of the same type of welds. This information can be mapped from quality management systems.

Table 6.1: Summarized historical quality performance of the corresponding ten types welds.

| Product Type | Number of Inspected Items $(n_i)$ | Number of Repaired Items $(X_i)$ | Fraction Nonconforming $P(p_i|X_i)$ |
|---|---|---|---|
| 1 | 21 | 2 | Beta (2.5, 19.5) |
| 2 | 23 | 1 | Beta (1.5, 22.5) |
| 3 | 7 | 0 | Beta (0.5, 7.5) |
| 4 | 14 | 2 | Beta (2.5, 12.5) |
| 5 | 17 | 2 | Beta (2.5, 15.5) |
| 6 | 37 | 3 | Beta (3.5, 34.5) |
| 7 | 10 | 1 | Beta (1.5, 9.5) |
| 8 | 41 | 4 | Beta (4.5, 37.5) |
| 9 | 55 | 3 | Beta (3.5, 52.5) |
| 10 | 51 | 6 | Beta (6.5, 45.5) |

Estimated and actual man-hour information of the ten welds are summarized in Table 6.2. Estimated man-hours represent the original estimation during project planning phase, while actual rework man-hours represent the tracked rework man-hours during the project execution phase. To demonstrate how various rework conditions impact actual rework cost control processes, three scenarios (i.e., no rework, rework under control, and rework over control) are provided. In practice, the estimated and actual rework man-hour information are available in companies' Enterprise Resource Planning (ERP) systems (e.g., Oracle JD Edwards). Here, it was assumed that the efficiency factor $\eta_i$ is 1.2, where the rework man-hours are 1.2 times that of the estimated man-hours for rework to account for extra work associated with the rework process, such as the disassembly of the nonconforming items.

Table 6.2: Estimated man-hours and actual rework man-hours (with three scenarios) of ten welds.

| Product ID | Estimated Man-hours | Actual Rework Man-hours (Hours) | | |
|---|---|---|---|---|
| | | No Rework | Rework Under Control | Rework Over Control |
| 1 | 3.00 | 0.00 | 0.00 | 0.00 |
| 2 | 2.50 | 0.00 | 0.00 | 0.00 |
| 3 | 3.00 | 0.00 | 0.00 | 0.00 |
| 4 | 1.50 | 0.00 | 1.80 | 1.80 |
| 5 | 2.00 | 0.00 | 0.00 | 2.40 |
| 6 | 1.00 | 0.00 | 1.20 | 1.20 |
| 7 | 3.00 | 0.00 | 0.00 | 0.00 |
| 8 | 3.00 | 0.00 | 0.00 | 0.00 |
| 9 | 2.50 | 0.00 | 0.00 | 0.00 |
| 10 | 2.00 | 0.00 | 2.40 | 0.00 |

Table 6.3 summarizes the quality inspection results during the project execution phase. Value "0" indicates that the weld has passed quality inspection, while value "1" indicates that the weld has failed the quality inspection and was reworked. The three scenarios of inspection results are consistent with the three scenarios of actual man-hour tracking. The quality inspection records were obtained from the quality management system, as per ISO 9000 requirements.

Table 6.3: Actual quality inspection results of ten welds.

| Product ID | Inspection Results (0 = Pass, 1 = Fail) | | |
|---|---|---|---|
| | No Rework | Rework Under Control | Rework Over Control |
| 1 | 0 | 0 | 0 |
| 2 | 0 | 0 | 0 |
| 3 | 0 | 0 | 0 |
| 4 | 0 | 1 | 1 |
| 5 | 0 | 0 | 1 |
| 6 | 0 | 1 | 1 |
| 7 | 0 | 0 | 0 |
| 8 | 0 | 0 | 0 |
| 9 | 0 | 0 | 0 |
| 10 | 0 | 1 | 0 |

Given these data, the proposed approach is applied to demonstrate the analytical decision-making processes for rework cost estimation and control purposes. The outcomes of these two decision-support metrics verify the functionalities of the proposed data-driven simulation model.

***Rework Cost Estimation***

For one iteration, the fraction nonconforming $p_i$ for weld $i$ is sampled from the beta distribution listed in Table 6.1. Then, the estimated rework man-hours for each state (i.e., product) and the total estimated rework man-hours are derived by Eq. (6.11) and Eq. (6.12), respectively.

To obtain the frequency histogram of the total estimated rework man-hours, the proposed absorbing Markov chain model was run for 1,000 iterations. The simulated frequency histogram for the total estimated rework man-hours is shown in Figure 6.3. Visually, this frequency histogram is a non-symmetrical right-skewed distribution. With this obtained frequency histogram, practitioners can incorporate a proper number of estimated rework man-hours to the estimated man-hours at the project planning stages based on their risk attitude by selecting an appropriate quantile. The quantile represents the probability of achieving the corresponding rework man-hours as determined from the simulated frequency histogram. For instance, practitioners may choose 10%, 50%, or 90% quantiles of the histogram to represent an aggressive, neutral, or conservative risk attitude, respectively. Here, all 10% quantiles are derived and provided in Table 6.4 for the generated frequency histogram of the total estimated rework man-hours. The 50% quantile (median) of this frequency histogram is 3.4, which indicates the expected rework man-hours are 3.4 with a neutral risk attitude.

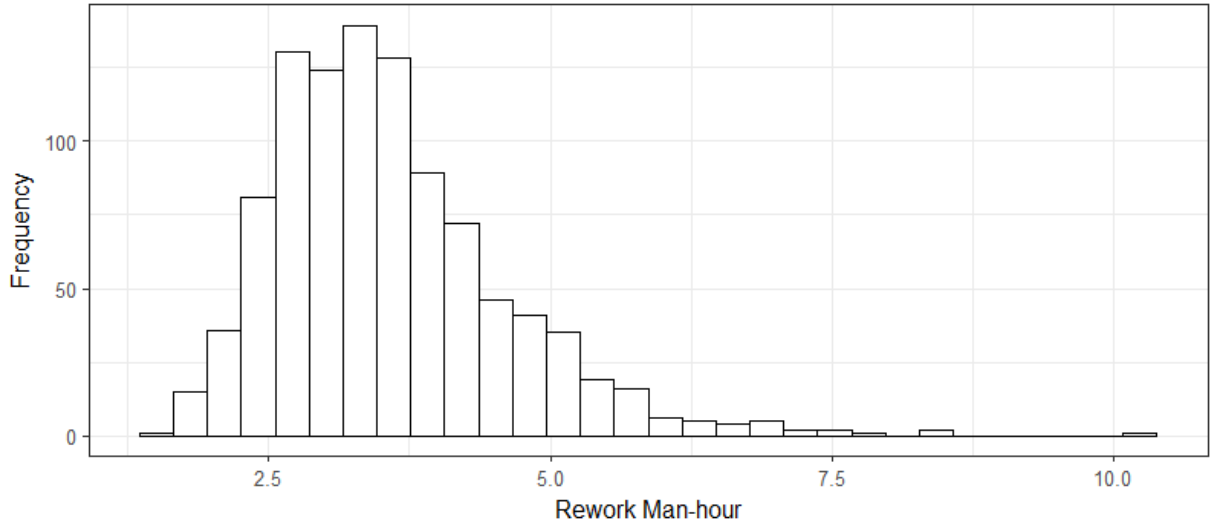

Figure 6.3: Frequency histogram of the total estimated rework man-hours.

Table 6.4: Statistical summary of the simulated histogram.

| Quantiles | 0% | 10% | 20% | 30% | 40% | 50% | 60% | 70% | 80% | 90% | 100% |
|---|---|---|---|---|---|---|---|---|---|---|---|
| Rework Man-hours | 1.4 | 2.4 | 2.7 | 2.9 | 3.2 | 3.4 | 3.6 | 3.9 | 4.3 | 4.9 | 10.2 |

From the simulated outcome, practitioners can increase their understanding of rework cost uncertainty, which can, in turn, improve overall estimation performance for bidding and cost control decision-support processes.

***Rework Cost Control***

During the project execution stage, uncertainty in rework man-hour estimation can be eliminated by incorporating the tracked, actual rework cost performance. For this illustrative example, the actual man-hours (Table 6.2) and inspection results (Table 6.3) for each weld are utilized to perform the rework cost control analysis. Following Eq. (6.13), control charts, as shown in Figure 6.4, are constructed for three different scenarios, namely, no rework, rework under control, and rework over control. The rework man-hour estimation is updated in each state, which is represented by an error bar (2.5% and 97.5% quantiles) with the median value (50% quantile) marked. As the project proceeds to completion, the uncertainty (i.e., error bar) of the estimated rework man-hours contracts towards the median value. In the final state, the final rework man-hours are the actual number determined from the accumulation of all, actual rework man-hours.

For scenario one, no rework was required for any of the ten welds. As expected, the rework man-hour decreases continuously to zero [Figure 6.4(a)]. From state 2, the estimated man-hour decreases below LCL, which indicates that the actual prefabrication rework performance is improved compared to the historical, performance-based forecast. The absence of any rework is used for illustrative purposes; it is an ideal condition and is not realistic in practice.

For scenario two, three welds are reworked (weld 4, 6 and 10). The rework man-hour values fluctuate within the control area [Figure 6.4(b)], indicating that the actual prefabrication process is under control, and no specific action is required.

For scenario three, due to the consecutive rework for weld 4, 5 and 6, the forecasted rework man-hours extend beyond the UCL at state 6 [Figure 6.4(c)]. The anomaly is detected and marked as a red dot at state 6. Although actions have been taken to improve rework performance, the forecasted rework man-hours are still higher than UCL at state 7 and 8. After that, the forecasted rework man-hours return within the control area. In real projects, such anomalies could arise as a consequence of unskilled operators, change of machines, or/and modified work procedures.

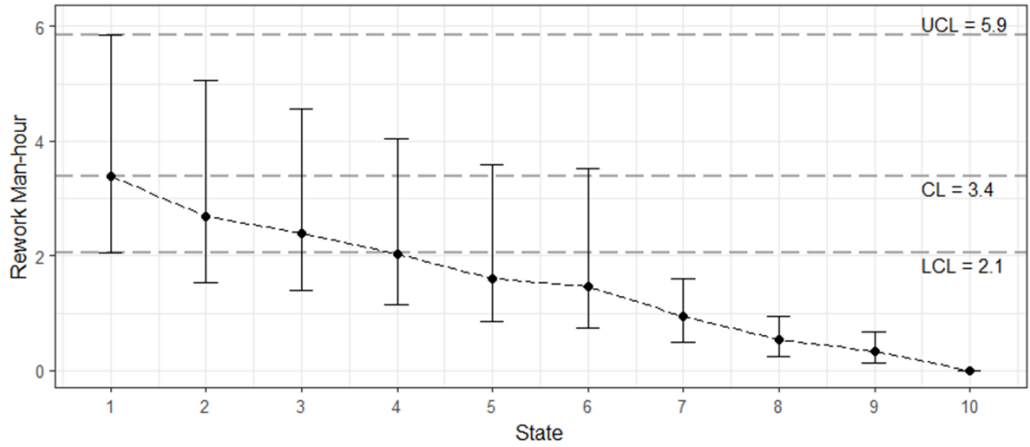

(a)

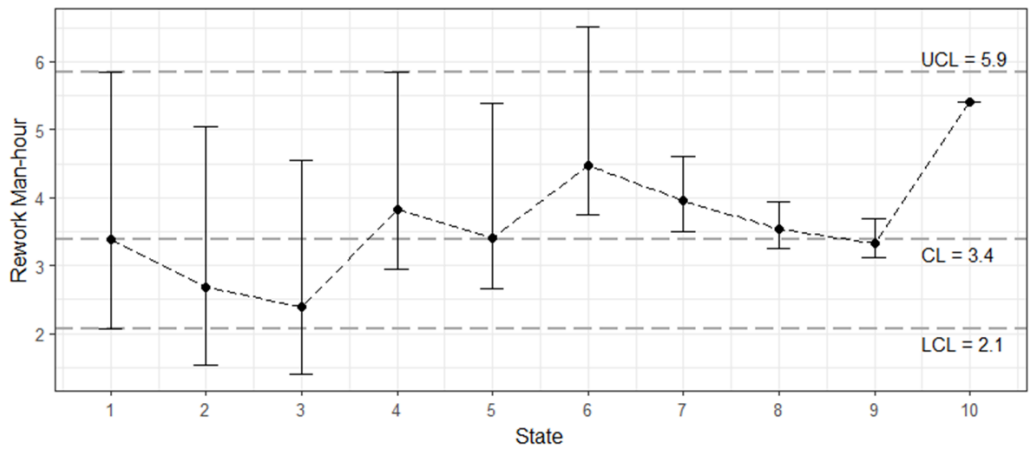

(b)

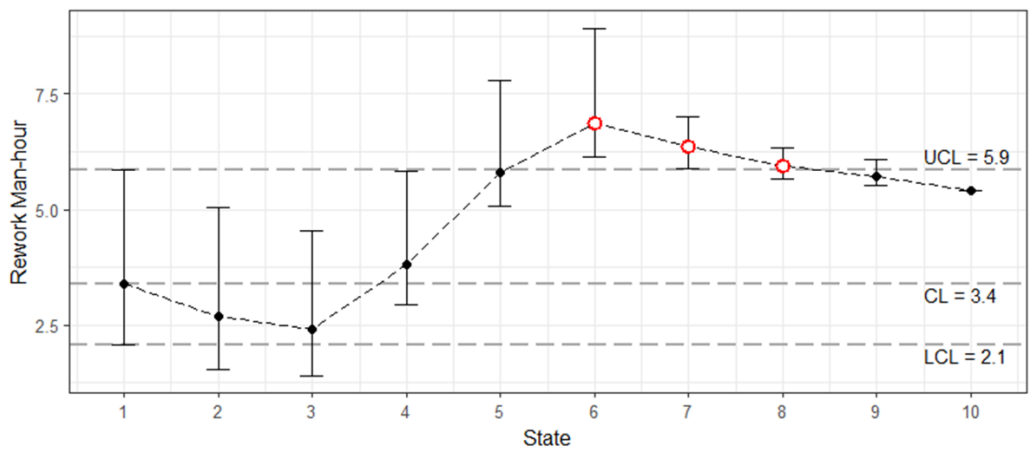

(c)

Figure 6.4: Rework man-hours control charts for three scenarios: (a) no rework, (b) rework under control, and (c) rework out of control.

This decision-support metric provides real-time updated rework cost prediction and monitoring, which can be utilized to detect issues during the product prefabrication process. It also provides a visualized tool for practitioners to further understand and investigate root causes of rework cost overrun.

## 6.7  Case Study

To validate the feasibility and applicability of the proposed data-driven simulation model, the pre-developed simulation-based analytics framework in Chapter 3 has been utilized as the simulation environment to incorporate the newly created absorbing Markov chain simulation model and decision-support metrics, as shown in Figure 6.5. The cost management system has been added into the data source module, while the data analysis and simulation modules have been reprogrammed using R to facilitate decision-support module function. A data adapter is developed to facilitate the processes of data connection, data wrangling, and data cleaning. This data adapter transforms multi-relational data into one, centralized dataset for further analytical processes. The data analysis module is comprised of algorithms that model fraction nonconforming, generate Bayesian-based input models, and create decision-support metrics. The simulation module utilizes input models for all types of products and creates simulated raw data for the establishment of decision-support metrics. Finally, the two types of decision-support metrics are generated by the decision-support module.

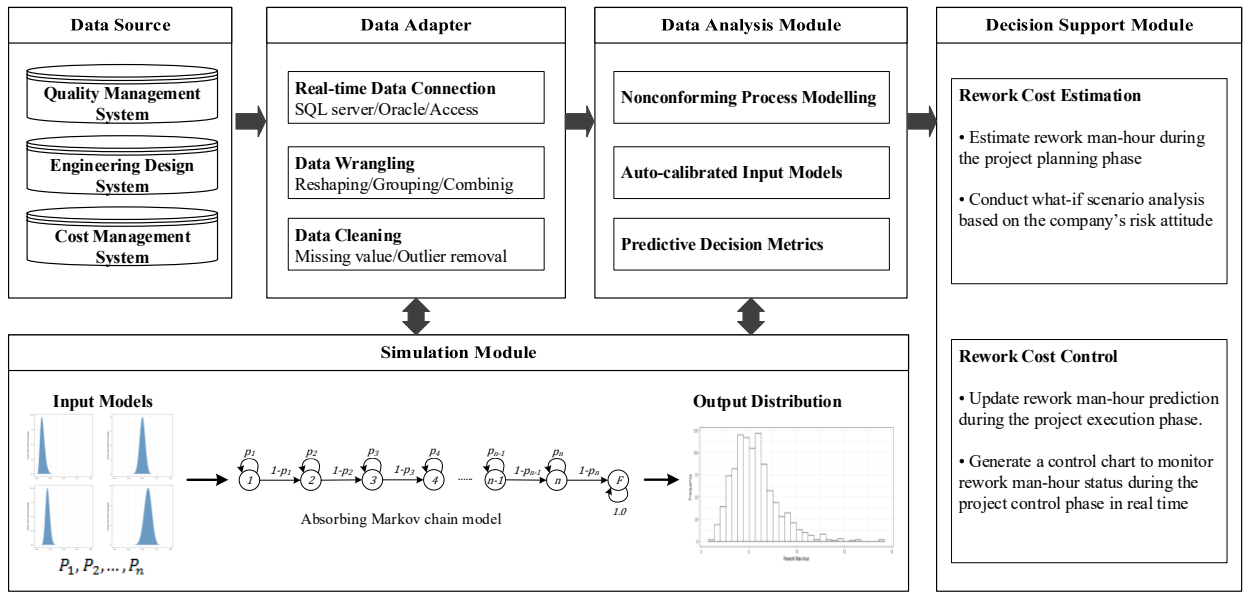

Figure 6.5: Simulation-based analytics for rework cost estimation and control.

An industrial pipe fabrication company in Edmonton, Canada, was studied to implement the proposed system. First, their engineering design system was used to extract pipe weld design attributes, such as NPS (i.e., outside diameter), pipe schedule (i.e., wall thickness), and material type for all welds. A combination of these design attributes was used to define a weld type; for instance, pipe (80, 4, C, BW) represents butt welds with NPS of 80, schedule of 4, and material C. Each combination is treated as a type of product in the proposed system. Then, required quality inspection and cost information were extracted from quality management (e.g., ArcuTrack) and cost management (e.g., Oracle JD Edwards) systems.

One historical pipe fabrication project from the studied company was selected to demonstrate the implementation of the proposed system. This particular project was chosen as it was considered an exceptional benchmark in terms of quality performance among historical projects. An improved rework cost performance, therefore, is expected for the selected project. For demonstration purposes, all 2,370 butt welds were used to perform the analytical analysis. From the cost management system, the original man-hour estimation of the project was 11,766 hours, while the actual rework man-hours were 1,164 hours. As per the quality inspection records, 96 out of 2,370 welds were reworked. Accordingly, the project fraction nonconforming is 96/2370 = 4.1%. The project design information was mapped to historical quality inspection results to derive the fraction nonconforming distributions for all types of welds. Using the modified simulation-based analytics framework, two types of decision-support metrics were generated.

### Rework Cost Estimation

Here, the rework cost of the selected project was estimated by running the specialized absorbing Markov chain model for 1,000 iterations. The running time for the entire automated analytical process was 218 seconds on a desktop with 3.20 GHz CPU and 24 GB random-access memory (RAM). The duration is acceptable for practical purposes. The simulated frequency histogram of estimated rework man-hours is a right-skewed, non-symmetrical distribution, as depicted in Figure 6.6. To quantitatively understand the simulation result, all 10% quantiles and the actual rework man-hours are summarized in

Table 6.5 for illustrative purposes. The median value is 1,554 hours. This simulated result can be used to determine the company's bidding strategy through a "what-if" scenario analysis. Once the appropriate quantile is selected, the corresponding rework man-hours can be derived and used to determine the rework-associated contingency for bid preparation.

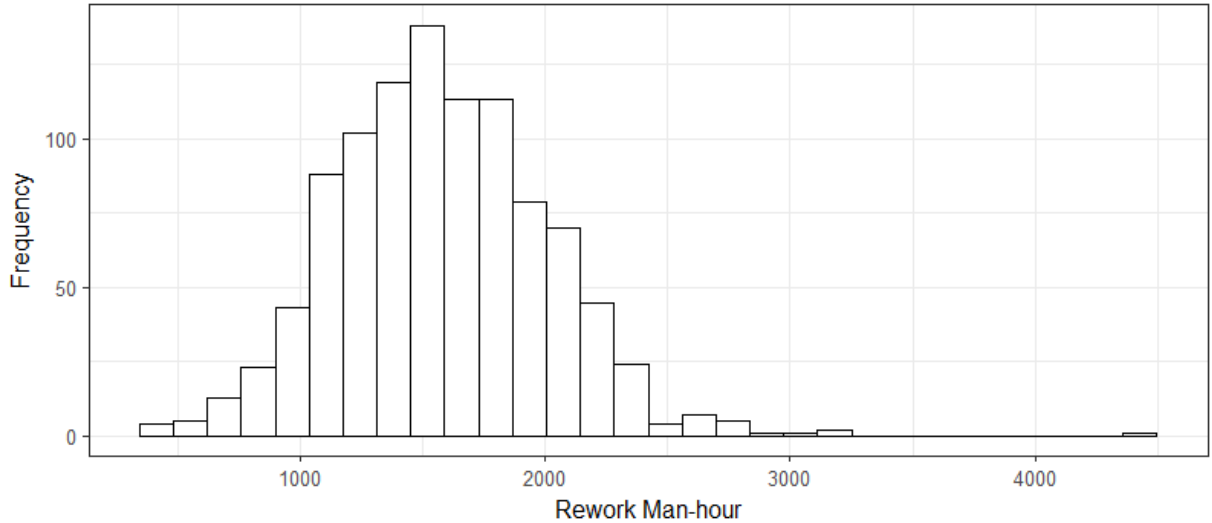

Figure 6.6: The frequency histogram of the total estimated rework man-hours of the pipe welding project.

Table 6.5: Summary of 10% quantiles of simulated rework man-hours and actual rework man-hours.

| | Simulated | | | | | | | | | | | Actual |
|---|---|---|---|---|---|---|---|---|---|---|---|---|
| | 0% | 10% | 20% | 30% | 40% | 50% | 60% | 70% | 80% | 90% | 100% | |
| Rework Man-hours | 394 | 1066 | 1209 | 1342 | 1457 | 1554 | 1666 | 1782 | 1923 | 2129 | 4405 | 1164 |

Actual rework man-hours were 1,164 hours and were within the 80% credible interval [1066, 2129] of the simulated rework man-hours. Compared to actual rework man-hours, the simulated median value was much greater than the actual rework man-hours, demonstrating that project quality performance was superior to the historical performance. From the median value, it can be concluded that 390 (1,554 − 1,164) man-hours were saved due to improved quality performance, in turn, contributing to increased project profitability. Notably, these man-hours can be transformed into a dollar value by incorporating operators' payment rates.

*Rework Cost Control*

Utilizing the estimated rework cost and actual rework cost information, a control chart of rework man-hours was constructed using following the propsed approach. All 2,370 production states are depicted in Figure 6.7 to demonstrate the dynamically updated rework man-hour predictions during the execution phase of the project. The solid black line represents the median values of the forecasted rework man-hours over states, while the grey area represents the 95% credible interval of the forecasted rework man-hours over states. Values of UCL, CL, and LCL are shown in the figure.

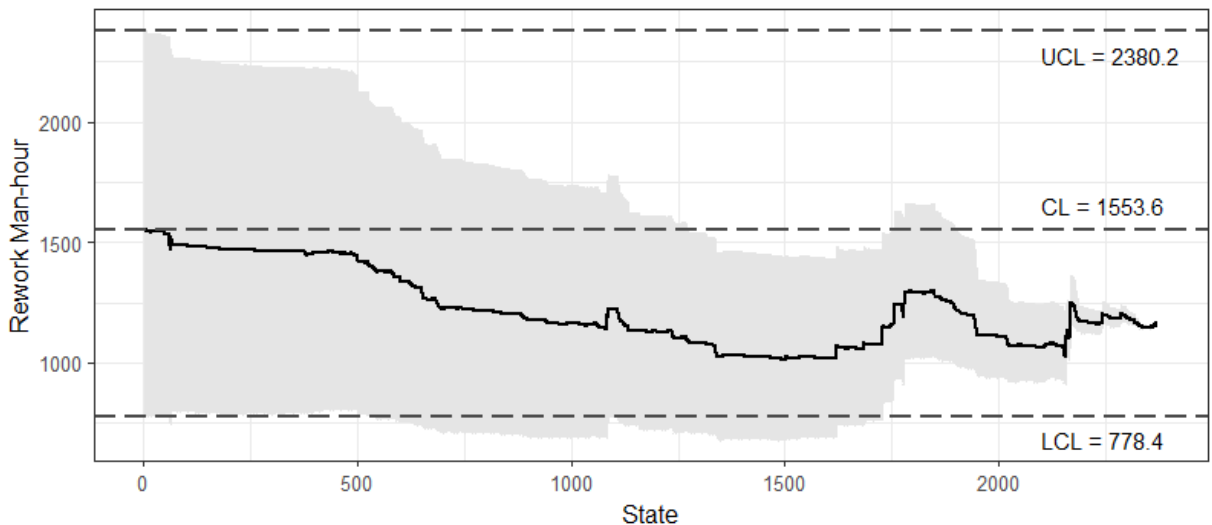

Figure 6.7: Rework man-hours control chart of the pipe welding project.

The overall trend of the state-dependent rework man-hour prediction decreases, and the credible interval for predicted rework man-hours converges towards zero, as the proportion of finished products increases. Two obvious increments, near states 1,750 and 2,150, are observed. Through the investigation of production dates and design information of these welds, the authors have hypothesized two root causes for these two increments. The period around state 1,750 included multiple consecutive rainy days. Professionals have confirmed that welding quality is directly impacted by the humidity of the welding environment. With regards to the increment around state 2,150, uncommon, specialty-shaped pieces were produced during that period, and welding operators may have lacked sufficient experience and skill for these types of welds.

## 6.8   Conclusion

The construction industry relies heavily on professionals' knowledge and experiences while performing cost estimation and control. Although many ERP systems have been widely implemented to facilitate data collection processes, practitioners still struggle with efficiently sourcing, interpreting, and transforming existing data into valuable information to support rework cost-related decision-making.

This research develops a data-driven simulation approach, which quantitatively addresses hard issues facing rework cost estimation and control by incorporating Bayesian-based quality performance measurements and a specialized absorbing Markov chain-based simulation model, to assist with decision-making regarding construction product prefabrication rework cost estimation and control. Utilizing the previously established Bayesian-based quality performance measurement, this approach achieves dynamic updating of model inputs, thereby providing more accurate and reliable decision-support metrics. Notably, the specialized model can be generalized for more complex production processes (e.g., CPM networks) to achieve rework cost estimation and control aims.

To better facilitate practical needs, two types of decision-support metrics are derived for serving rework cost estimation and control purposes. These metrics provide quantifiable and interpretable outcomes to augment the decision-making processes for practitioners. Companies can use this approach as a capable decision-support tool to perform rework cost estimation during the project planning phase and to monitor rework cost statuses during the project execution phase.

To assess the functionalities of the proposed analytical model, an illustrative example is demonstrated in detail. This novel model has also been integrated into the pre-developed simulation-analytics framework to support decision-making processes for a pipe fabrication company in Edmonton, Canada. This practical application has demonstrated the feasibility and applicability of the proposed approach by providing expected, reasonable decision metrics of a project that was determined to be a benchmark project in rework performance.

Although the proposed model can substantially enhance rework cost estimation and control practices, its accuracy and reliability largely depend on the quality and availability of companies' data. For instance, more detailed design information (e.g. shapes of pipe spools) is not well documented in the studied company. With the implementation of mature quality, design, and cost management systems, better modelling inputs will be available to enhance simulation performance. In the future, the ability of additional and alternative data sources to improve input modelling strategies and to generalize the absorbing Markov chain model for more complex production processes should be explored. Also, the computational efficiency should be enhanced by deploying advanced data mining techniques.

## 6.9    Acknowledgements

This research is funded by the NSERC Collaborative Research and Development Grant (CRDPJ 492657). The authors would like to acknowledge Rob Reid, Doug McCarthy, Jason Davio, and Christian

Jukna at Falcon Fabricators and Modular Builders Ltd. for sharing their knowledge and expertise of industrial pipe welding quality and cost management.

## 6.10 References

Abdul-Rahman, H. (1995). "The cost of non-conformance during a highway project: a case study." *Construction Management and Economics*, 13(1), 23–32.

Barber, P., Graves, A., Hall, M., Sheath, D., and Tomkins, C. (2000). "Quality failure costs in civil engineering projects." *International Journal of Quality & Reliability Management*, 17(4/5), 479–492.

Bishop, C. M. (2006). *Pattern Recognition and Machine Learning*. Springer.

Burati Jr, J. L., Farrington, J. J., and Ledbetter, W. B. (1992). "Causes of quality deviations in design and construction." *Journal of Construction Engineering and Management*, 118(1), 34-49. https://doi.org/10.1061/(ASCE)0733-9364(1992)118:1(34)

Hwang, B.-G., Thomas, S. R., Haas, C. T., and Caldas, C. H. (2009). "Measuring the impact of rework on construction cost performance." *Journal of Construction Engineering and Management*, 135(3), 187-198. https://doi.org/10.1061/(ASCE)0733-9364(2009)135:3(187)

Ji, W., and AbouRizk, S. M. (2017a). "Credible interval estimation for fraction nonconforming: Analytical and numerical solutions." *Automation in Construction*, 83, 56–67.

Ji, W., and AbouRizk, S. M. (2017b). "Simulation-based analytics for quality control decision support: a pipe welding case study." *Journal of Computing in Civil Engineering*, https://doi.org/10.1061/(ASCE)CP.1943-5487.0000755. [Accepted: Oct. 23, 2017]

Josephson, P.-E., and Hammarlund, Y. (1999). "The causes and costs of defects in construction: A study of seven building projects." *Automation in construction*, 8(6), 681–687.

Love, P. E. D. (2002). "Influence of project type and procurement method on rework costs in building construction projects." *Journal of Construction Engineering and Management*, 128(1), 18-29, https://doi.org/10.1061/(ASCE)0733-9364(2002)128:1(18)

Love, P. E. D., Edwards, D. J., Watson, H., and Davis, P. (2010). "Rework in civil infrastructure projects: determination of cost predictors." *Journal of Construction Engineering and Management*, 136(3), 275-282. https://doi.org/10.1061/(ASCE)CO.1943-7862.0000136

Montgomery, D. C. (2007). *Introduction to Statistical Quality Control*. John Wiley & Sons.

Potts, Keith, and Nii Ankrah. (2014). *Construction Cost Management: Learning from Case Studies*. Routledge.

Resnick, S. I. (2013). *Adventures in Stochastic Processes*. Springer Science & Business Media.

# 7    CHAPTER 7: CONCLUSION

## 7.1    Research Conclusions

This research outlines the development of a simulation-based analytics decision-support system for nonconforming quality-associated decision supports.

Chapter 2 proposes a Bayesian statistics-based analytical solution and an MCMC-based numerical solution to estimate the credible interval for fraction nonconforming. Both solutions provide a more accurate, reliable, and interpretable estimation of sampling uncertainty and can be used to improve the functionality of automated, nonconforming quality management systems. An industrial case study, from a pipe fabrication company in Alberta, Canada, is presented to demonstrate the feasibility and applicability of the proposed credible interval estimation methods.

Chapter 3 develops a quantitatively-driven analytics approach that allows simulation models to be adjusted by real-time data and measurements. The approach implements a Bayesian statistics-based fraction nonconforming estimation to recalibrate and realign models with real-time data, which are generated by actual quality control systems. The approach also develops descriptive and predictive analytical metrics, namely operator quality performance measurements and project quality performance forecasts, for supporting and improving decision-making processes. For practical purposes, a C#-based prototype is deployed to facilitate implementation at an industrial company in Edmonton, Canada.

Chapter 4 proposes an integrated, data-driven approach to quantify pipe welding operators' quality performances for industrial construction projects by implementing a Markov Chain

Monte Carlo (MCMC)-based fraction nonconforming estimation and an A/B testing algorithm. The existing quality management data and engineering design data from a pipe fabrication company in Edmonton, Canada, are processed and analyzed to demonstrate the feasibility and applicability of the proposed approach. Potential applications of the research findings are also discussed from the perspectives of production planning, employee training, and strategic recruiting.

Chapter 5 proposes a hybrid data mining approach to quantitatively analyze product complexity of prefabricated construction components from product nonconforming quality performance data. The proposed model is constructed in three steps, which (1) measure product complexity by introducing a Bayesian-based nonconforming quality performance indicator; (2) score each type of product complexity by developing a Hellinger distance-based distribution similarity measurement; and (3) cluster products into homogeneous complexity groups by using the agglomerative hierarchical clustering technique. An industrial company in Edmonton, Canada, is conducted to validate the feasibility and applicability of the proposed model.

Chapter 6 develops a data-driven simulation model to quantitatively assist decision support in quality-induced rework cost estimation and control for construction product prefabrication. Two types of decision-support metrics are developed to support decision-making processes, namely (1) rework cost estimation during the project planning phase and (2) rework cost control during the project execution phase. The proposed approach is also integrated into the previously developed simulation-based analytics framework and is implemented at an industrial pipe fabrication company in Edmonton, Canada.

## 7.2 Academic Contributions

The research outcomes have resulted in several academic contributions:

- Provision of a novel Bayesian-based approach for fraction nonconforming uncertainty modelling to address hard issues in updating input models of simulation models to realign with the dynamic, real-time data that is generated by the actual system.

- Advancement of conventional statistical quality control approaches using simulation-based techniques to provide predicting and updating functionalities.

- Provision of valuable insights regarding the implementation of MCMC-based numerical methods for complex and arbitrary probability distribution approximation, particularly, when closed-form solutions are difficult to derive or do not exist.

- Development of an integrated dynamic simulation environment that utilizes real-time data to enhance the predictability of simulation models in the construction domain. This system integrates the transformation of real-time data, the creation of analytical models, and the generation of decision-support metrics.

- Advancement of uncertain data clustering techniques using Hellinger distance-based distribution similarity measurement, which leads to improved computational efficiency for uncertain data clustering algorithms in the area of data mining.

- Definition of construction product complexity and the proposal of a systematic approach for analyzing product complexity using the indicator of product quality performance.

- Creation of a novel, absorbing Markov chain model for simulating construction product fabrication processes associated with rework. The proposed model has the potential to be generalized for more complex processes (e.g., CPM network).

## 7.3    Industrial Contributions

The industrial contributions from the collaborative research efforts are as follows:

- Development of a simulation-based analytics decision-support system to enhance several quality-associated decision-support processes, including project quality performance forecasting, operator quality performance measurement, construction product complexity determination, and rework cost management. Notably, the proposed system can be generalized for all nonconforming quality-related products across multiple industries.

- Creation of reliable and interpretable decision-support metrics at product-, project-, and operator-levels for quality performance measurement, construction product complexity assessment, and rework cost management. These metrics reduce the data interpretation load of practitioners, allowing users to discover valuable knowledge and information from existing data sources.

- This research also creates meaningful simulation results, which assist practitioners in performing quality and rework cost risk analyses during both the planning and execution phases of construction.

## 7.4    Research Limitations

Although the research findings in above chapters support the developed approaches, certain limitations of this research should be noted and explored.

- The proposed fraction nonconforming distribution and credible interval estimation techniques in Chapter 2 are limited by the assumption of a fixed prior distribution. Prior determination is a complex problem that requires incorporation of historical data, professional experience, and existing knowledge. To improve current computing performance, a systematic approach of determining the prior distribution should be further developed.

- The proposed simulation-based analytics system in Chapter 3 is limited to quality processes that utilize binary variables (i.e., fraction nonconforming process). Furthermore, while the proposed system can be used to inform decision-making processes, professional experience and knowledge are still required to ensure effective decisions are made.

- The developed operator performance measurement approach in Chapter 4 and the created product complexity analysis approach in Chapter 5 do not appropriately take other types of performance factors into account, such as productivity performance and safety performance. In future research, additional types of data should be investigated to improve the research work.

- The accuracy and reliability of the developed simulation model in Chapter 6 still largely depend on the quality and availability of companies' data. With the implementation of mature cost management systems (e.g., Oracle JD Edwards), better modelling inputs will be available to enhance simulation performance.

## 7.5   Envisioned Future Research

This section reveals the possible future directions based on this doctoral research work.

- To develop an efficient, data-driven approach to systematically devise strategic quality inspection plan (i.e., sampling size) for various product types and operators, and consequently improve fabrication quality management performance.

- To investigate the modelling of individuals' critical decision-making behaviors to incorporate impacts of human factors, thereby reducing simulated system variability.

- To investigate and incorporate additional types of data sources, such as safety management systems, productivity management systems, and human resource systems, to comprehensively enhance the functionalities (e.g. operator competency assessment and product complexity analysis) of the proposed simulation-based analytics decision-support system.

- To develop more integrated simulation models using distributed simulation technologies to systematically simulate the complex interactions between various project performance aspects (e.g. schedule, cost, quality, and safety) and critical impact factors (e.g. weather, labour, equipment, and human behavior).

- To address the issue of automating the process of model topology (i.e. structure) updating to reflect complex changes in the actual project execution processes, thereby reducing the modification efforts of practitioners and modellers.

# REFERENCES

Abdul-Rahman, H. (1995). "The cost of non-conformance during a highway project: A case study." *Construction Management and Economics*, Taylor & Francis, 13(1), 23–32.

AbouRizk, S., Hague, S., Ekyalimpa, R., and Newstead, S. (2016). "Simphony: A next generation simulation modelling environment for the construction domain." *Journal of Simulation*, 10(3), 207-215.

Antani, K. R. (2014). "A study of the effects of manufacturing complexity on product quality in mixed-model automotive assembly." Doctor of Philosophy (PhD), CLEMSON UNIVERSITY.

ASME (2005). *Process Piping : ASME Code for Pressure Piping, B31*, New York : American Society of Mechanical Engineers.

Baccarini, D. (1996). "The concept of project complexity—a review." *International Journal of Project Management*, 14(4), 201-204.

Baldwin, C. Y., and Clark, K. B. (2000). *Design Rules: The Power of Modularity*, MIT press.

Barrie, D. S., and Paulson, B. C. (1992). *Professional Construction Management: Including CM, Design-Construct, and General Contracting*, McGraw-Hill Science/Engineering/Math.

Barber, P., Graves, A., Hall, M., Sheath, D., and Tomkins, C. (2000). "Quality failure costs in civil engineering projects." *International Journal of Quality & Reliability Management*, 17(4/5), 479–492.

Barton, D., and Court, D. (2012). "Making advanced analytics work for you." *Harvard business review*, 90(10), 78-83.

Bassioni, H. A., Price, A. D., and Hassan, T. M. (2004). "Performance measurement in construction." *Journal of management in engineering*, 20(2), 42-50.

Battikha, M. (2002). "QUALICON: Computer-based system for construction quality management." *Journal of Construction Engineering and Management*, 10.1061/(ASCE)0733-9364(2002)128:2(164), 164-173.

Berger, J. O. (1985). *Statistical Decision Theory and Bayesian Analysis*, Springer Science & Business Media.

Bishop, C. (2006). *Pattern Recognition and Machine Learning*, Springer, New York.

Black, T. C., and Thompson, W. J. (2001). "Bayesian data analysis." *Computing in Science & Engineering*, CRC Press Boca Raton, FL, 3(4), 86–91.

Borwein, J. M., and Crandall, R. E. (2013). "Closed forms: What they are and why we care." *Notices of the AMS*, 60(1), 50-65.

Brown, L. D., Cai, T. T., and DasGupta, A. (2001). "Interval estimation for a binomial proportion." *Statistical Science*, 101-117.

Burati Jr, J. L., Farrington, J. J., and Ledbetter, W. B. (1992). "Causes of quality deviations in design and construction." *Journal of Construction Engineering and Management*, 118(1), 34-49. https://doi.org/10.1061/(ASCE)0733-9364(1992)118:1(34)

Carlo, M., Methods, S., Markov, U., and Applications, T. (1970). "Monte Carlo sampling methods using Markov chains and their applications." *Biometrika*, 57(1), 97–109.

Casella, G., and Berger, R. L. (2002). *Statistical Inference*, Duxbury Pacific Grove, CA.

Chan, A. P., Scott, D., and Chan, A. P. (2004). "Factors affecting the success of a construction project." *Journal of Construction Engineering and Management*, 130(1), 153-155.

Chen, L., and Luo, H. (2014). "A BIM-based construction quality management model and its applications." *Automation in Construction*, 46, 64-73.

Chin, S., Kim, K., and Kim, Y.-S. (2004). "A process-based quality management information system." *Automation in Construction*, 13(2), 241-259.

Chini, A. R., and Valdez, H. E. (2003). "ISO 9000 and the U.S. Construction Industry." *Journal of Management in Engineering*, 19(2), 69–77.

Darema, F. (2004). "Dynamic data driven applications systems: A new paradigm for application simulations and measurements." *Computational Science-ICCS 2004*, 662-669.

Dean, J. (2014). *Big Data, Data Mining, and Machine Learning: Value Creation For Business Leaders And Practitioners*, John Wiley & Sons.

De Veaux, R. D., Velleman, P. F., Bock, D. E., Vukov, A. M., and Wong, A. C. (2005). *Stats: data and models*, Pearson/Addison Wesley Boston.

Dube, P., Gonçalves, J. P., Mahatma, S., Barahona, F., Naphade, M., and Bedeman, M. (2014). "Simulation based analytics for efficient planning and management in multimodal freight transportation industry." *Proc., 2014 Winter Simulation Conf.*, IEEE, Piscataway, NJ,, 1943-1954.

Firouzi, A., Yang, W., and Li, C.-Q. (2016). "Prediction of Total Cost of Construction Project with Dependent Cost Items." *Journal of Construction Engineering and Management*, 10.1061/(ASCE)CO.1943-7862.0001194, 04016072.

Galvin, P., and Morkel, A. (2001). "The effect of product modularity on industry structure: the case of the world bicycle industry." *Industry and Innovation*, 8(1), 31.

Gelman, A., Carlin, J. B., Stern, H. S., and Rubin, D. B. (2014). *Bayesian Data Analysis*. CRC press.

Gui, H., Xu, Y., Bhasin, A., and Han, J. (2015). "Network A/B testing: From sampling to estimation." *Proceedings of the 24th International Conference on World Wide Web - WWW*

*'15*, ACM Press, 399–409.

Han, J., Pei, J., and Kamber, M. (2011). *Data Mining: Concepts and Techniques*, Elsevier.

Hastings, W. K. (1970). "Monte Carlo sampling methods using Markov chains and their applications." *Biometrika*, 57(1), 97-109.

Hellinger, E. (1909). "Neue Begründung der Theorie quadratischer Formen von unendlichvielen Veränderlichen." *Journal für die reine und angewandte Mathematik*, 136, 210-271.

Henclewood, D., Hunter, M., and Fujimoto, R. (2008). "Proposed methodology for a data-driven simulation for estimating performance measures along signalized arterials in real-time." *Proc., 2008 Winter Simulation Conf.*, IEEE, Piscataway, NJ,, 2761-2768.

Hitchcock, D. B. (2003). "A history of the Metropolis-Hastings Algorithm." *The American Statistician*, 57(4), 254-257.

Hoyle, D. (2001). *ISO 9000: Quality Systems Handbook*, Butterworth-Heinemann.

Huang, Y., and Verbraeck, A. (2009). "A dynamic data-driven approach for rail transport system simulation." *Proc., 2009 Winter Simulation Conf.*, IEEE, Piscataway, NJ, 2553-2562.

Hwang, B.-G., Thomas, S. R., Haas, C. T., and Caldas, C. H. (2009). "Measuring the impact of rework on construction cost performance." *Journal of Construction Engineering and Management*, 135(3), 187-198. https://doi.org/10.1061/(ASCE)0733-9364(2009)135:3(187)

Jaafari, A. (2000). "Construction business competitiveness and global benchmarking." *Journal of Management in Engineering*, 16(6), 43-53.

Ji, W., and AbouRizk, S. M. (2016). "A Bayesian inference based simulation approach for estimating fraction nonconforming of pipe spool welding processes." *Proc., 2016 Winter Simulation Conf.*, IEEE, Piscataway, NJ, 2935-2946.

Ji, W., and AbouRizk, S. M. (2017a). "Credible interval estimation for fraction nonconforming:

Analytical and numerical solutions." *Automation in Construction*, 83, 56–67.

Ji, W., and AbouRizk, S. M. (2017). "Implementing a data-driven simulation method for quantifying pipe welding operator quality performance." *Proc., 2017 Joint Conference on Computing in Construction (JC3)*, ITCdl, 79-86.

Ji, W., and AbouRizk, S. M. (2017b). "Simulation-based analytics for quality control decision support: a pipe welding case study." *Journal of Computing in Civil Engineering*, https://doi.org/10.1061/(ASCE)CP.1943-5487.0000755. [Accepted: Oct. 23, 2017]

Josephson, P.-E., and Hammarlund, Y. (1999). "The causes and costs of defects in construction: A study of seven building projects." *Automation in Construction*, 8(6), 681–687.

Jiang, B., Pei, J., Tao, Y., and Lin, X. (2013). "Clustering uncertain data based on probability distribution similarity." *IEEE Transactions on Knowledge and Data Engineering*, 25(4), 751-763.

Jun, D. H., and El-Rayes, K. (2011). "Fast and accurate risk evaluation for scheduling large-scale construction projects." *Journal of Computing in Civil Engineering,* 10.1061/(ASCE)CP.1943-5487.0000106, 407-417.

LaValle, S., Lesser, E., Shockley, R., Hopkins, M. S., and Kruschwitz, N. (2011). "Big data, analytics and the path from insights to value." *MIT Sloan Management Review*, 52(2), 21.

Liese, F., and Vajda, I. (2006). "On divergences and informations in statistics and information theory." *IEEE Transactions on Information Theory*, 52(10), 4394-4412.

Lu, M., and AbouRizk, S. M. (2000). "Simplified CPM/PERT simulation model." *Journal of Construction Engineering and Management*, 10.1061/(ASCE)0733-9364(2000)126:3(219).

Luo, L., He, Q., Jaselskis, E. J., and Xie, J. (2017). "Construction project complexity: Research trends and implications." *Journal of Construction Engineering and Management*, https://doi.org/10.1061/(ASCE)CO.1943-7862.0001306, 04017019.

Love, P. E. D. (2002). "Influence of project type and procurement method on rework costs in building construction projects." *Journal of Construction Engineering and Management*, 128(1), 18-29, https://doi.org/10.1061/(ASCE)0733-9364(2002)128:1(18)

Love, P. E. D., Edwards, D. J., Watson, H., and Davis, P. (2010). "Rework in civil infrastructure projects: determination of cost predictors." *Journal of Construction Engineering and Management*, 136(3), 275-282. https://doi.org/10.1061/(ASCE)CO.1943-7862.0000136

Metropolis, N., Rosenbluth, A. W., Rosenbluth, M. N., Teller, A. H., and Teller, E. (1953). "Equation of state calculations by fast computing machines." *The Journal of Chemical Physics*, 21(6), 1087-1092.

Moller, J., and Waagepetersen, R. P. (2003). *Statistical Inference and Simulation for Spatial Point Processes*, CRC Press.

Montgomery, D. C. (2007). *Introduction to Statistical Quality Control*, John Wiley & Sons.

Morey, R. D., Hoekstra, R., Rouder, J. N., Lee, M. D., and Wagenmakers, E.-J. (2016). "The fallacy of placing confidence in confidence intervals." *Psychonomic Bulletin & Review*, 23(1), 103-123.

Nicholson, B. J. (1985). "On the F-distribution for calculating bayes credible intervals for fraction nonconforming." *IEEE Transactions on Reliability*, R-34(3), 227-228.

Novak, S., and Eppinger, S. D. (2001). "Sourcing by design: Product complexity and the supply chain." *Management Science*, 47(1), 189-204.

O'Connor, J. T., O'Brien, W. J., and Choi, J. O. (2016). "Industrial project execution planning:

modularization versus stick-built." *Practice Periodical on Structural Design and Construction*, 21(1), 4015014.

Pei, J., Jiang, B., Lin, X., and Yuan, Y. (2007). "Probabilistic skylines on uncertain data." *Proc., Proceedings of the 33rd international conference on Very large data bases*, VLDB Endowment, 15-26.

Potts, Keith, and Nii Ankrah. (2014). *Construction Cost Management: Learning from Case Studies*. Routledge.

Resnick, S. I. (2013). *Adventures in Stochastic Processes*. Springer Science & Business Media.

Ripley, B., Lapsley, M., and Ripley, M. B. (2016). "Package 'RODBC'", <https://cran.r-project.org/web/packages/RODBC/index.html > (Sept. 19, 2017).

Robert, C., and Casella, G. (2011). "A short history of Markov Chain Monte Carlo: subjective recollections from incomplete data." *Statistical Science*, 102-115.

Robinson, D. (2017). *Introduction to Empirical Bayes: Examples from Baseball Statistics*.

Salem, O., Solomon, J., Genaidy, A., and Minkarah, I. (2006). "Lean construction: from theory to implementation." *Journal of Management in Engineering*, 22(4), 168-175.

Senescu, R. R., Aranda-Mena, G., and Haymaker, J. R. (2012). "Relationships between project complexity and communication." *Journal of Management in Engineering*, https://doi.org/10.1061/(ASCE)ME.1943-5479.0000121, 183-197.

Song, L., Wang, P., and AbouRizk, S. (2006). "A virtual shop modeling system for industrial fabrication shops." *Simulation Modelling Practice and Theory*, 14(5), 649-662.

Schubert, E., Koos, A., Emrich, T., Züfle, A., Schmid, K. A., and Zimek, A. (2015). "A framework for clustering uncertain data." *Proceedings of the VLDB Endowment*, 8(12), 1976-1979.

Shewchuk, J. P., and Guo, C. (2012). "Panel stacking, panel sequencing, and stack locating in residential construction: lean approach." *Journal of Construction Engineering and Management*, 138(9), 1006-1016.

Tierney, L. (1994). "Markov chains for exploring posterior distributions." *the Annals of Statistics*, 1701-1728.

Touran, A. (1993). "Probabilistic cost estimating with subjective correlations." *Journal of Construction Engineering and Management*, 10.1061/(ASCE)0733-9364(1993)119:1(58), 58-71.

Wang, P., Mohamed, Y., Abourizk, S., and Rawa, A. (2009). "Flow production of pipe spool fabrication: simulation to support implementation of lean technique." *Journal of Construction Engineering and Management*, 135(10), 1027-1038.

Weaver, B. P., and Hamada, M. S. (2016). "Quality quandaries: A gentle introduction to Bayesian statistics." *Quality Engineering*, 28(4), 508–514.

Weiss, N. A. (2012). *Introductory Statistics*, Pearson Education USA.

Willmott, C. J., and Matsuura, K. (2005). "Advantages of the mean absolute error (MAE) over the root mean square error (RMSE) in assessing average model performance." *Climate research*, 30(1), 79-82.

Wickham, H. (2014). "Tidy data." *Journal of Statistical Software*, 59(10), 1-23.

Wickham, H., Henry, L., and RStudio (2017). "tidyr: Easily Tidy Data with spread () and gather () Functions.", < https://cran.r-project.org/web/packages/tidyr/tidyr.pdf> (Sept. 14, 2017).

Wickham, H., Francois, R., Henry, L., Müller, K., and RStudio (2017). "dplyr: A grammar of data manipulation.", < https://cran.r-project.org/web/packages/dplyr/dplyr.pdf > (Sept. 14, 2017).

Williams, T. M. (1999). "The need for new paradigms for complex projects." *International journal of project management*, 17(5), 269-273.

Yates, J., and Aniftos, S. (1997). "International standards and construction." *Journal of Construction Engineering and Management*, 123(2), 127-137.

Züfle, A., Emrich, T., Schmid, K. A., Mamoulis, N., Zimek, A., and Renz, M. (2014). "Representative clustering of uncertain data." *Proc., Proceedings of the 20th ACM SIGKDD international conference on Knowledge discovery and data mining*, ACM, 243-252.

**APPENDIX A: DATA ADAPTER DEVELOPMENT**

In practice, most information cannot be adequately represented by independent data tables; rather, multiple types of objects are linked together through various types of linkages. Such data are usually stored in relational databases. While multi-relational databases can provide richer information for data mining, the transformation of relational data into a single table largely improves the efficiency of data mining. A data adapter integrates real-time information from various data sources into one, centralized dataset. This is particularly important for data that are collected from a variety of sources or databases.

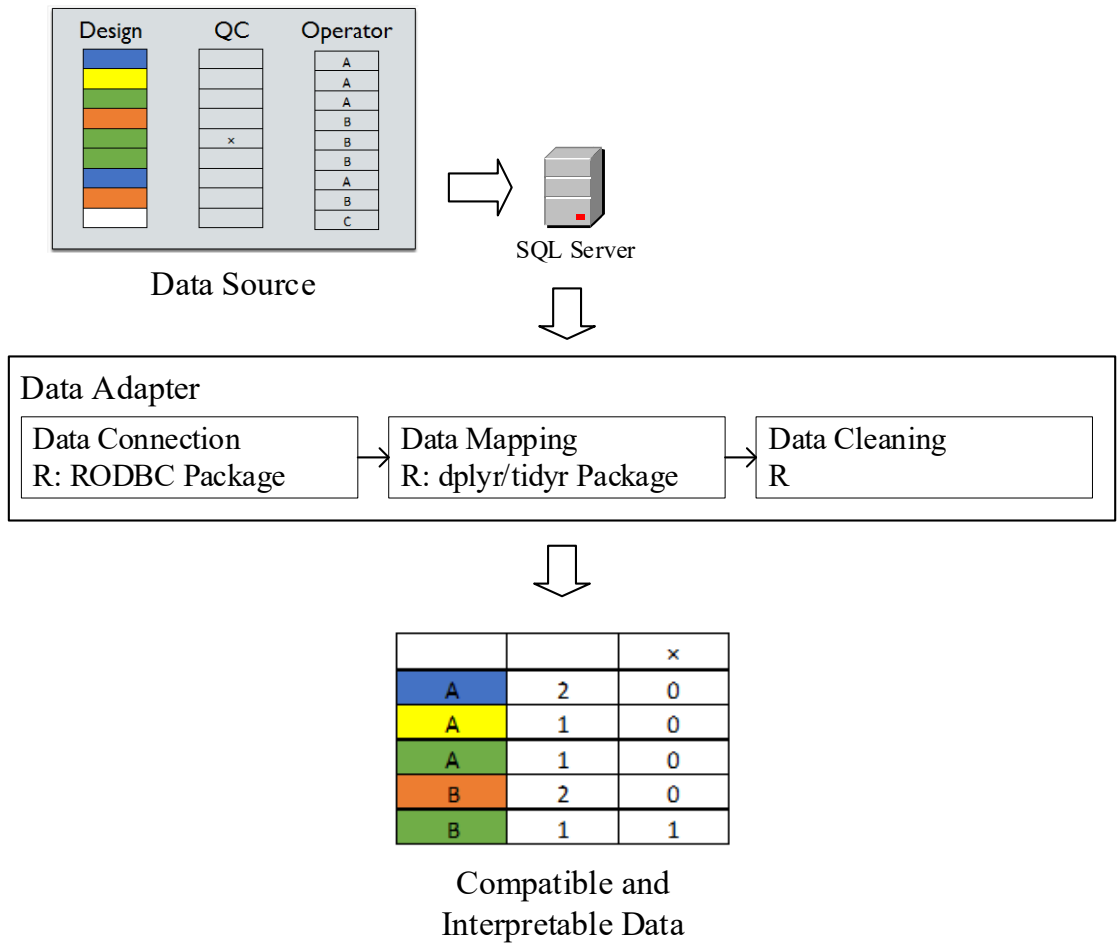

Figure A.1: The workflow of the developed data adapter.

In this research, a data adaptor was built using R for the purpose of multi-relational data connection, mapping, and cleaning tasks to generate a compatible and interpretable dataset, so it can be efficiently used in further data mining and simulation analysis. The developed data adapter is specific to the partner company involved in this research. As depicted in Figure A.1, isolated information, such as design attributes, quality inspection results, and operator information are stored in the company's SQL server. Through the developed data adapter, the isolated tables are integrated into a compatible table, which summarizes each operator's quality performance for a certain combination of design attributes. Here, detailed R codes for each data adapter function development is elaborated as follows.

### Data Connection

For real-time data connection, the R package for Open Database Connectivity (RODBC) was used to connect to the SQL Server of the quality management and engineering design systems. The RODBC package provides access to databases (including Microsoft Access, Microsoft SQL Server, or Oracle Database) through an Open Database Connectivity (ODBC) interface. RODBC has two group of functions: the main internal odbc* commands, which implement low level access to the ODBC functions, and the sql* functions, which operate at a higher level to read, save, copy, and manipulate data between data frames and SQL tables. The data connection through RODBC was periodically redone to ensure that accessed data was up-to-date, thereby achieving real-time computation. The programming code for performing data connection and high-level data manipulation is listed as follows.

```
library(RODBC)
#Connection to the Database
Data <- odbcConnect("DatabaseName")
#Use SQL to select columns from multi-relational tables
A <- sqlQuery(Data,
            "select wd.welder_id, pw.weld_abbreviation, ps.schedule,
pd.diameter_imperial,pmc.category,p.proj_type,dbo.check_weld_detail_inspected
(wd.weld_id,wd.detail_id,wd.proj_id,4) as inspection_result from welds w
            left join spools s on w.spool_id = s.spool_id and w.proj_id =
s.proj_id
            left join iso_dwgs iso on iso.iso_dwg_id = w.iso_dwg_id and
iso.proj_id = w.proj_id
            left join iso_dwg_main isom on isom.iso_dwg_main_id =
iso.iso_dwg_main_id and isom.proj_id = iso.proj_id
            left join line_classes lc on w.class_id = lc.class_id
            left join pipe_welds pw on w.weld_type_id = pw.weld_id
            left join pipe_schedules ps on ps.schedule_id =
w.pipe_schedule_id
            left join pipe_diameters pd on w.pipe_size_id = pd.diameter_id
            left join pipe_materials pm on w.material_id = pm.material_id
            left join weld_details wd on wd.weld_id = w.weld_id and
wd.proj_id = w.proj_id
            left join pipe_material_categories pmc on pm.category_id =
pmc.category_id
            left join pipe_material_groups pmg on pm.group_id = pmg.group_id
            left join bays b on b.bay_id = s.bay_id
            left join projects p on w.proj_id = p.proj_id
            left join employees e on wd.welder_id = e.empl_id
            left join weld_detail_inspections wdi on wdi.proj_id = wd.proj_id
and wdi.weld_id = w.weld_id and wdi.detail_id = wd.detail_id and
wdi.inspection_id = 4 and wdi.latest = 1")
```

The initially connected data (top 20 rows of 675,512 entries) is shown as Figure A.2. Welder
information and detailed design information are associated with each produced weld. In the next
step, this initial data is cleaned, reshaped and summarized to support further analytical tasks.

Figure A.2: The initially connected dataset.

### *Data Mapping*

The term "tidy data" has been used to refer to data that are maintained in a table form where each variable is saved in its own column and each observation is saved in its own row (Wickham 2014). Tidy datasets are easy to manipulate, model, and visualize (Wickham 2014). Here, the dplyr and tidyr packages were used to perform data wrangling tasks. By making use of multiple processors, these packages can perform data wrangling tasks in relatively little time, which is critical for processing large-sized datasets, such as the ones used in the present study (Wickham 2017; Wickham and Francois 2017). Here, the following code demonstrates the data mapping process for the functionality of operator quality performance measurement.

```
library(dplyr)
library(tidyr)
#Delete rows with NA and blank value
A <- na.omit(A)
A <- A[!A$schedule == "", ]
#Delete rows with inspection result values other than 0 and 1
A <- A[A$inspection_result == '0' | A$inspection_result == '1', ]
A <- mutate(A, Count = 1)
#Select welds with type of BW
A <- filter(A, weld_abbreviation == 'BW')
#Select welds from pipe shop
A <- filter(A, proj_type == '0')
#Treat Pipe Size as factor
A$diameter_imperial <- as.factor(A$diameter_imperial)
A <- dplyr::rename(A, Welder = welder_id, Schedule = schedule, PipeSize =
diameter_imperial, Material = category, ProjectType = proj_type, WeldType =
weld_abbreviation, InspectionResult = inspection_result)
B <- summarise_each(group_by(A, Welder, WeldType, Schedule, PipeSize,
Material, WeldType, InspectionResult),funs(sum))
#Replace numbers of inspection result with ABC
Status = data.frame(s = c("A", "B", "C"),v = c(0, 1, 2))
B$InspectionResult <- with(Status, s[match(B$InspectionResult, v)])
B <- tidyr::spread(B, InspectionResult, Count)
B[is.na(B)] <- 0  #Replace NA with 0
B <- mutate(B, TotalWelds = A + B + C, InspectedWelds = B + C, RepairedWelds
= C)
B <- select(B, Welder, Schedule, PipeSize, Material, TotalWelds,
InspectedWelds, RepairedWelds)
#Select weld attributes and welders
C <- filter(B, Schedule == 'STD')
C <- filter(C, PipeSize == '2')
C <- filter(C, Material == 'Material A')
C <- filter(C, InspectedWelds > 99)
C <- data.frame(C)
D <- select(C, Welder, InspectedWelds, RepairedWelds)
```

Here, the initially connected dataset is processed by performing the following cleaning and selection tasks. The transformed dataset A (top 20 rows of 224,294 entries) is shown as Figure A.3.

- Omit all rows with "NA" or " ";

- Delete rows with inspection results other than "0" and "1";

- Select "BW" as the weld type;

- Select "0" as the project type, which means fabrication project.

Figure A.3: The transformed dataset A.

Then, dataset A is processed to summarize the number of total welds, inspected welds, and required welds for each welder for a certain combination of design attributes. The summarized dataset B (top 20 rows of 9,709 entries) is depicted as Figure A.4.

Figure A.4: The transformed dataset B.

In consistent with Chapter 4, weld type (STD, 2, A, BW) is selected. Welders, who were inspected more than 100 times, were selected for quality performance measurement. The processed dataset C is represented as Figure A.5 (17 entries).

| | WeldType | Welder | Schedule | PipeSize | Material | TotalWelds | InspectedWelds | RepairedWelds |
|---|---|---|---|---|---|---|---|---|
| 1 | BW | 7 | STD | 2 | Material A | 289 | 208 | 9 |
| 2 | BW | 15 | STD | 2 | Material A | 261 | 147 | 5 |
| 3 | BW | 213 | STD | 2 | Material A | 400 | 120 | 6 |
| 4 | BW | 296 | STD | 2 | Material A | 579 | 355 | 11 |
| 5 | BW | 309 | STD | 2 | Material A | 150 | 119 | 8 |
| 6 | BW | 311 | STD | 2 | Material A | 353 | 100 | 8 |
| 7 | BW | 844 | STD | 2 | Material A | 606 | 264 | 9 |
| 8 | BW | 3404 | STD | 2 | Material A | 447 | 111 | 11 |
| 9 | BW | 4755 | STD | 2 | Material A | 318 | 175 | 25 |
| 10 | BW | 6739 | STD | 2 | Material A | 205 | 175 | 9 |
| 11 | BW | 8954 | STD | 2 | Material A | 181 | 123 | 5 |
| 12 | BW | 9211 | STD | 2 | Material A | 422 | 307 | 27 |
| 13 | BW | 11952 | STD | 2 | Material A | 261 | 207 | 16 |
| 14 | BW | 13051 | STD | 2 | Material A | 235 | 104 | 7 |
| 15 | BW | 13214 | STD | 2 | Material A | 176 | 175 | 11 |
| 16 | BW | 13937 | STD | 2 | Material A | 162 | 139 | 13 |
| 17 | BW | 13966 | STD | 2 | Material A | 674 | 316 | 16 |

Showing 1 to 17 of 17 entries

Figure A.5: The transformed dataset C.

Finally, the columns of "WelderID", "InspectedWelds", and "RepairedWelds" are selected as the finalized dataset for the purpose of welder quality performance measurement, as shown in Figure A.6.

Figure A.6: The finalized dataset D.

## Data Cleaning

Omitted values, noise, and inconsistencies render data inaccurate and unreliable. Data cleaning, which can be used to improve data accuracy and reliability, is an essential step of the proposed approach. Although many mining routines have developed procedures for dealing with these types of data, these procedures are not always robust (Han et al. 2011). Therefore, specific data cleaning rules should be specified for each particular case.

In this example, missing values are omitted by using the following code.

```
#Delete rows with NA and blank value
A <- na.omit(A)
A <- A[!A$schedule == "", ]
```

Inconsistencies in quality inspection records are removed by using the following code.

```r
#Delete rows with inspection result values other than 0 and 1
A <- A[A$inspection_result == '0' | A$inspection_result == '1', ]
```